\documentclass[aps, pre, twocolumn, superscriptaddress, showkeys, showpacs]{revtex4-1}

\newcommand{\Prob}{\mathbb{P}}

\usepackage{etex}
\usepackage[figuresright]{rotating}
\usepackage{amsmath}
\usepackage{amssymb}
\usepackage{bbm}
\usepackage{subfigure}
\usepackage{pstricks,pst-node,pst-plot}
\usepackage{color}
\usepackage{pst-all}
\usepackage{pspicture}
\usepackage{booktabs}
\usepackage{wasysym}
\usepackage[all]{xy}
\usepackage{epsfig,graphicx}
\usepackage{theorem}
\usepackage{natbib}
\usepackage{tikz}
\usepackage{algorithm}
\usepackage{algorithmic}
\usepackage{multirow}
\usepackage{array}
\usepackage{caption}
\usepackage[a4paper]{geometry}

\newcounter{count}
\newcounter{asscount}

\newenvironment{theorem}[1][Theorem \arabic{count}]{\vspace{1em}\refstepcounter{count}\begin{trivlist}
\item[\hskip \labelsep {\bfseries #1}]\em}{\end{trivlist}\vspace{1em}}

\newenvironment{lemma}[1][Lemma \arabic{count}]{\vspace{1em}\refstepcounter{count}\begin{trivlist}
\item[\hskip \labelsep {\bfseries #1}]\em}{\end{trivlist}\vspace{1em}}

\newenvironment{corollary}[1][Corollary \arabic{count}]{\vspace{1em}\refstepcounter{count}\begin{trivlist}
\item[\hskip \labelsep {\bfseries #1}]\em}{\end{trivlist}\vspace{1em}}

\newenvironment{example}[1][Example \arabic{count}]{\vspace{1em}\refstepcounter{count}\begin{trivlist}
\item[\hskip \labelsep {\bfseries #1}]}{\end{trivlist}\vspace{1em}}

\newenvironment{proof}[1][Proof]{\vspace{1em}\refstepcounter{count}\begin{trivlist}
\item[\hskip \labelsep {\bfseries #1}]}{\hfill$\Box$\end{trivlist}\vspace{1em}}

\newenvironment{remark}[1][Remark]{\vspace{1em}\refstepcounter{count}\begin{trivlist}
\item[\hskip \labelsep {\bfseries #1}]}{\end{trivlist}\vspace{1em}}

\begin{document}


\title{Mean Field Dynamics of Graphs I: Evolution of Probabilistic Cellular Automata for Random and Small-World Graphs}



\author{Lourens J. Waldorp}
\email[Corresponding author: ]{waldorp@uva.nl}
\author{Jolanda J. Kossakowski}
\affiliation{Department of Psychology\\
		 University of Amsterdam\\
		 Nieuwe Achtergracht 129-B\\
		 1018 XE Amsterdam \\
		 The Netherlands}

\date{\today}

\begin{abstract}
It was recently shown how graphs can be used to provide descriptions of psychopathologies, where symptoms of, say, depression, affect each other and certain configurations determine whether some- one could fall into a sudden depression. To analyse changes over time and characterise possible future behaviour is rather difficult for large graphs. We describe the dynamics of networks using one-dimensional discrete time dynamical systems theory obtained from a mean field approach to (elementary) probabilistic cellular automata (PCA). Often the mean field approach is used on a regular graph (a grid or torus) where each node has the same number of edges and the same probability of becoming active. We show that we can use variations of the mean field of the grid to describe the dynamics of the PCA on a random and small-world graph. Bifurcation diagrams for the mean field of the grid, random, and small-world graphs indicate possible phase transitions for certain parameter settings. Extensive simulations indicate for different graph sizes (number of nodes) that the mean field approximation is accurate. The mean field approach allows us to provide possible explanations of `jumping' behaviour in depression.
\end{abstract}


\maketitle


\section{Introduction}
In psychopathology sudden changes from 'normal' to depressed moods can occur. Such 'discontinuities' can be the result of a relatively small change in the environment or person. Recently, mental disorders have been described as a network of interacting symptoms \cite{Borsboom:2011,Borkulo:2014}, which provides a framework where an explanation for  such sudden changes in mood could be found. A mental  disorder can be viewed as a network of symptoms, each symptom influencing other symptoms. For instance, lack of sleep during the night could lead to poor concentration during the day, which in turn could lead to lack of sleep again by worrying that your job may be on the line. Here we use this idea and model the dynamics of psychopathology networks as probabilistic cellular automata. Then to analyse the dynamics we use a mean field approach where each node is similar in behaviour to all others. We extend known results for the mean field approach in this context to other types of graphs (random and small-world graphs), where the mean field can be interpreted as a weighted average of all nodes in the graph. We also give concentration inequalities for approximations using variations of the grid mean field.  

Cellular automata are discrete dynamical systems that have deterministic, local rules to move from one generation to the next \cite{Wolfram:1984b,Sarkar:2000}. Introduced by \citeauthor{Neumann:1951}, the most famous version is Conways game of life, popularised by \citeauthor{Gardner:1970}, and has found many applications from computer science \cite{Wolfram:1984} to neuronal population modelling \cite{Kozma:2005} to epdemiology \cite{Kleczkowski:1999}. In a cellular automaton each cell or node in a finite grid (usually a subset of $\mathbb{Z}^{2}$) can be 'active' or 'inactive' (1 or 0) and if, for instance, two (direct) neighbours are active, then the node will become active at the next time step. Another example of a cellular automaton is bootstrap percolation, where each node can only become active and cannot be inactivated by its neighbours, and the objective is to determine the initial configuration of active nodes that result in all nodes being active \cite{Janson:2012}. In general, a new generation in a cellular automaton is determined by a local and homogeneous update rule $\phi$. For each node $x$ in the graph this induces a sequence of states, an orbit. A (random) configuration at time 0 then determines whether all nodes in the network will be active, inactive, or whether the network will demonstrate periodic behaviour. A generalisation of a cellular automaton is to introduce a probability $p_{\phi}$ to decide whether or not a node will become active or not determined by a node's neighbours. One such rule is the majority rule which gives the probability to switch depending on whether the majority of its neighbours are active. Such a system is called a probabilistic cellular automaton (PCA). Here we will investigate the dynamic behaviour of the proportion of active nodes (density) for PCA with a majority rule that are defined on toroidal, random and small-world graphs. 

Many versions of PCA exist and of particular interest are those that behave similar to the Ising network. The reason is that the Ising network is often used to model realistic phenomena, like magnetisation \cite{Kindermann:1980,Sethna2004} or psychopathologies \cite{Borkulo:2014}. We have in mind the application to psychopathology here. In such systems the symptoms of disorders are the nodes in the graph and edges between the symptoms are estimated from data using the Ising mdoel \cite{Borkulo:2014} or from verbal accounts. In our companion paper in this issue we elaborate on real data analysis using results from discrete Markov chains. 

\citeauthor{Watts:1999} showed that a one-dimensional, large-scale cellular automaton (deterministic) where the connectivity between nodes was arranged as a small-world, could perform the density (all zeroes) and synchronisation (alternating all zeroes and all ones) tasks. \citeauthor{Newman:1999} gave approximations for path length and clustering on a small-world, to obtain an analytic solution to the threshold above which a large number of nodes are active. \citeauthor{Callaway:2000} also studied percolation in different graph topologies in deter, focussing on the consequences of (randomly) deleting nodes. Here again the objective was to concentrate on stable solutions of the graphs. 
 In a probabilistic version, \citeauthor{Tomassini:2005} investigated a one-dimensional PCA on a regular and small-world graph in terms of its performance on the density and synchronisation tasks. They determined by using evolutionary algorithms that a small-world topology is most efficient to solving both tasks, corresponding to the results of \citeauthor{Watts:1999} in a deterministic version. Their objective was different from ours in that here we are interested in all types of dynamic behaviour (stable or not), and specifically representing this behaviour for the PCA by the mean field. 

Our starting point is the work by \citeauthor{Balister:2006} and \citeauthor{Kozma:2005} where a two-dimensional (toroidal) grid on a PCA is defined. The mean field is then used to determine the unconditional probability distribution of the density (relative number of active nodes).
\citeauthor{Balister:2006} show that the mean field model predicts a bifurcation for small values of the probability of a node switching to another state and determine its critical point for a neighbourhood of size five \cite[see also][]{Kozma:2004,Kozma:2005}. This is of particular interest in our case as it may explain mood disorders (e.g., depression or manic-depression) from symptoms and their connectivity. To apply these results to random and small-world graphs we determine the marginal distribution across the possible degree probabilities given the topology of the random or small-world graph. Extending results of homogeneous graphs has been applied to social networks \cite{Barrat:2008} and to cellular automata \cite{Janson:2015}.

We first introduce probabilistic cellular automata in Section \ref{sec:pca}. Then in Section \ref{sec:mf} we show how the traditional version of a PCA on a grid can be reduced to a single discrete time dynamical system, called the mean field. In Section \ref{sec:mfrg} we show that for the random graph we can use a variation on the formulation for the grid of the dynamical system to describe dynamics. We use these results on the random graph to show in Section \ref{sec:mfsw} that we can obtain a similar approximation for the small-world graph, again using the formulation for the grid. Having shown that these approximations are appropriate, we see in Section \ref{subsec:dynamics} what the dynamics of the process is for the different topologies. We follow these theoretical results by extensive simulations to verify the accuracy of the mean field in Section \ref{sec:numerical}.

\section{Probabilistic cellular automata}\label{sec:pca}
A cellular automaton is a dynamical system of nodes in a fixed, finite grid where directly connected nodes determine the state of a node at each subsequent time step \cite{Wolfram:1984}. Each node $x$ in a node set $V=\{1,2,\ldots,n\}$ is at time $t$ in one of the states of a finite alphabet $\Sigma$. The nearest neighbours in the graph $G=(V,E)$ are given by the edges in $E$. Often the graph $G$ is the square lattice $\mathbb{Z}^{2}$, where each node has exactly four neighbours \cite[see e.g., ][]{Grimmett:2010}. A local rule determines based on the direct neighbours what the value of the alphabet of node $x\in V$ will be at time $t+1$. Let the neighbourhood of $x$ be the set of nodes that are directly connected to $x$, $\Gamma(x)=\{y\in V: (x,y)\in E\}$.  A local rule $\phi: \Gamma \to \Sigma$ assigns for each configuration of the neighbourhood of $x$ a value $a\in \Sigma$. If we additionally introduce a probability for each local configuration $\phi$, we obtain a probabilistic cellular automaton (PCA). The probability is a function $p_{a}: \Sigma^{|\Gamma|}\times \Sigma \to [0,1]$ such that a probability is assigned to each node $x$ to have label $a$ for a configuration $\phi$ dependent on the neighbourhood $\Gamma(x)$, with $\sum_{a\in\Sigma}p_{a}=1$. The local rule $\phi$ is applied iteratively to each result, and hence induces a stochastic process with sequence $\Phi_{t}:V\to \Sigma$ for each time step $t$. For node $x\in V$ we write $\Phi_{t}(x)=\phi^{t}(x_{0})$, where $x_{0}$ is the value at time $t=0$, and $\Phi_{t}(V)$ is the image for all nodes simultaneously. Each node therefore has an orbit \cite{Hirsch:2004}, which is the sequence $(\Phi_{t}(x),t\ge 0)=(x_{0},\phi(x_{0}),\phi(\phi(x_{0})),\ldots)$.

\begin{example}\label{ex:majority} (Majority rule on 0/1) 
Let the alphabet be $\Sigma=\{0,1\}$ and take a finite subset of the square lattice $V\subset \mathbb{Z}^{2}$. In this lattice $V$ each node has 4 (nearest) neighbours. We use the majority rule, which says that if $|\phi_{x}^{-1}(1)| = |\{y\in \Gamma(x): \Phi_{t}(y)=1\} |$ is greater than $2$, then the node $x$ will be 1 with probability $1-p$, and otherwise 0 with probability $p$. The sequence $(\Phi_{t}(x),t\ge 0)$ is any orbit of 0s and 1s $(0,1,1,\ldots)$.
\end{example}

\noindent
In Example \ref{ex:majority} the probability $p$ determined by the neighbourhood $\Gamma(x)$ is independent of the state of the node $x$ itself. Such a model is called {\em totalisitc} \cite{Balister:2006}. Additionally, the probabilities for 0 and 1 were defined by the same parameter $p$, which is then called {\em symmetric}, i.e., $p_{1}=1-p_{0}$. Here we focus  on totalistic and symmetric models with size 2 alphabet $\Sigma=\{0,1\}$.

\section{Mean field approximation on graphs}\label{sec:mf}\noindent
The key ingredient of the mean field approximation, shown by \citeauthor{Balister:2006}, is that the properties of interest are uniform over the graph. For the (toroidal) grid topology this is easy to see: Any node $x\in V$ has the same number of neighbours $|\Gamma|=4$, where each node in the neighbourhood becomes 0 or 1 by the same local rule. It follows that any four nodes in the grid could serve as part of the neighbourhood for $x$. In a probabilistic automaton, therefore, the local rule depends only on the number of 1s in any random draw of $4$ nodes from all nodes $V$. We first consider the case for a grid and then move onto the random and small-world graph.

\subsection{Mean field on a grid}\label{sec:mf-grid}\noindent
Let the graph $G_{\rm grid}(n,\Gamma)$ be a grid with $n$ nodes and boundary conditions such that each node has exactly four neighbours. We consider the density $\rho_{t}$ defined by $|\Phi_{t}^{-1}(1)|/n$, where the set of nodes that are 1 is $\Phi_{t}^{-1}(1)=\{ y\in V: \phi^{t}(y)=1 \}$. It follows that we require the probability of $\{\Phi_{t}(x)=1\}$ given a certain number of nodes in state 1 at the previous time point. 
The probability of switching to state 1, $\xi_{|\Gamma|}(r)$, is conditional on $r$ of the neighbours that are $1$. Let $|\phi^{-1}(1)|=|\{ y\in \Gamma:\phi(y)=1 \}|$ be the number of 1s in the neighbourhood of $x$. Then we define the probability of state 1 given that $r$ neighbours are 1 as
\begin{align}\label{eq:pr1}
\xi_{|\Gamma|}(r)=\Prob(\Phi_{t}(x)=1\mid |\phi^{-1}(1)|=r)
\end{align}
One of the possibilities to define $\xi_{|\Gamma|}$ is the majority rule: if the majority of neighbours in $\Gamma$ are 1, then the node will be 1 with probability $1-p$ at $t+1$. The majority rule is defined as
\begin{align}\label{eq:psym}
\xi_{|\Gamma|}(r)=
\begin{cases}
p	&\text{if } r\le |\Gamma|/2\\
1-p &\text{if } r> |\Gamma|/2
\end{cases}
\end{align}
To obtain the probability of $\Phi_{t}(x)=1$, the state of $x$ being in state 1, we need to determine the probability of a neighbourhood having $r$ active nodes. In the mean field we have that the probability of a 1 is homogeneous and so we obtain a binomial distribution for the number of 1s in the neighbourhood $\Gamma$. The probability that the number of nodes in the neighbourhood  equals $r$ is in the mean field the same as $|\Gamma|$ Bernoulli trials each with probability of success $\rho_{t}$. Hence
\begin{align}\label{eq:pr1}
\Prob(|\phi^{-1}(1)|=r\mid \rho_{t})=
\binom{|\Gamma|}{r}\rho_{t}^{r}(1-\rho_{t})^{|\Gamma|-r}
\end{align}
where $r=0,1,\ldots,|\Gamma|$. Then the probability of the event $\{ \Phi_{t}(x)=1 \mid \rho_{t}\}$, that node $x$ will have state 1 at time $t+1$ given the density $\rho_{t}$ at time $t$, is
\begin{align}\label{eq:pr1}
p_{\rm grid}(\rho_{t})=
\sum_{r=0}^{|\Gamma|}\xi_{|\Gamma|}(r)\binom{|\Gamma|}{r}\rho_{t}^{r}(1-\rho_{t})^{|\Gamma|-r}.
\end{align}
Let 
\begin{align*}
q_{|\Gamma|/2}(\rho_{t})=\sum_{r=0}^{|\Gamma|/2}\binom{|\Gamma|}{r}\rho_{t}^{r}(1-\rho_{t})^{|\Gamma|-r}
\end{align*}
Combining this probability with the majority rule (\ref{eq:psym}) gives
\begin{align*}
p_{\rm grid}(\rho_{t})=
pq_{|\Gamma|/2}(\rho_{t})
+ (1-p)\left(1-q_{|\Gamma|/2}(\rho_{t})\right)
\end{align*}
which is used in \citeauthor{Kozma:2005}. This mean field result follows directly from Theorem 2.1 in \citeauthor[][]{Balister:2006}. Let $B(n,p)$ denote a binomial random variable with $n$ Bernoulli trials each with success probability $p$.
\begin{lemma}\label{cor:balister}\cite[][Theorem 2.1]{Balister:2006}
Let $G_{\rm grid}(n,\Gamma)$ be a grid with a PCA as defined above with $\xi_{|\Gamma|}(r)$ according to the majority rule in (\ref{eq:psym}). Then the evolution of the number of active nodes $n\rho_{t}$ is
\begin{align}\label{eq:rho-evo}
n\rho_{t+1} = B(n,p_{\rm grid}(\rho_{t})).
\end{align}
The mean and variance for the density $\rho_{t}$, respectively, $\mu_{\rm grid}=p_{\rm grid}$ and $\sigma^{2}_{\rm grid}=p_{\rm grid}(1-p_{\rm grid})/n$.
\end{lemma}
It follows that the conditional probability of $n\rho_{t+1}$ active nodes in the grid on $\{n\rho_{t}=k\}$ at the previous time step is 
\begin{align}\label{eq:trans-prob-grid}
&\Prob(n\rho_{t+1}=r\mid n\rho_{t}=k)=\notag \\
&\qquad \binom{n}{r}p_{\rm grid}(\rho_{t})^{r}(1-p_{\rm grid}(\rho_{t}))^{n-r}
\end{align}
It is easily seen that this is a discrete time Markov process on a finite state space of size $n$, since the probability of $n\rho_{t+1}$ depends only on $\rho_{t}$. Equation (\ref{eq:trans-prob-grid}) is the transition probability of the discrete time Markov process for the number of active nodes in the graph. 

Because $\phi_{t}(x)$ is Bernoulli distributed $B(1,p_{\rm grid}(\rho_{t}))$ for all nodes $x\in V$, and the number of active nodes $|\Phi^{-1}(1)|$ is the sum of these Bernoulli trials, we can apply the law of large numbers so that for large $n$, $\rho_{t+1}$ is close to $\mu_{t}:=\mu(\rho_{t})$ with high probability. Indeed, we can use Chernov's bound to suggest that using $p_{\rm grid}$ is good enough for large graphs. 
\begin{lemma}(Accuracy bound of density)\label{lem:chernov}
Let $n\rho_{t}=\sum_{x\in V}\phi^{t}(x)$ be the sum of $n$ Bernoulli trials given by (\ref{eq:rho-evo}), with mean of the density $p_{\rm grid}(\rho_{t})$. For every $0< \varepsilon<\min\{p_{{\rm grid}},1-p_{\rm grid}\}$, let $\delta=2\exp(-\varepsilon^{2}/2\sigma^{2}_{{\rm grid}})$. We then have with probability at least $1-\delta$
\begin{align}
|\rho-p_{{\rm grid}}|\le \sqrt{\frac{p_{\rm grid}(1-p_{\rm grid})}{n}2\log(2/\delta)}
\end{align}
\end{lemma}
A proof is in the Appendix. So we can use the mean field $p_{\rm grid}(\rho_{t})$ for grids of large size $n$. With $\delta=0.05$, we obtain the interval with probability at least 0.95 of $[\mu_{\rm grid}-2.72\sigma_{\rm grid},\mu_{\rm grid}-2.72\sigma_{\rm grid}]$. Another interval can be obtained from the DeMoivre-Laplace central limit theorem. This theorem tells us that for large enough $n$, $z_{\rm grid}=(\rho_{t}-p_{{\rm grid}})/\sigma_{{\rm grid}}$ is distributed as $N(0,1)$. In fact, if the third order moment of $z_{\rm grid}$ is $c<\infty$, the Berry-Esseen theorem says that the order of approximation of the distribution of $\rho_{t}$ to the normal distribution is $O(3c/\sqrt{n})$ \cite{Venkatesh:2013}. This provides an interval for $\rho_{t+1}$ as a measure of accuracy with $[\mu_{{\rm grid}}-1.96\sigma_{{\rm grid}},\mu_{{\rm grid}}+1.96\sigma_{{\rm grid}}]$ with probability $0.95$.
Clearly, in both limit laws the size of the network $n$ determines the accuracy of the approximation.

\subsection{Mean field on a random graph}\label{sec:mfrg}
In the original setting of a grid (with boundary conditions, so a torus) the number of neighbours is fixed and it was seen that the mean field approximation $p_{\rm grid}$ was accurate for the density because each node is identical with respect to a change depending on its neighbours. Here we introduce the neighbourhood size $|\Gamma|$ as a random variable and then determine the probability of $\Phi_{t}(x)=1$ given $\rho_{t}$ by averaging over all possible sizes of neighbourhoods weighted by its probability for neighbourhood size.
This is done in a random graph where each node has a binomial number of neighbours. 
Let $G_{\rm rg}(n,p_{e})$ be a random graph with $n$ nodes and (constant) probability $p_{e}$ of an edge being present \cite{Bollobas:2001,Durrett:2007}. Let the size of the neighbourhood $|\Gamma|$ be a binomial random variable with maximal value $n-1$  neighbours and probability $p_{e}$, that is, $B(n-1,p_{e})$ and 
\begin{align*}
\Prob_{p_{e}}(|\Gamma|=k)=\binom{n-1}{k}p_{e}^{k}(1-p_{e})^{n-k-1}.
\end{align*}
Then the probability of obtaining an active node can be defined conditionally on the event $\{|\Gamma|=k\}$, the neighbourhood having size $k$. Then marginalising over the possible the neighbourhood size, we obtain $p_{\rm rg}$ for the probability of a node being active in the binomial process. Proofs can be found in the Appendix.

\begin{lemma}(Probability on a random graph)\label{lem:prg}
Consider a PCA on a random graph $G_{\rm rg}(n,p_{e})$ with edge probability $p_{e}$ and a local rule $\phi$ determined by the majority rule for the neighbourhood $|\Gamma|=k$ according to $\xi_{k}$ as in (\ref{eq:pr1}). Then the probability of obtaining an active node at time $t+1$ is
\begin{align}\label{eq:pr-rg}
&p_{\rm rg}(\rho_{t}) =\notag \\ 
&\sum_{k=0}^{n-1}\sum_{r=0}^{k}\xi_{k}(r)\binom{k}{r}\rho_{t}^{r}(1-\rho_{t})^{k-r}\Prob_{p_{e}}(|\Gamma|=k).
\end{align}
\end{lemma}
Intuitively we would expect that if we restrict the neighbourhoods in (\ref{eq:pr-rg}) to the expected neighbourhood size $p_{e}(n-1)$ for each node, then the approximation should be reasonably close. This would make it possible to analytically determine fixed points more easily and simplify computation for large graphs considerably. We next show that this is a reasonable approach. 
\begin{lemma}(Simple probability on a random graph)\label{lem:prg-ext}
If the neighbourhood in $G_{\rm rg}(n,p_{e})$ of each node in a PCA as in Lemma \ref{lem:prg} is fixed with the expected number of nodes under the random graph $\nu=\lfloor p_{e}(n-1)\rfloor$ such that $k=\nu$ in $p_{\rm grid}$, then the probability $p_{\rm rg}$ reduces to
\begin{equation}\label{eq:pr-rand}
p_{\rm grid}^{\nu}(\rho_{t})= \sum_{r=0}^{\nu}\xi_{\nu}(r)\binom{\nu}{r}\rho_{t}^{r}(1-\rho_{t})^{\nu-r}.
\end{equation}
The approximation error is
\begin{align*}
&|p_{\rm rg}-p_{\rm grid}^{\nu}|\le |p-1/2| \notag \\
&  \times2\exp(-(n-1) \varepsilon^{2}/p_{e}(1-p_{e})+\log(n)),
\end{align*}
as $\varepsilon$ decreases to 0 and with $0<p_{e}<1$.
\end{lemma}
This implies of course that the number of active nodes $n\rho_{t}$ in the random graph with probability $p_{\rm rg}$ in (\ref{eq:pr-rg}) and the number of active nodes with probability $p_{\rm grid}^{\nu}$ in (\ref{eq:pr-rand}) converge in probability with exponential rate with graph size $n$. 
\begin{figure*}
\includegraphics[width=.45\textwidth]{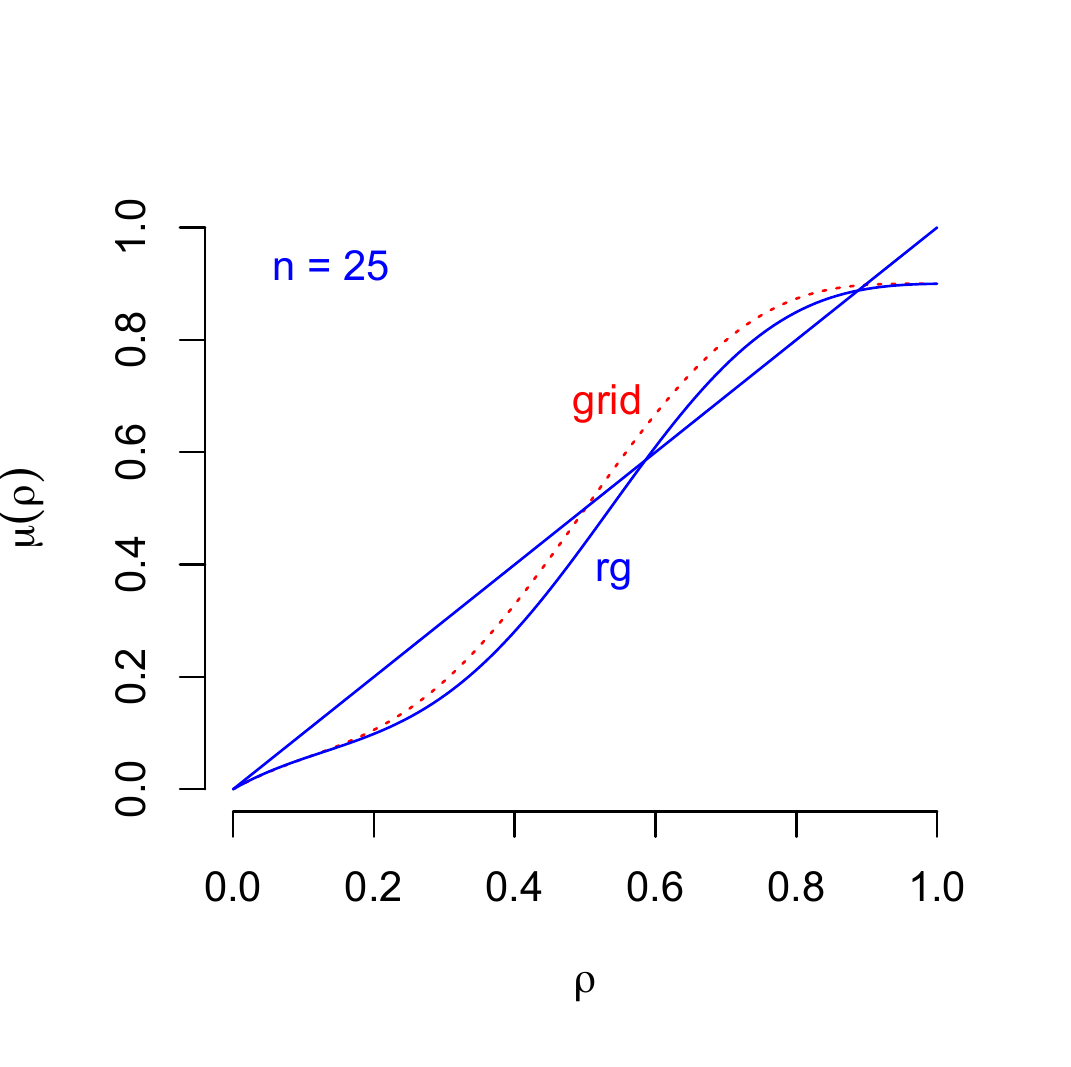}
\includegraphics[width=.45\textwidth]{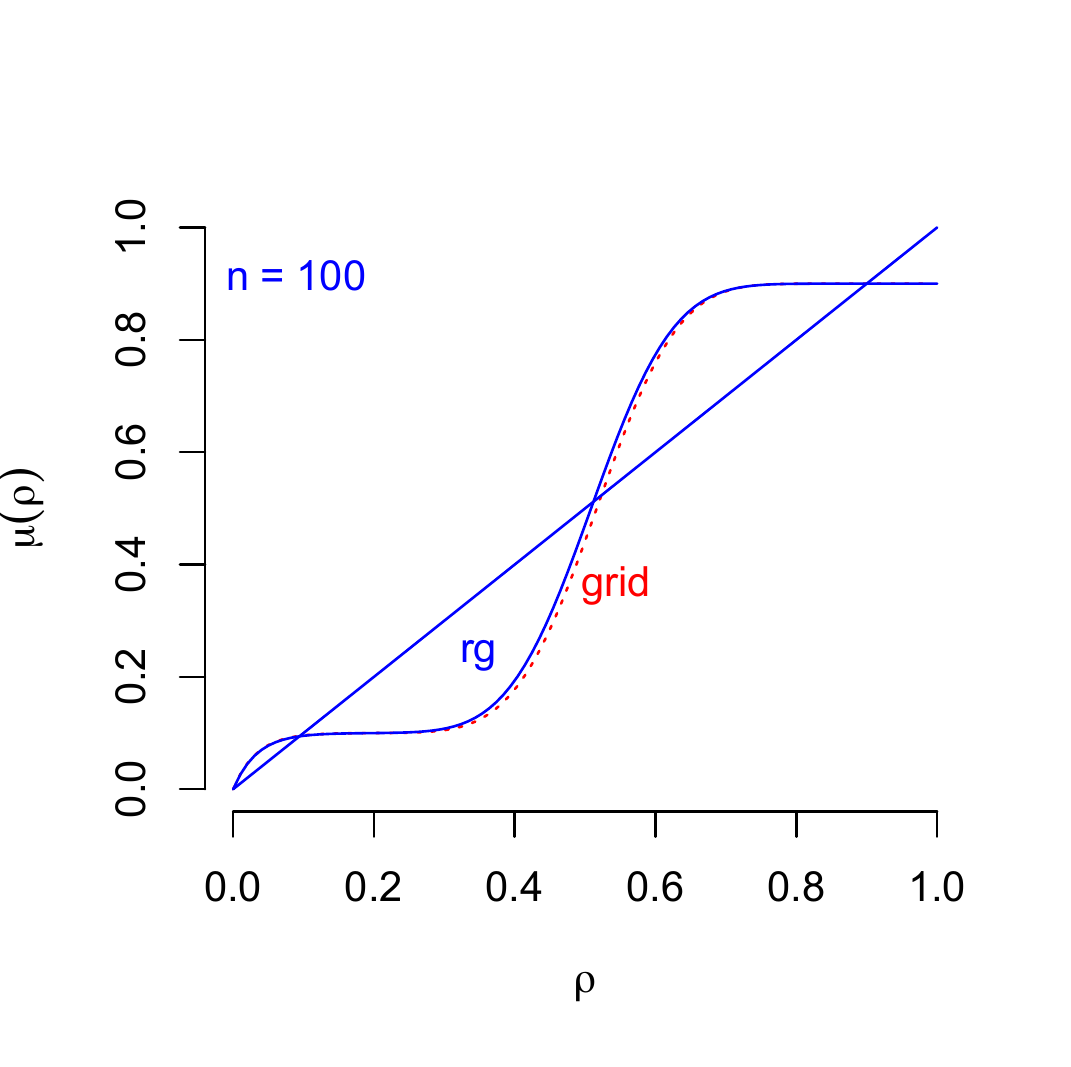}
\caption{The expectation $p_{\rm rg}$ (blue, solid curve) of equation (\ref{eq:pr-rg}) and $p_{rm rand}$ (red, dotted curve) of equation (\ref{eq:pr-rand}) with $p=0.1$ and $p_{e}=0.3$. Left panel shows the curves for a graph of size $n=25$, showing a clear difference between the curves, and the right panel for graph size $n=100$. Note that the difference between the curves at the crossings with the $45^{\circ}$ line is small.}
\label{fig:pr-rg-rand}
\end{figure*}
\begin{remark}
To retain a probability of an edge in $p_{\rm grid}^{\nu}$ leads to a larger approximation error, i.e., using 
\begin{align}\label{eq:pr-grid-pe}
&p_{\rm grid}^{n-1}(\rho_{t})= \notag \\
&\sum_{r=0}^{n-1}\xi_{\nu}(r)\binom{n-1}{r}(\rho_{t}p_{e})^{r}(1-\rho_{t}p_{e})^{n-r-1}
\end{align}
leads to an error of at most $|p-1/2|$, which makes it non ignorable (see the Appendix). 
\end{remark}
We now have expression (\ref{eq:pr-rand}) similar to (\ref{eq:pr1}) for a random graph with the probability of an active node at time $t$ determined by both the density $\rho_{t}$ and an edge being present $p_{e}$ in the size of the neighbourhood. From (\ref{eq:pr-rg}) in Lemma \ref{lem:prg} and Corollary \ref{cor:balister} and Lemma \ref{lem:prg-ext} the evolution equation for the random graph follows. 
\begin{theorem}(Evolution on a random graph)\label{thm:evo-rg}
Let a PCA with local rule $\phi$ be defined on a random graph $G_{\rm rg}(n,p_{e})$, and let the probability of obtaining an active node in $G_{\rm rg}(n,p_{e})$ be $p_{\rm grid}^{\nu}$ as defined in (\ref{eq:pr-rand}). Then the evolution equation for the random graph with large $n$ is
\begin{align}
n\rho_{t+1}=B(n,p_{\rm grid}^{\nu}(\rho_{t})).
\end{align}
The mean and variance of the density $\rho_{t}$ respectively, $\mu_{\rm grid}^{\nu}=p_{\rm grid}^{\nu}$ and $\sigma_{\rm grid,\nu}^{2}=p_{\rm grid}^{\nu}(1-p_{\rm grid}^{\nu})/n$. 
\end{theorem}
We immediately have that the probability $p_{\rm grid}^{\nu}$ is close to the density $\rho$ for each time point $t$ for large graph size $n$. In fact, we find by the triangle inequality
\begin{equation*}
|\rho-p_{\rm grid}^{\nu}|\le |\rho-p_{\rm rg}| + |p_{\rm rg}-p_{\rm grid}^{\nu}|,
\end{equation*}
and both terms converge to 0. The first term $|\rho-p_{\rm rg}|$ converges to 0 by Lemma \ref{lem:chernov} with $p_{\rm rg}$ and using Lemma \ref{lem:prg}, and $|p_{\rm rg}-p_{\rm grid}^{\nu}|$ converges to 0 by Lemma \ref{lem:prg-ext}.

The process $n\rho_{t}$ on a random graph is also a discrete time Markov process, as before, and has transition probability (\ref{eq:trans-prob-grid}) with $p_{\rm grid}^{\nu}$.
We can then apply a similar analysis of dynamics to $\mu_{\rm grid}^{\nu}=p_{\rm grid}^{\nu}$ as before. 

Note that we require for obvious reasons that the graph is connected. It follows that we need a minimum probability $p_{e}$ such that the graph is connected. The probability that a random graph $G_{\rm rg}$ is connected is $\exp(-\exp(-\lambda))$, where $p_{e}=(\log n+\lambda +o(1))/n$ with $\lambda$ fixed \citep[][Theorem 7.3]{Bollobas:2001}. For instance, if we choose the probability of $G_{\rm rg}$ being connected to be 0.99 and we use $n=50$, then we obtain $\lambda = 4.6$ and hence $p_{e}=0.17$. We can therefore not go below 0.17 for a graph with $n=50$ nodes. 
\subsection{Small-world graph}\label{sec:mfsw}
A small-world graph is one which has high average clustering and low average path length, relative to a random graph with the same number of nodes and edges. These graphs have been shown to model realistic networks like those of working relations between actors and the nerve cells in the worm C. elegans \citep{Watts:1998}, and subsequently the small-world has been shown to apply to many different networks, like the (parcellated) brain \citep{Sporns:2006}. And most recently, the network of symptoms as defined by the diagnostic statistical manual (a compendium to diagnose patients) has been found to be a small-world. This finding is a possible explanation for the correlations between pairs of symptoms found in different subpopulations \citep{Borsboom:2011}.

\begin{figure*}
\includegraphics[width=.45\textwidth]{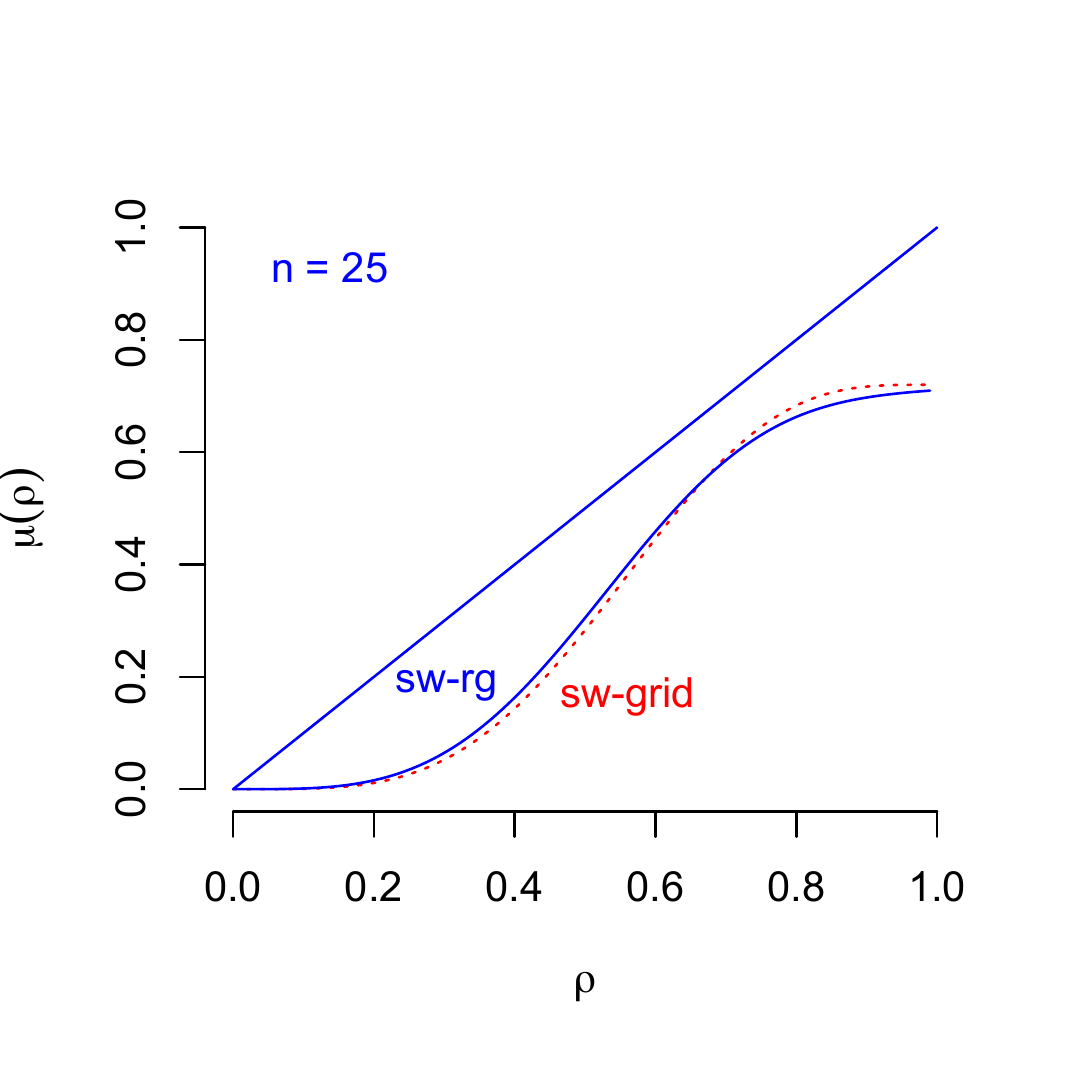}
\includegraphics[width=.45\textwidth]{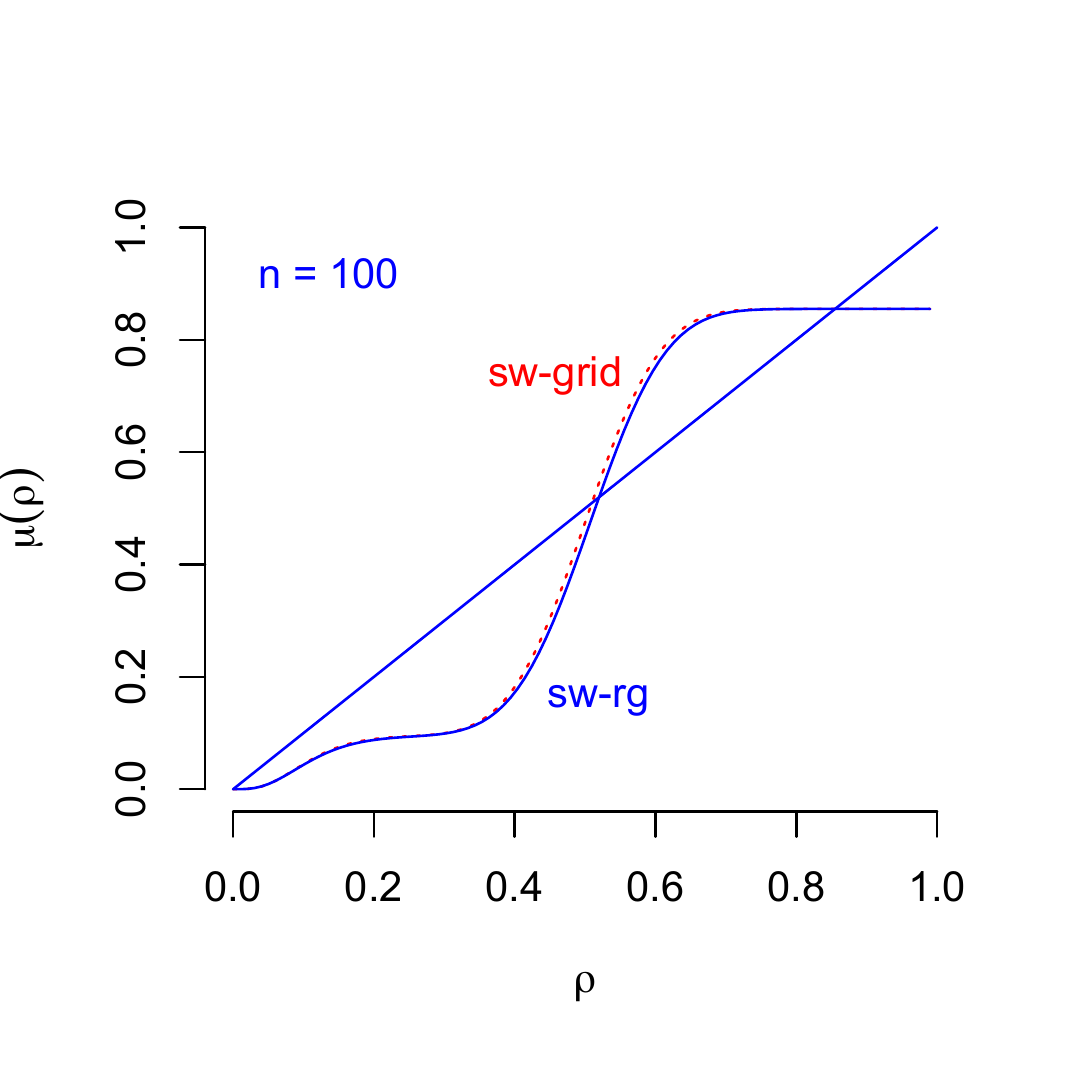}
\caption{The expectation $p_{\rm sw}$ (blue, solid curve) of equation (\ref{eq:pr-sw-rg}) and $p_{\rm sw}^{\nu}$ (red, dotted curve) of equation (\ref{eq:pr-rand}) with $p=0.1$ and $p_{w}=0.3$. Left panel shows the curves for a graph of size $n=25$, showing a clear difference between the curves, and the right panel for graph size $n=100$. Note that the difference between the curves at the crossings with the $45^{\circ}$ line is small.}
\label{fig:pr-sw-rg-rand}
\end{figure*}

Here we use the modified Newman-Watts (NW) small-world of \citet{Newman:1999}, where for a given grid structure where each node has neighbourhood $\Gamma$, a set of $(n-1)p_{w}$ edges is on average independently added to the graph, where $p_{w}$ is the probability of two nodes being wired. Such a graph is denoted by $G_{\rm sw}(n,\Gamma,p_{w})$.
The same idea as with the random graph, where the probability for an active node was corrected by the probability of the degree of a node, averaged over all possible neighbourhood sizes, can be used for the random part in the NW small-world. In the NW small-world we start with a grid with neighbourhood size $|\Gamma|$, which is fixed, and augment the graph randomly with edges according to a binomial variable with probability $p_{w}$. Let 
\begin{align*}
\Prob_{p_{w}}(|\Gamma|=k)=\binom{n-1}{k}p_{w}^{k}(1-p_{w})^{n-k-1}
\end{align*}
We then obtain 
\begin{align}\label{eq:pr-sw}
&p_{\rm sw}(\rho_{t}) = \notag \\
&\sum_{k=|\Gamma|}^{n-1}\sum_{r=0}^{k-|\Gamma|}\xi_{k}(r)\binom{k}{r}\rho_{t}^{r}(1-\rho_{t})^{k-r}\Prob_{p_{w}}(|\Gamma|=k).
\end{align}
We could define the small-world probability using this definition. But we can split up $p_{\rm sw}$ in two terms, one involving the fixed neighbourhood $\Gamma$ of the grid, and one random neighbourhood consisting of the possible shortcuts. We therefore start with the probability in a grid $p_{\rm grid}$ corrected by the $(1-p_{w})^{n-|\Gamma|}$ requiring that no possible randomly added edges are present, i.e., we obtain 
\begin{equation}\label{eq:pr-grid-sw}
p_{\rm grid}^{\rm sw}=p_{\rm grid}(1-p_{w})^{n-|\Gamma|}
\end{equation}
for the first part of the fixed grid. Then, in accordance with the random part of the NW small-world, a probability is added to emulate the possible additional neighbours in the random part of the graph, ignoring the first $|\Gamma|$ neighbours from the grid. 
Define the probability
\begin{align}
&p_{\rm rg,\Gamma}(\rho_{t}) = \notag \\ 
&\sum_{k=|\Gamma|+1}^{n-1}\sum_{r=0}^{k-|\Gamma|}\xi_{k}(r)\binom{k}{r}\rho_{t}^{r}(1-\rho_{t})^{k-r}\Prob_{p_{w}}(|\Gamma|=k),
\end{align}
where the first $|\Gamma|$ neighbours are ignored since they were included already as neighbours in the grid structure in $p_{\rm grid}^{\rm sw}$. Then we can write the small-world probability as $p_{\rm sw}=p_{\rm grid}(1-p_{w})^{n-|\Gamma|}+p_{\rm rg,\Gamma}$. For the second part, however, we have the approximation as before from the random graph, leaving out the first $\Gamma$ nodes from the grid. This leads to the simplification using the grid probability only
\begin{equation}\label{eq:pr-rand-sw}
p_{{\rm grid},\Gamma}^{\nu}(\rho_{t})= \sum_{r=|\Gamma|+1}^{\nu}\xi_{\nu}(r)\binom{\nu}{r}(\rho_{t})^{r}(1-\rho_{t})^{\nu-r},
\end{equation}
where $\nu = \lfloor p_{w}(n-|\Gamma|)\rfloor$. 
The error of approximation using $p_{{\rm grid}, \Gamma}^{\nu}$ instead of $p_{{\rm rg}, \Gamma}$ follows immediately from Lemma \ref{lem:prg-ext} for fixed grid neighbourhood $\Gamma$, except that the first $|\Gamma|$ nodes in the grid are taken out. 
\begin{corollary}\label{cor:pr-sw}
Let $G_{\rm sw}(n,\Gamma,p_{w})$ be the NW small-world graph of size $n$ with $|\Gamma|$ nodes in the fixed neighbourhood for each node. Furthermore, let $0<p_{w}<1$ be the wiring probability and $\nu=\lfloor p_{w}(n-|\Gamma|)\rfloor$. Then the approximation error for the probability using the grid structure $p_{\rm grid}^{\nu}$ in (\ref{eq:pr-rand-sw}) in the random part is 
\begin{align*}
&|p_{\rm rg,\Gamma}-p_{\rm grid,\Gamma}^{\nu}|\le |p-1/2| \\ 
&\times 2\exp(-(n-|\Gamma|) \varepsilon^{2}/p_{w}(1-p_{w})\\
&\quad +\log(n-|\Gamma|+1)),
\end{align*}
for $\varepsilon>0$.
\end{corollary}
Equations (\ref{eq:pr-sw}) to (\ref{eq:pr-rand-sw}) and Corollary \ref{cor:pr-sw} prove the following equation for the evolution on an NW small-world.
\begin{theorem}(Evolution on a small-world)\label{thm:evo-sw}
Define a PCA on a Newman-Watts small-world graph $G_{\rm sw}$ of size $n$ with $|\Gamma|$ neighbours for the initial graph and wiring probability $p_{w}$. Then with probabilities $p_{\rm grid}^{\rm sw}$ in (\ref{eq:pr-grid-sw}) and $p_{\rm grid,\Gamma}^{\nu}$ in (\ref{eq:pr-rand-sw}), the evolution of the number of active nodes is
\begin{align}\label{eq:evo-sw}
n\rho_{t+1} &= B(|\Gamma|,p_{\rm grid}^{\rm sw}(\rho_{t})) \notag \\
	&\quad+ B(n-|\Gamma|,p_{{\rm grid},\Gamma}^{\nu}(\rho_{t}))
\end{align}
with mean and variance for the density $\rho_{t}$, respectively, 
\begin{equation}
\mu_{\rm sw}=\frac{|\Gamma|}{n}p_{\rm grid}^{\rm sw}+\frac{n-|\Gamma|}{n}p_{{\rm grid},\Gamma}^{\nu}
\end{equation}
and 
\begin{align}
\sigma^{2}_{\rm sw}&=\frac{|\Gamma|}{n^{2}}p_{\rm grid}^{\rm sw}(1-p_{\rm grid}^{\rm sw}) \notag \\
&\quad + \frac{n-|\Gamma|}{n^{2}}p_{{\rm grid},\Gamma}^{\nu}(1-p_{{\rm grid},\Gamma}^{\nu}).
\end{align}
\end{theorem}
We write $p_{\rm sw}^{\nu}=p_{\rm grid}^{\rm sw}(|\Gamma|/n)+p_{\rm grid,\Gamma}^{\nu}((n-|\Gamma|)/n)$ for the NW small-world probability based on the approximation with the grid. Figure \ref{fig:pr-sw-rg-rand} shows two examples of the approximation $p_{\rm grid,\Gamma}^{\nu}$ for the random part in the NW small-world. It is clear from the corollary that convergence is a bit slow for small graphs since the difference of nodes in the fixed neighbourhood $\Gamma$ and in the expected $p_{w}(n-|\Gamma|)$ neighbours in the random part, determines the rate. 
Again, we can use $p_{\rm sw}^{\nu}$ to determine the dynamics of the mean field. 
%

\section{Dynamics of the mean field}\label{subsec:dynamics}
To investigate the dynamics we treat the mean field function $p_{\rm grid}$ for the grid, $p_{\rm grid}^{\nu}$ for the random graph, and $p_{\rm sw}$ for the NW small-world as a discrete dynamical system. We can then determine in principle the fixed points and describe its behaviour in the long term. In general, however, obtaining the fixed points is not trivial. Indeed, \citet{Balister:2006} provide an analytical solution for the fixed points for a specific case in $G_{\rm grid}$, but mention that a general solution is difficult. \citet{Janson:2015} give analytical solutions for the fixed points when leaving out the majority rule, making it a deterministic system. Here we keep the majority rule sacrificing the possibility of determining the critical points analytically. We therefore describe the qualitative behaviour of $p_{\rm grid}$, $p_{\rm grid}^{\nu}$, and $p_{\rm sw}$.

\subsection{Dynamics of the mean field in a grid}\label{subsec:dyn-grid}
\begin{figure*}
\includegraphics[width=.45\textwidth]{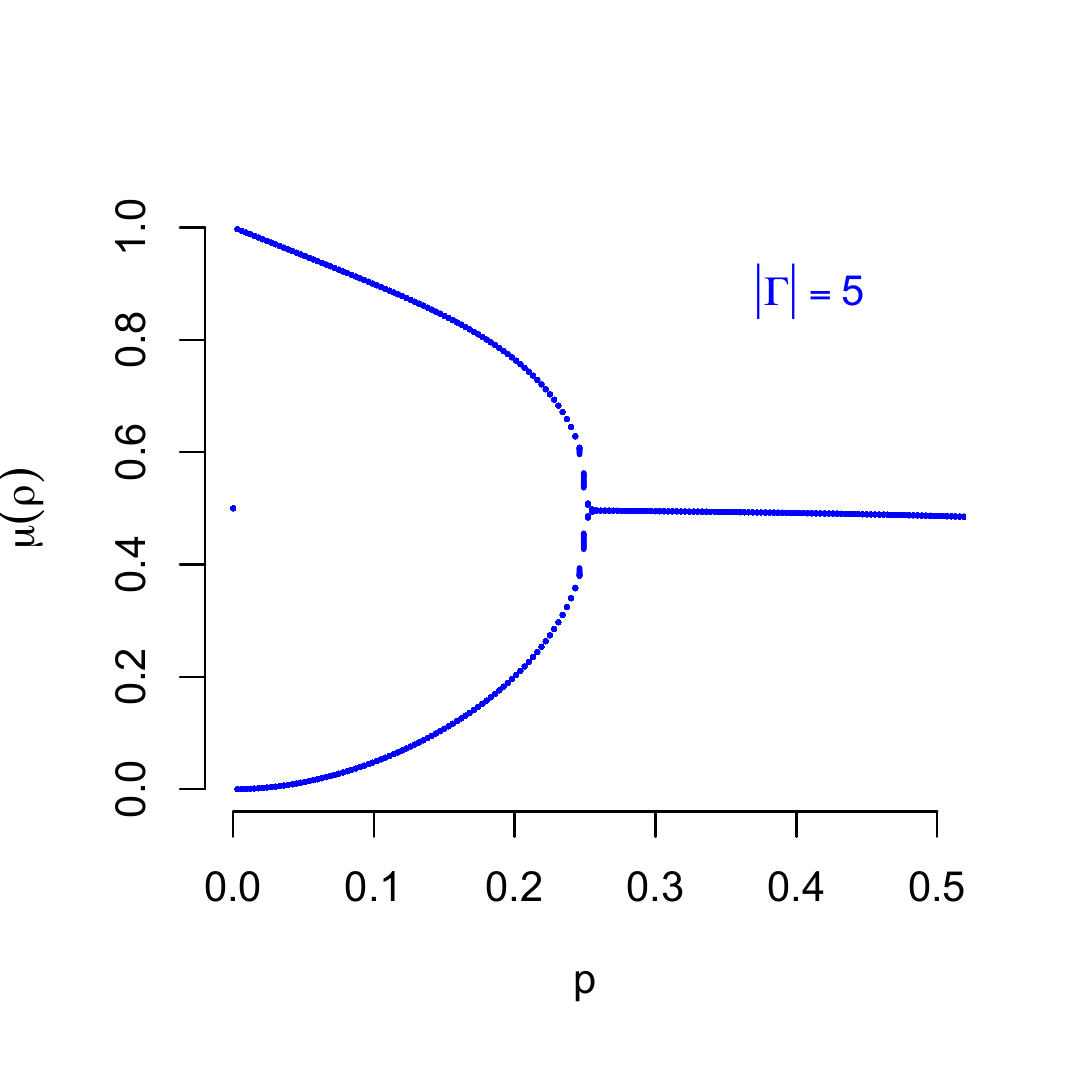}
\includegraphics[width=.45\textwidth]{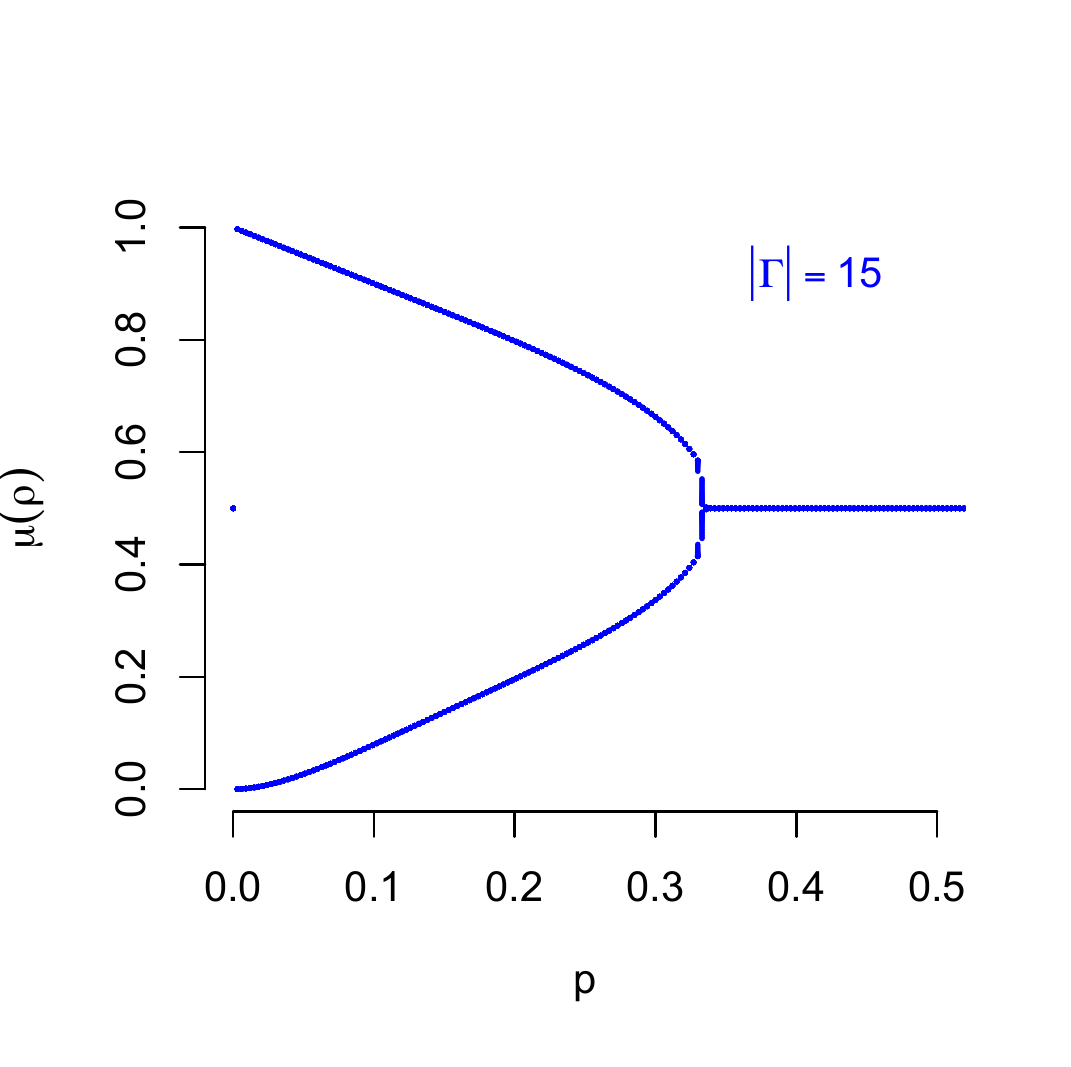}
\caption{Bifurcation plots of  $p_{\rm grid}$ in a graph of size $n=100$ for $|\Gamma|=5$ and $|\Gamma|=15$ neighbours.}
\label{fig:bif-grid}
\end{figure*}

\noindent
The dynamics of the mean field in the grid $G_{\rm grid}$ from (\ref{eq:pr1}) have been described in \citet{Balister:2006} and \citet{Kozma:2005} for a neighbourhood size of $|\Gamma|=5$. The function $\mu_{\rm grid}=p_{\rm grid}$ is continuous and since $[0,1]$ is closed and bounded, we find that $\mu_{\rm grid}$ has at least one fixed point in $[0,1]$ \citep{Holmgren:1996,Hirsch:2004}. A fixed point is one where we find $p_{\rm grid}(\rho_{t})=\rho_{t}$. Finding the fixed points for $p_{\rm grid}$ in general is not trivial. \citet{Balister:2006} showed that if $|\Gamma|=5$ in the finite grid, then $p=7/30\approx 0.233$ is a critical point, such that if $p$ is in $[7/30,1/2]$ then there is a stable fixed point at $\rho=0.5$, but when $p<7/30$ then $\rho=0.5$ is unstable and there are two other stable fixed points. This can be seen in Figure \ref{fig:bif-grid}, which shows two bifurcation plots, where for each value of $0<p\le 0.5$ the function $\mu_{\rm grid}=p_{\rm grid}$ is iteratively applied for about 1000 steps, and only the last 50 are plotted. Figure \ref{fig:bif-grid} shows that for $|\Gamma|=5$ neighbours the fixed point at is stable for $p\in [0,7/30)$ and bistable for $p\in [7/30,0.5]$, as predicted. 
Since $p_{\rm grid}$ is continuous, stability can be checked by considering the derivative $\partial \mu_{\rm grid}/\partial \rho=\dot{\mu}_{\rm grid}$. If $|\dot{\mu}_{\rm grid}|$ is bounded by 1, then the fixed point $\rho$ is attractive, otherwise it is repellent. 
The derivative with respect to $\rho_{t}$ is
\begin{align*}
&\dot{\mu}_{\rm grid}(\rho_{t})= \\
&\sum_{r=0}^{|\Gamma|}\xi_{|\Gamma|}(r)\binom{|\Gamma|}{r}(r-\rho_{t}|\Gamma|)\rho_{t}^{r-1}(1-\rho_{t})^{|\Gamma|-r-1}
\end{align*}
For example, the derivative for $p=0.15$ is not bounded by 1 for all values of $\rho$; the fixed point $\rho=0.5$ is repellent since at this point $\dot{\mu}(0.5)\approx 1.359$, and so iteration of $\mu$ will lead away from 0.5. The derivative for $p=0.35$ is smaller than 1 (0.672) and so in this case $\rho=0.5$ is an attractive fixed point.  It can be seen that for $|\Gamma|=5$ the critical point is at 0.233, as predicted by theory \citep{Balister:2006}. It can also be seen that increasing the neighbourhood size to $|\Gamma|=15$ (right panel) increases the critical point. This increase in critical point corresponds to the simulations in \citet{Kozma:2005} where ('long range') edges were added to the nodes, which increased the neighbourhood size. 

\subsection{Dynamics of the mean field in a random graph}\label{subsec:dyn-rg}
\begin{figure*}
\includegraphics[width=.45\textwidth]{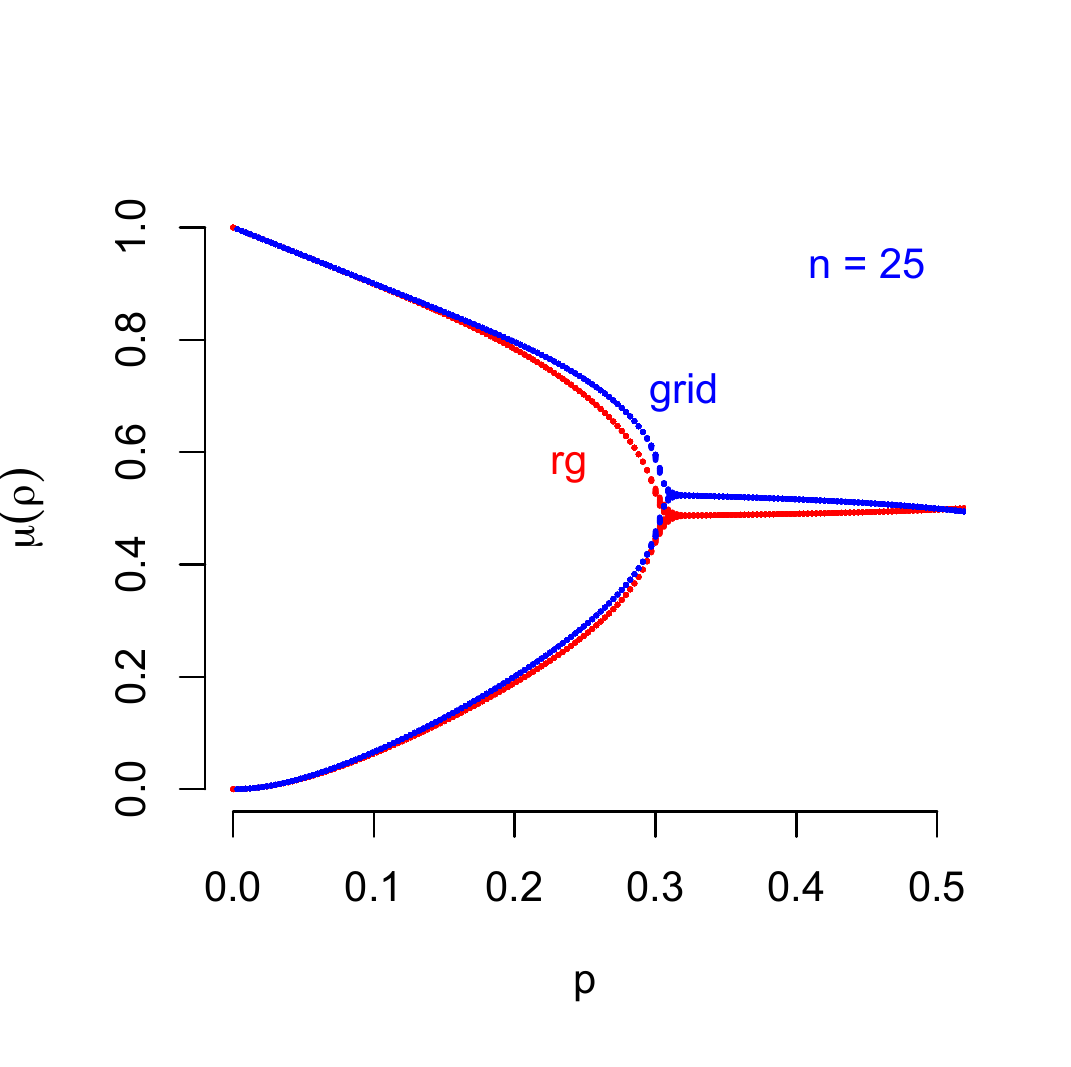}~
\includegraphics[width=.45\textwidth]{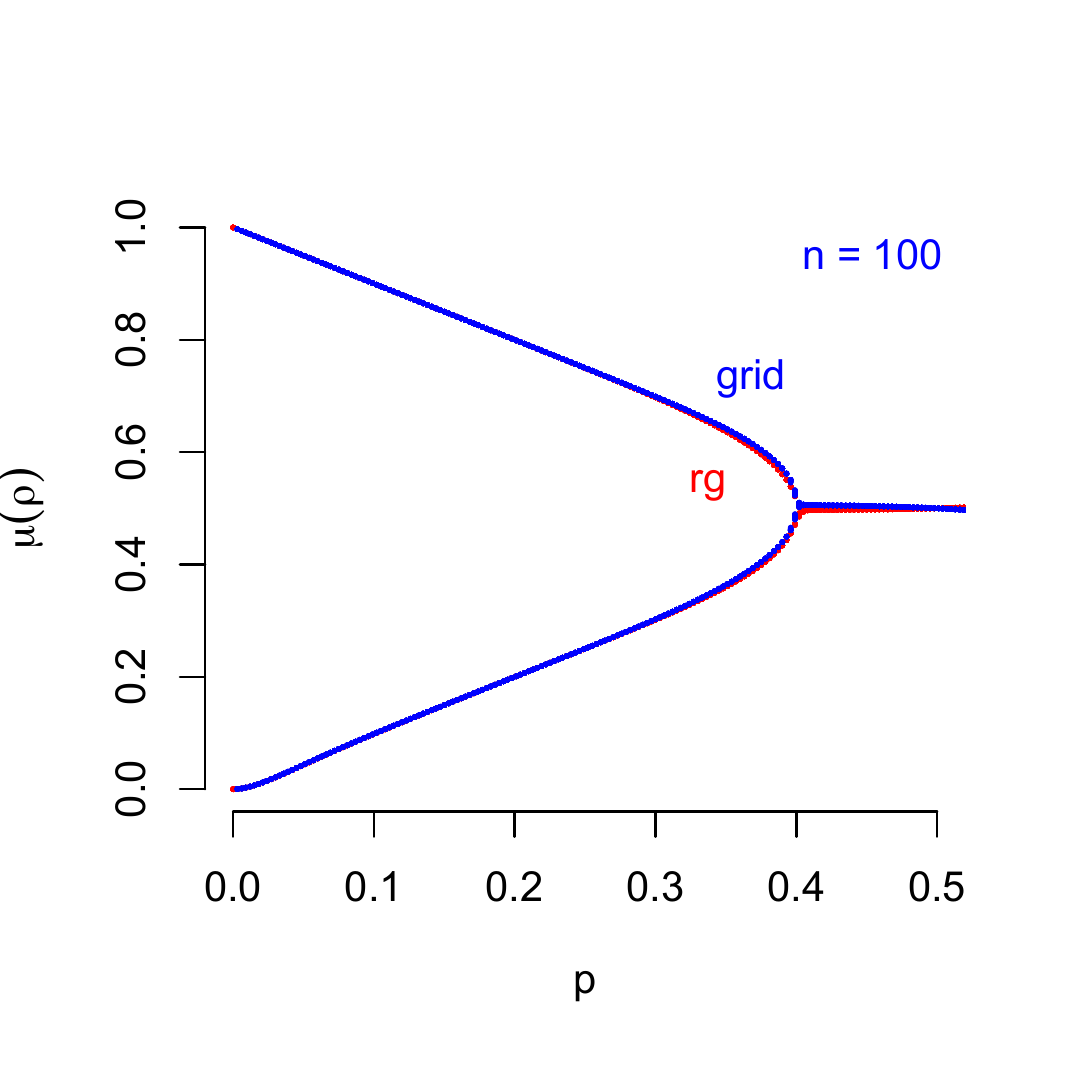}~
\caption{Plots of $\mu_{\rm rg}$ (red) and $\mu_{\rm grid}^{\nu}$ (blue) as a function of $p$. In the left panel the bifurcation plots are given for a random graph of size $n=25$ and in the right panel for a random graph of size $n=100$. All plots are obtained with edge probability $p_{e}=0.4$.}
\label{fig:bif-rg}
\end{figure*}

The dynamics of $p_{\rm grid}^{\nu}$ in the random graph $G_{\rm rg}$ are similar to that of the grid. The main difference is that the critical point of the bifurcation is closer to $p=1/2$. As is clear from the definition of $p_{\rm grid}^{\nu}$ in (\ref{eq:pr-rand}), the only difference with that of the grid is the neighbourhood size which is increased to $\nu=\lfloor p_{e}(n-1)\rfloor$. Figure \ref{fig:bif-rg} shows the result for a graph with $n=25$ nodes (left panel) and for a graph with $n=100$ nodes. The approximation of $p_{\rm grid}^{\nu}$ is quite accurate, also for the location of the critical point. With a graph of size $n=100$ the accuracy is such that $p_{\rm rg}$ and $p_{\rm grid}^{\nu}$ are nearly indistinguishable, which corresponds to the result in Lemma \ref{lem:prg-ext}. 
%

\subsection{Dynamics of the mean field in a small-world}\label{subsec:dyn-grid}
\begin{figure*}
\includegraphics[width=.45\textwidth]{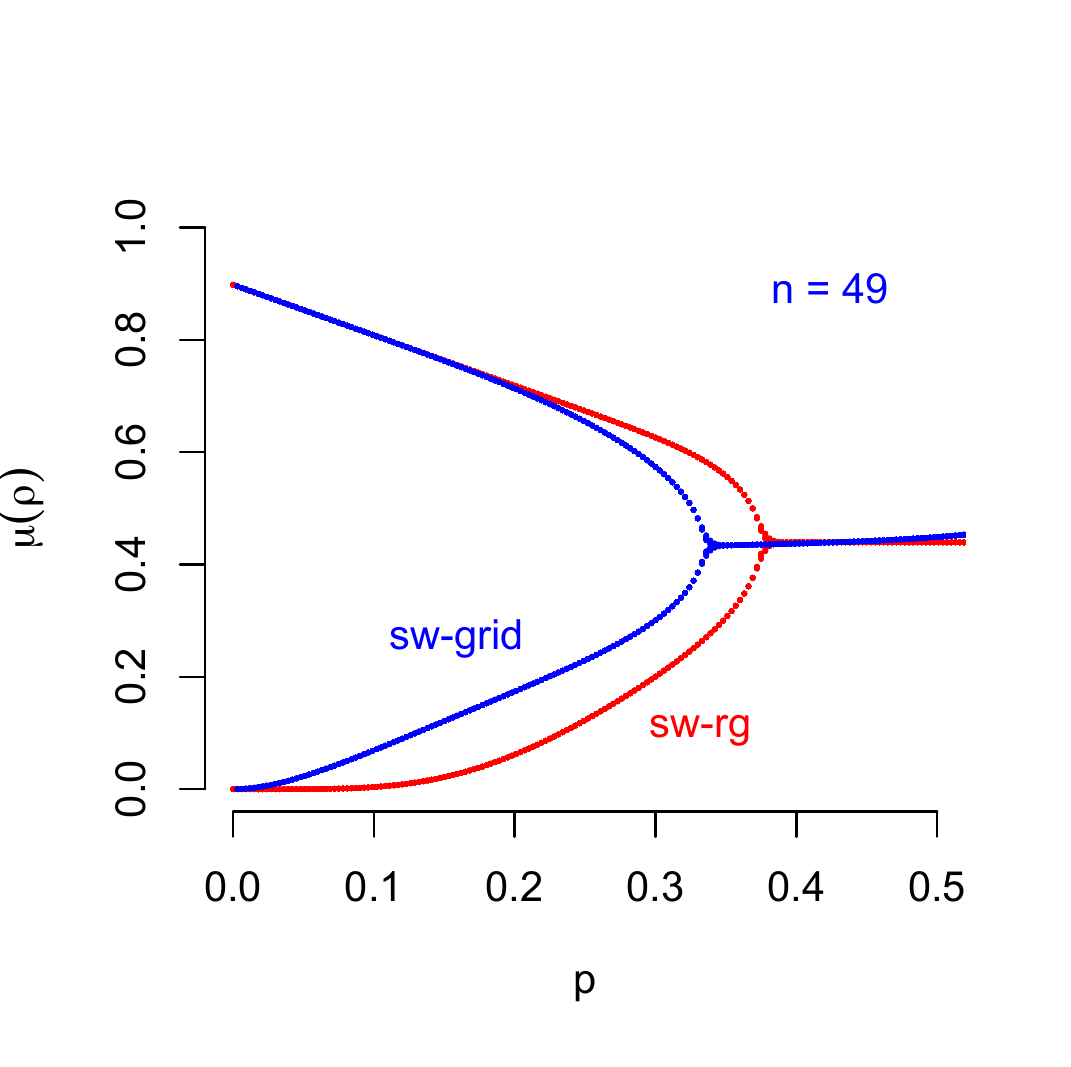}~
\includegraphics[width=.45\textwidth]{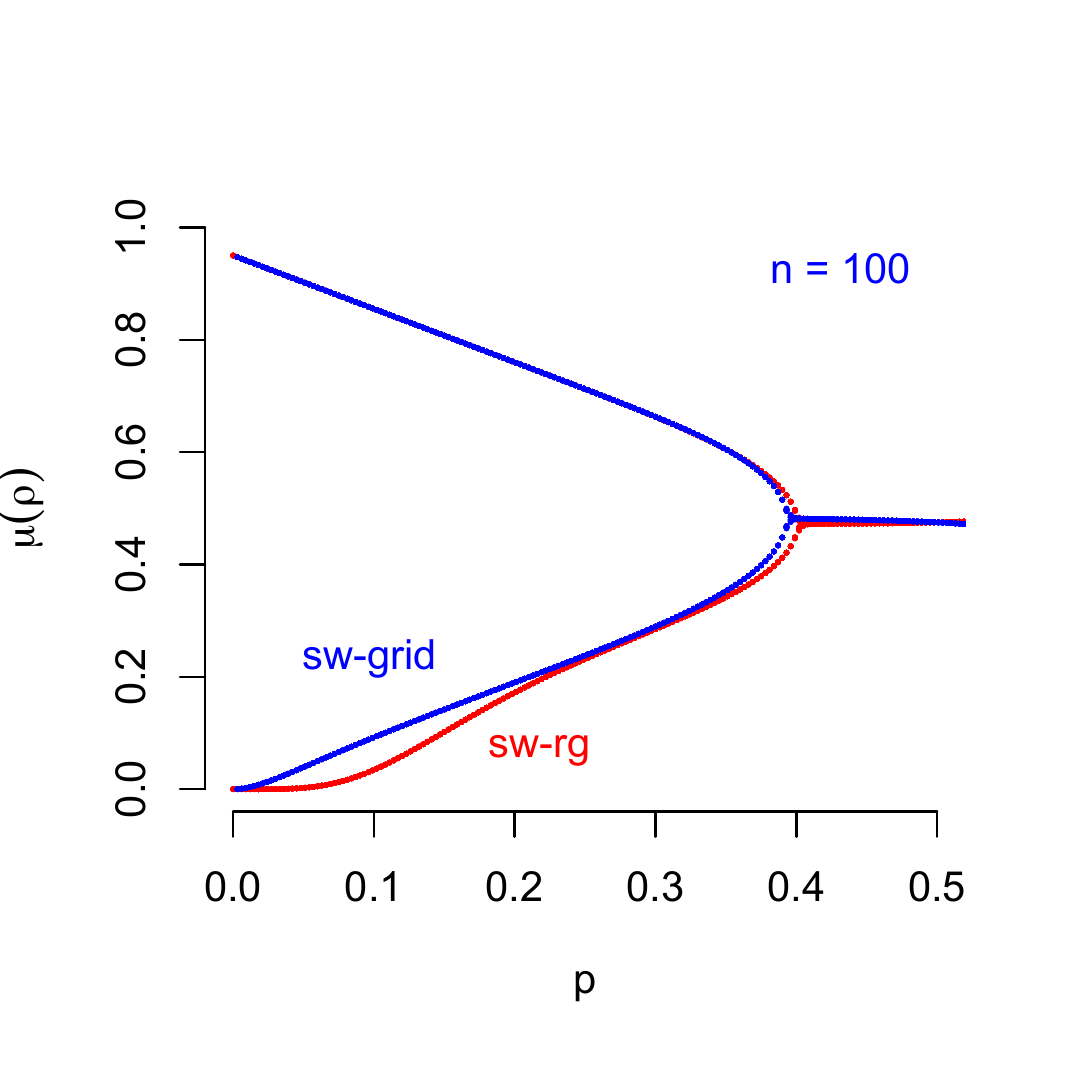}
\caption{Bifurcation plots of the small-world mean field $\mu_{\rm sw}^{\nu}(\rho)$ based on the grid with $\nu=\lfloor p_{e}(n-1)\rfloor$ neighbours (blue) and the mean field $\mu_{\rm sw}(\rho)$ based on the all possible neighbours in the random graph (red). In the left panel a small-world of $n=25$ nodes and in the right panel a graph of $n=100$ nodes; all graphs are obtained with the probability of wiring (adding edges) in the NW small-world of $p_{w}=0.4$.}
\label{fig:sw-bifurc}
\end{figure*}

The dynamic behaviour of $p_{\rm sw}$ is shown in Figure \ref{fig:sw-bifurc}. Generally, the behaviour is similar to that on the random graph. In Figure \ref{fig:sw-bifurc} the left panel shows a bifurcation plot of $p_{\rm sw}$ and and $p_{\rm sw}^{\nu}$ on $G_{\rm sw}(49,0.4)$. The accuracy of $p_{\rm sw}^{\nu}$ improves greatly for larger $n$, as seen in the right panel of Figure \ref{fig:sw-bifurc} for $G_{\rm sw}(100,0.4)$.
For low values of new edges in the NW small-world $p_{w}$, the probability $p_{\rm sw}^{\nu}$ is smaller than in the grid. This is because the probability $p_{\rm grid}^{\rm sw}=p_{\rm grid}(1-p_{w})^{n-|\Gamma|}$ is corrected by the number of edges not added to the graph. 

\section{Numerical evaluation of the mean field}\label{sec:numerical}
To evaluate the accuracy of the predictions of the mean field in the grid, random, and small-world graph, we simulated networks of different sizes in  the topology of a grid, a random graph, and a small-world graph. For each simulation run 0/1 data were generated according to one of the three types of graph using the {\tt\small R} package {\em IsingSampler} \cite{Borkulo:2014}. The in combination with the majority rule the PCA was run for a certain duration $T$ and the average states of the last section of the time series was determined to see if it matches that of the predictions of the mean field. 
To determine the accuracy we used both $90\%$ and $95\%$ confidence intervals obtained from the central limit theorem for each of the three different graphs (see Section \ref{sec:mf-grid}). 

\begin{figure*}
\includegraphics[width=0.2\textwidth]{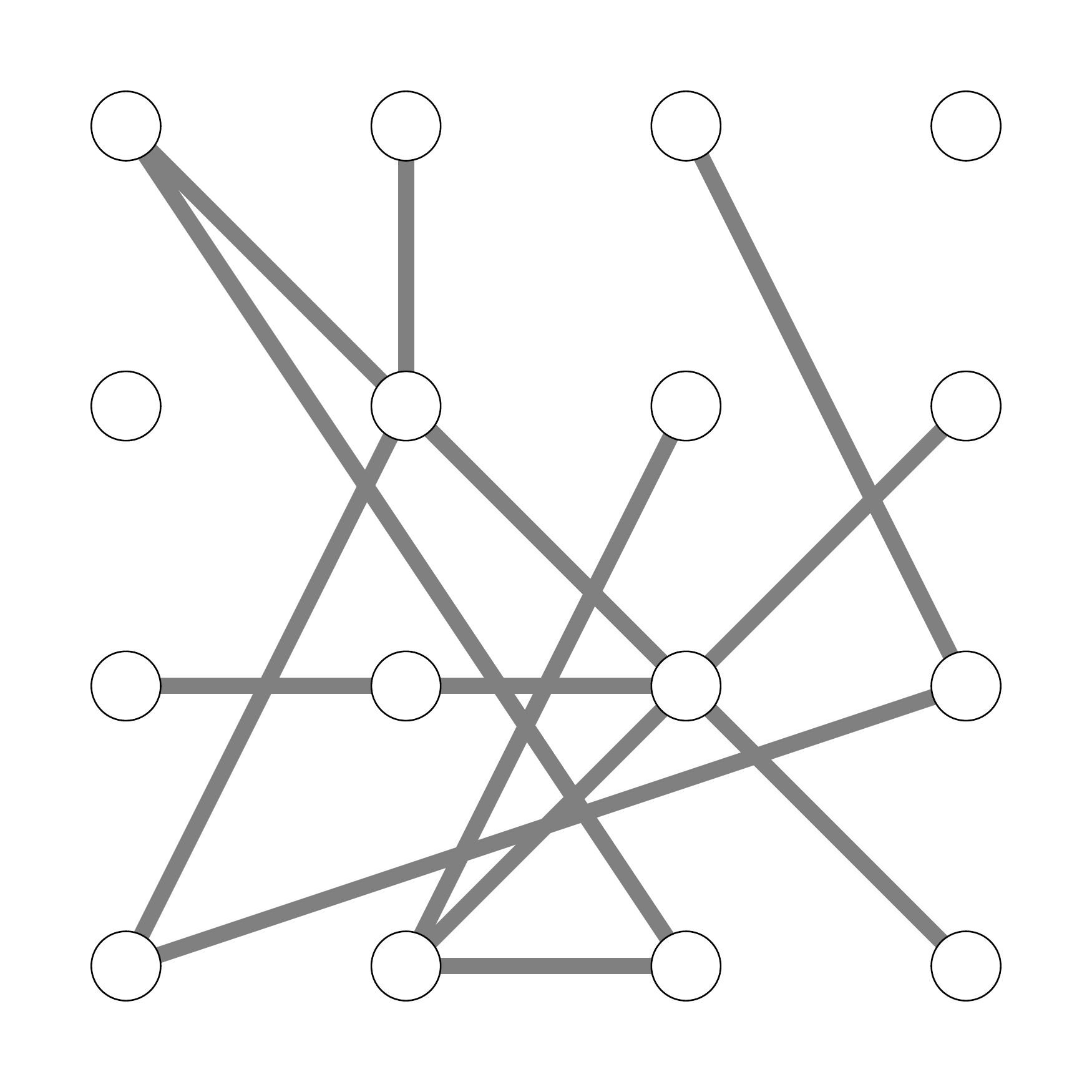}
\put(-55, 90){$n = 16$} 
\put(-120, 40){$p_e = 0.1$} 
\includegraphics[width=0.2\textwidth]{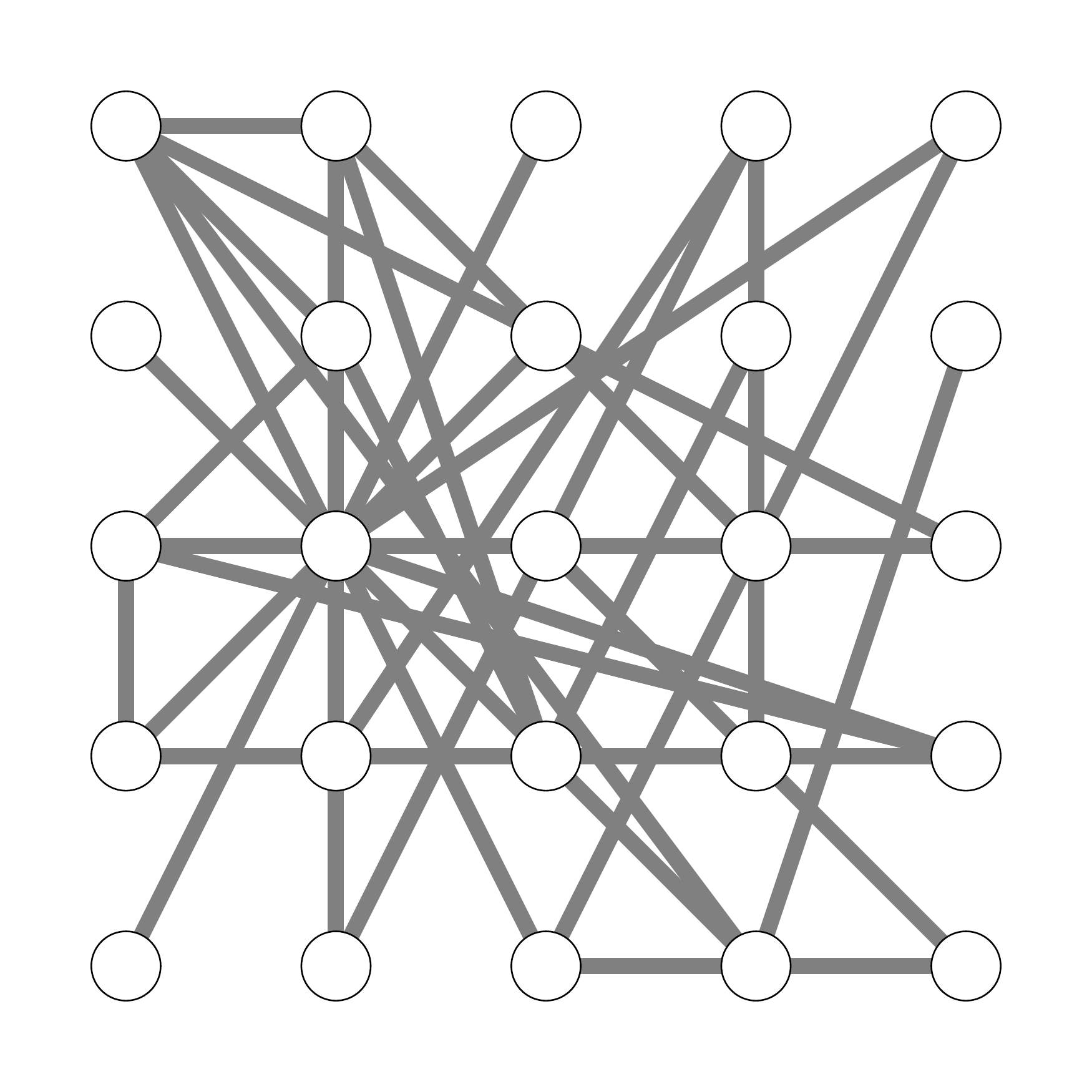}
\put(-55, 90){$n = 25$}   
\includegraphics[width=0.2\textwidth]{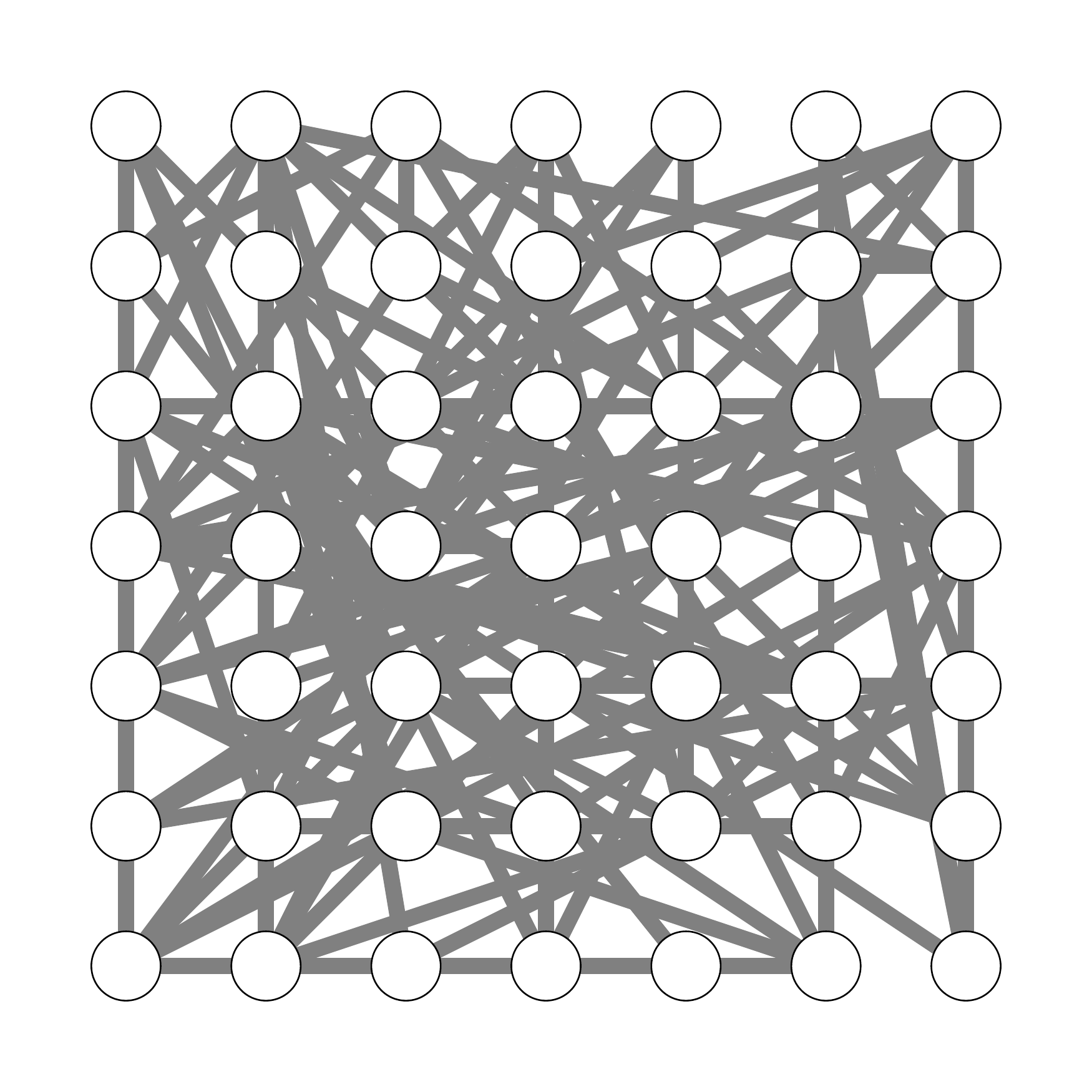}
\put(-55, 90){$n = 49$}    
\includegraphics[width=0.2\textwidth]{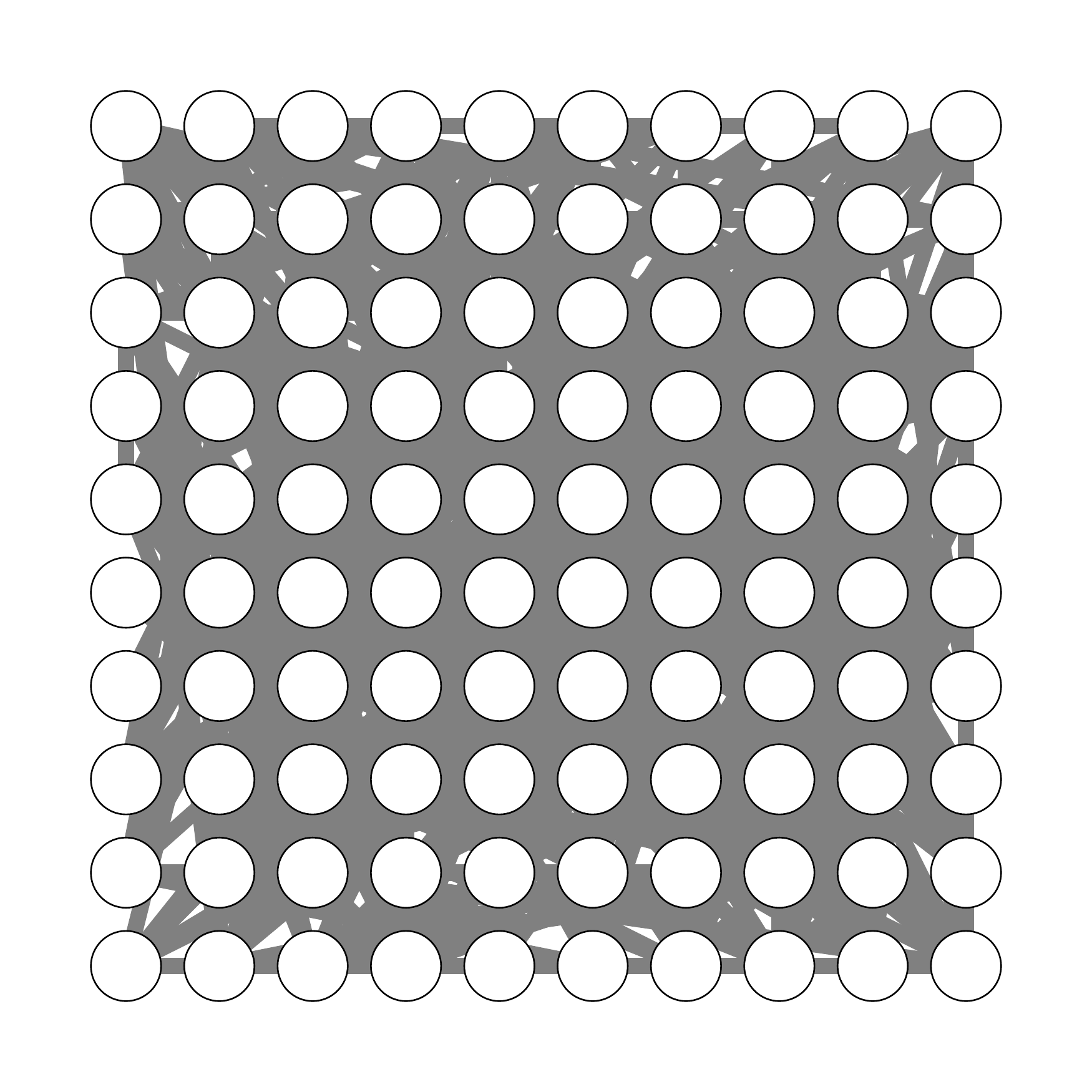} 
\put(-55, 90){$n = 100$}  \\

\includegraphics[width=0.2\textwidth]{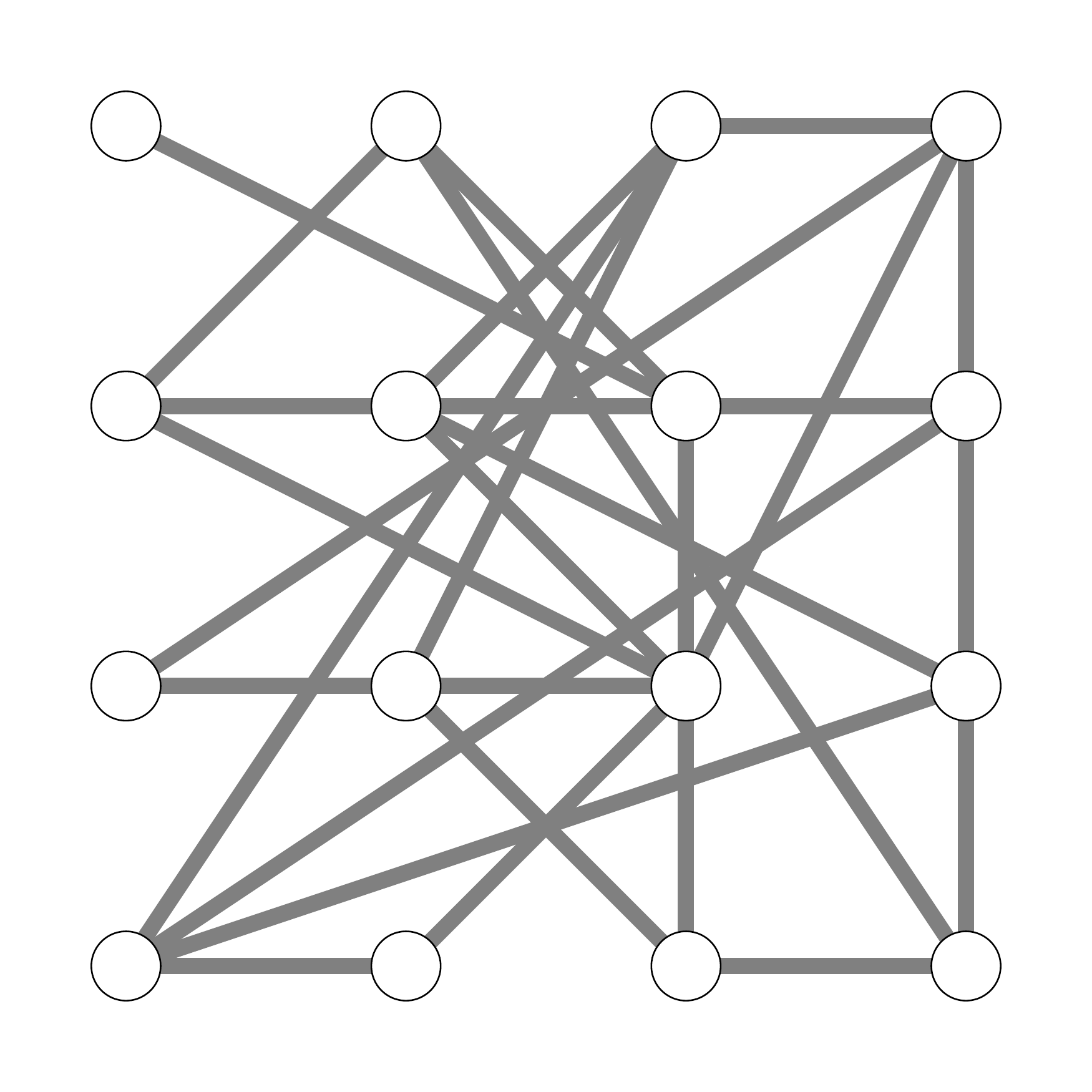}
\put(-120, 40){$p_e = 0.2$}  
\includegraphics[width=0.2\textwidth]{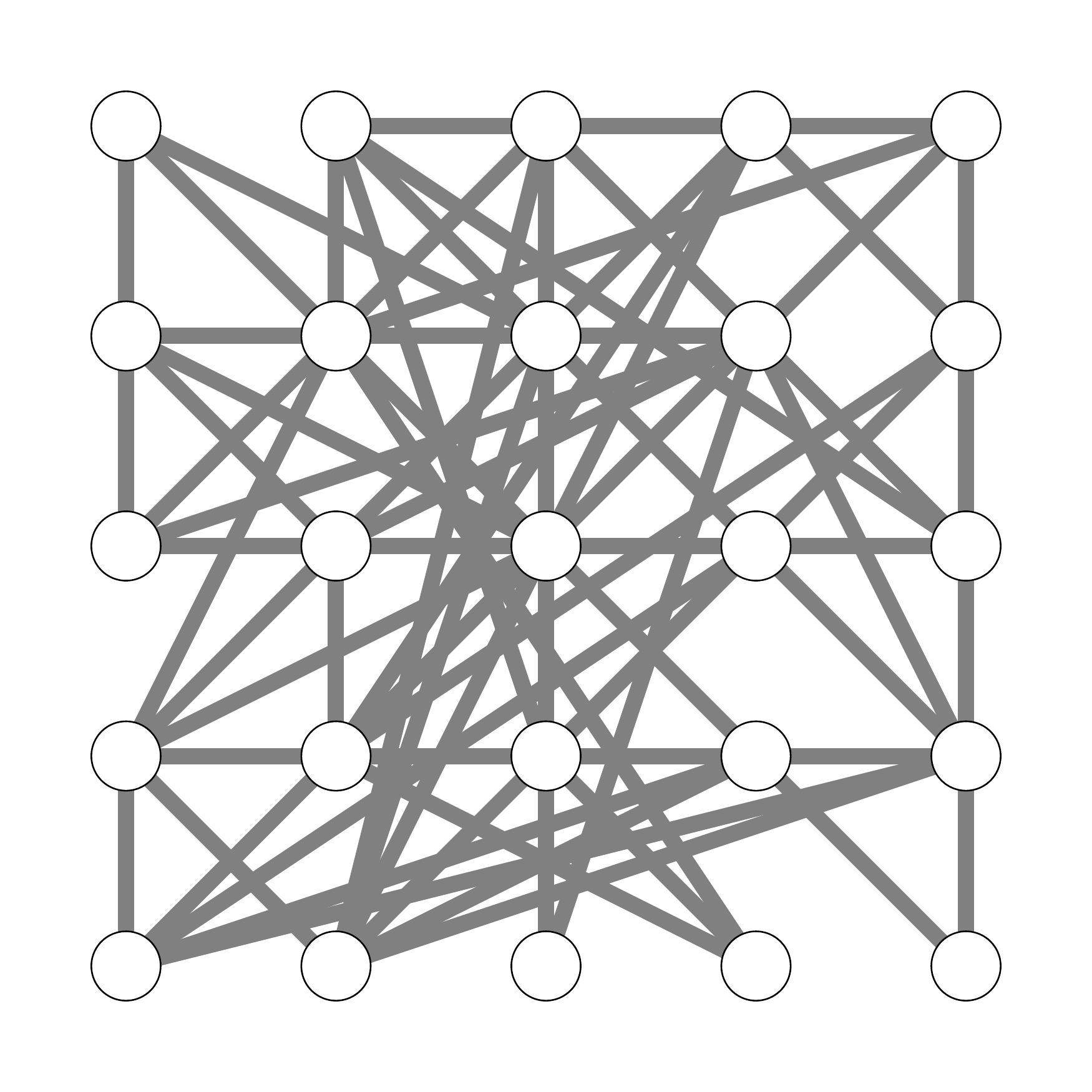}   
\includegraphics[width=0.2\textwidth]{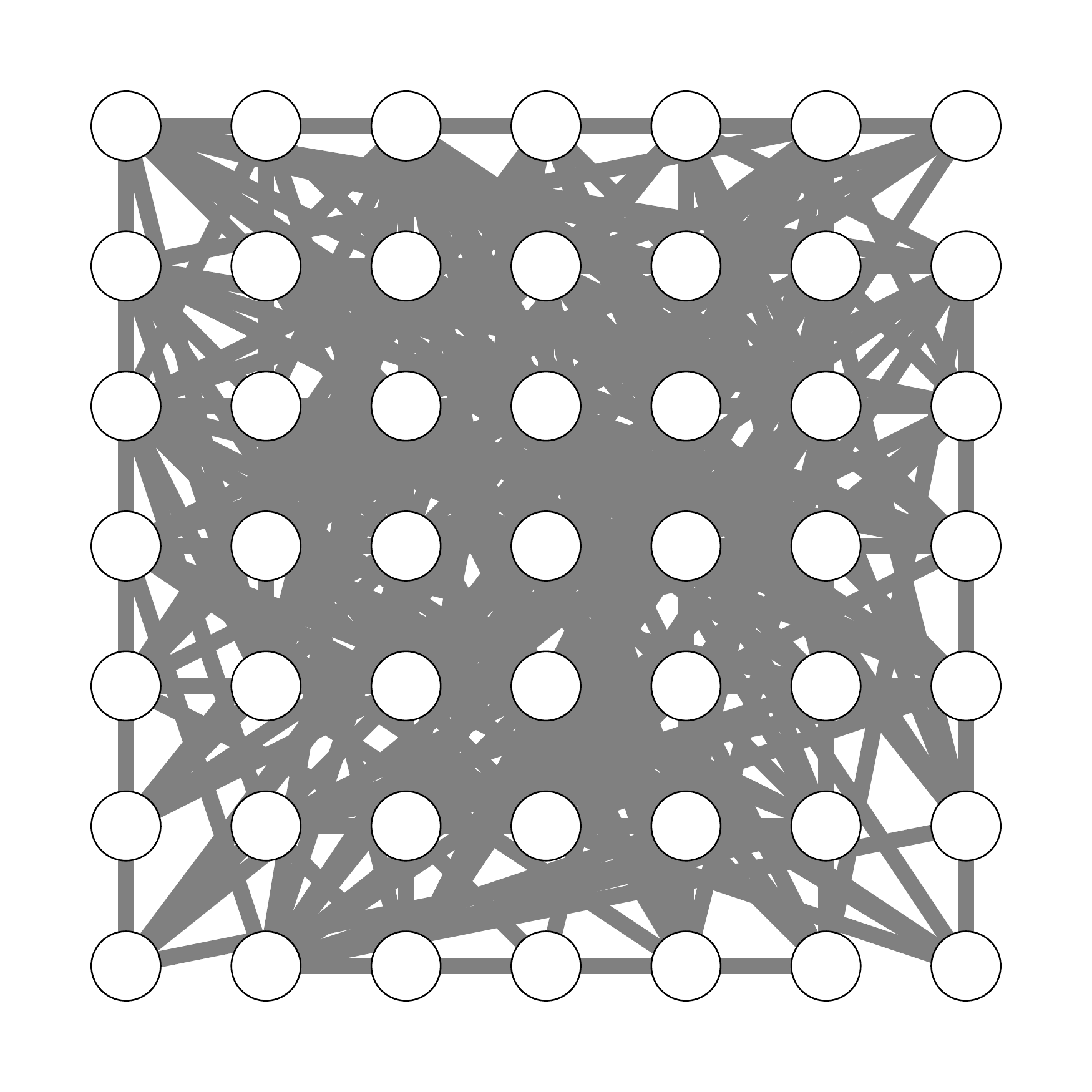} 
\includegraphics[width=0.2\textwidth]{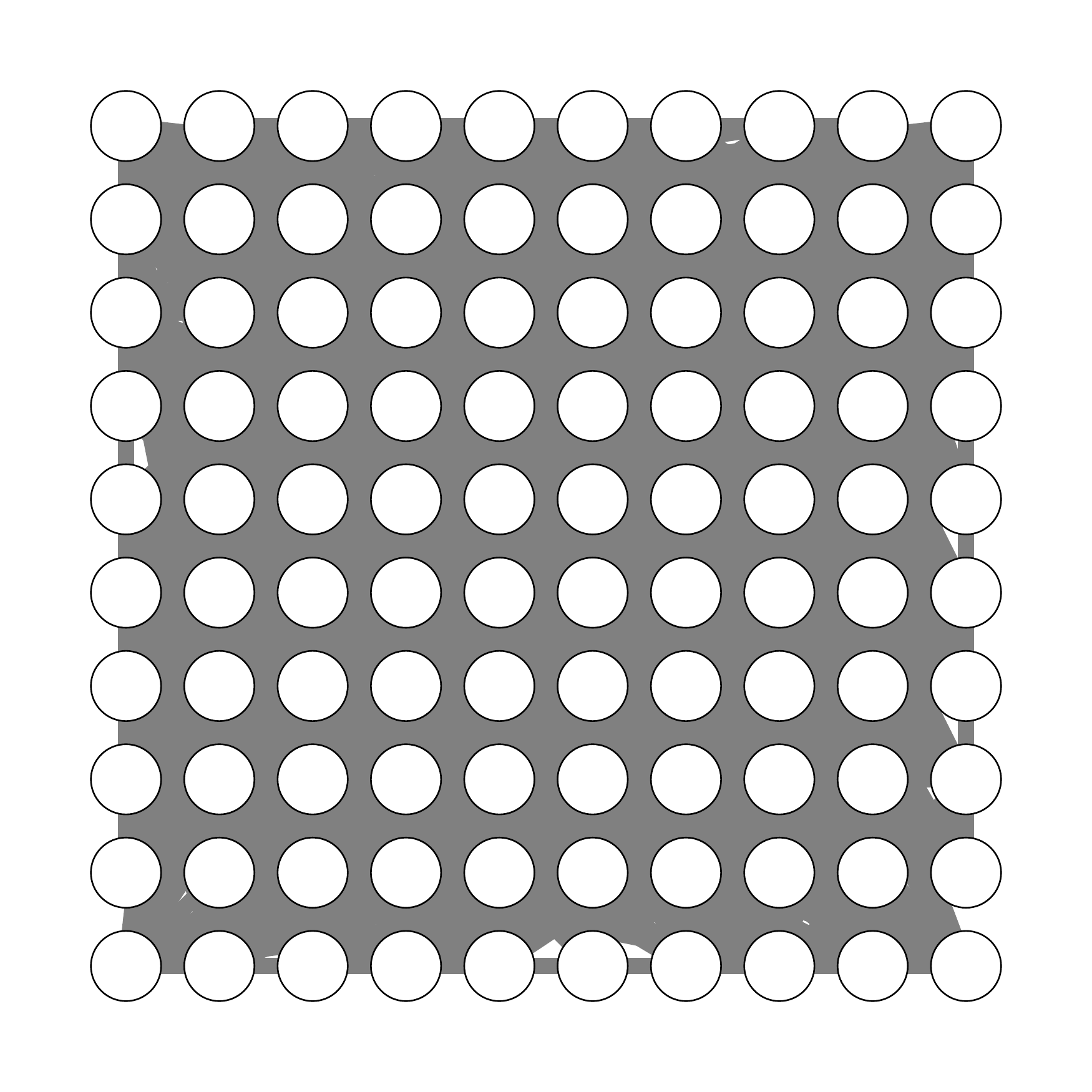} \\ 

\includegraphics[width=0.2\textwidth]{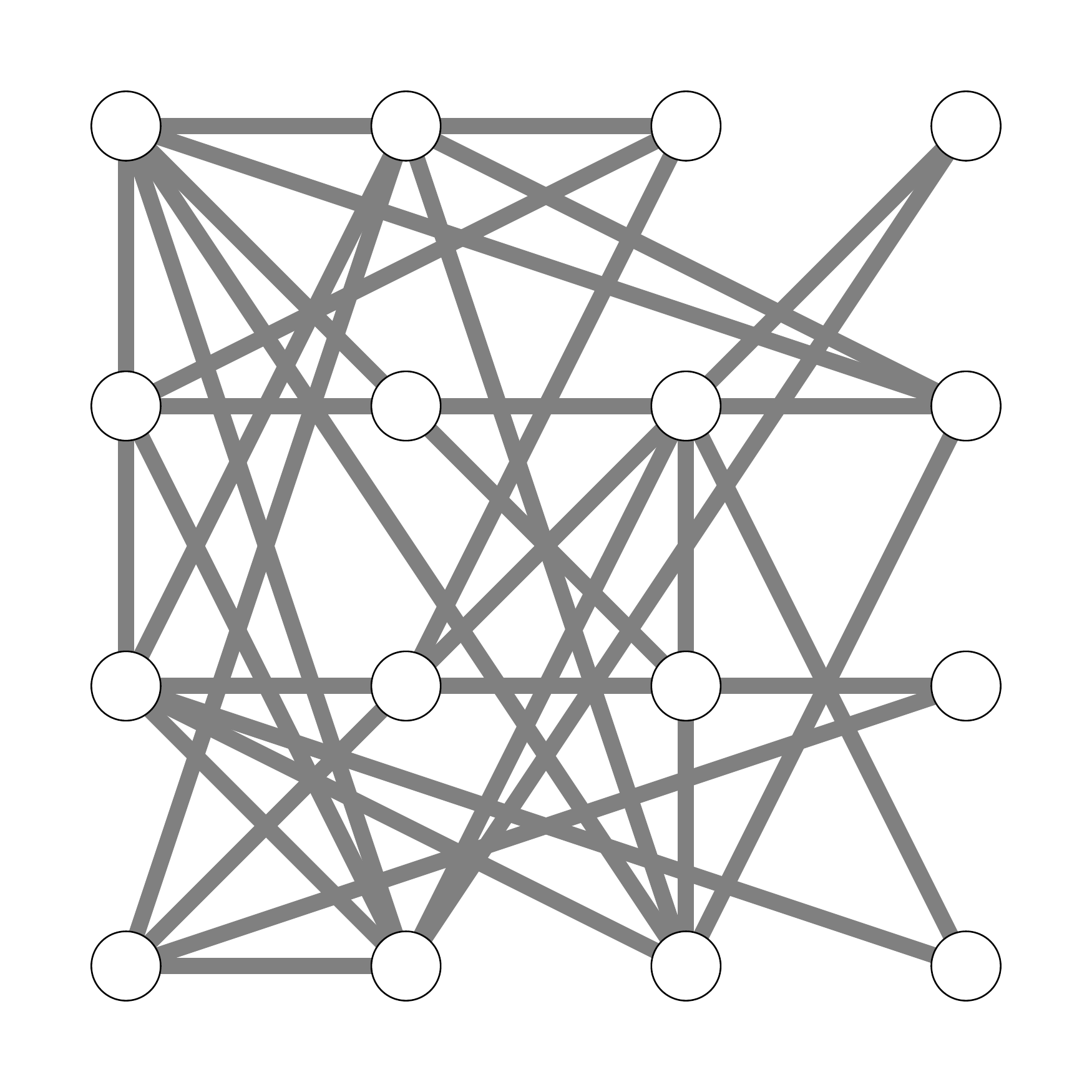}
\put(-120, 40){$p_e = 0.3$}  
\includegraphics[width=0.2\textwidth]{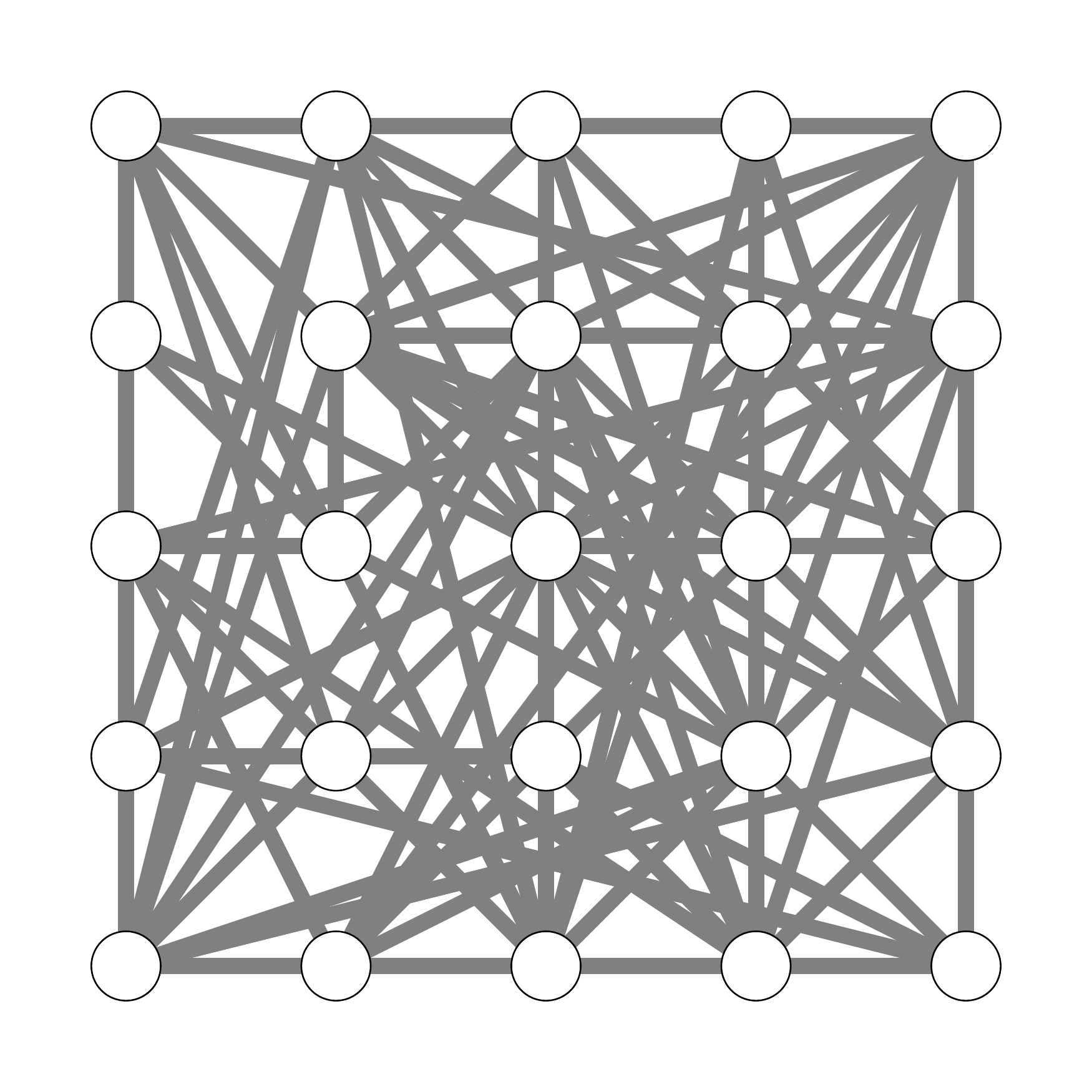} 
\includegraphics[width=0.2\textwidth]{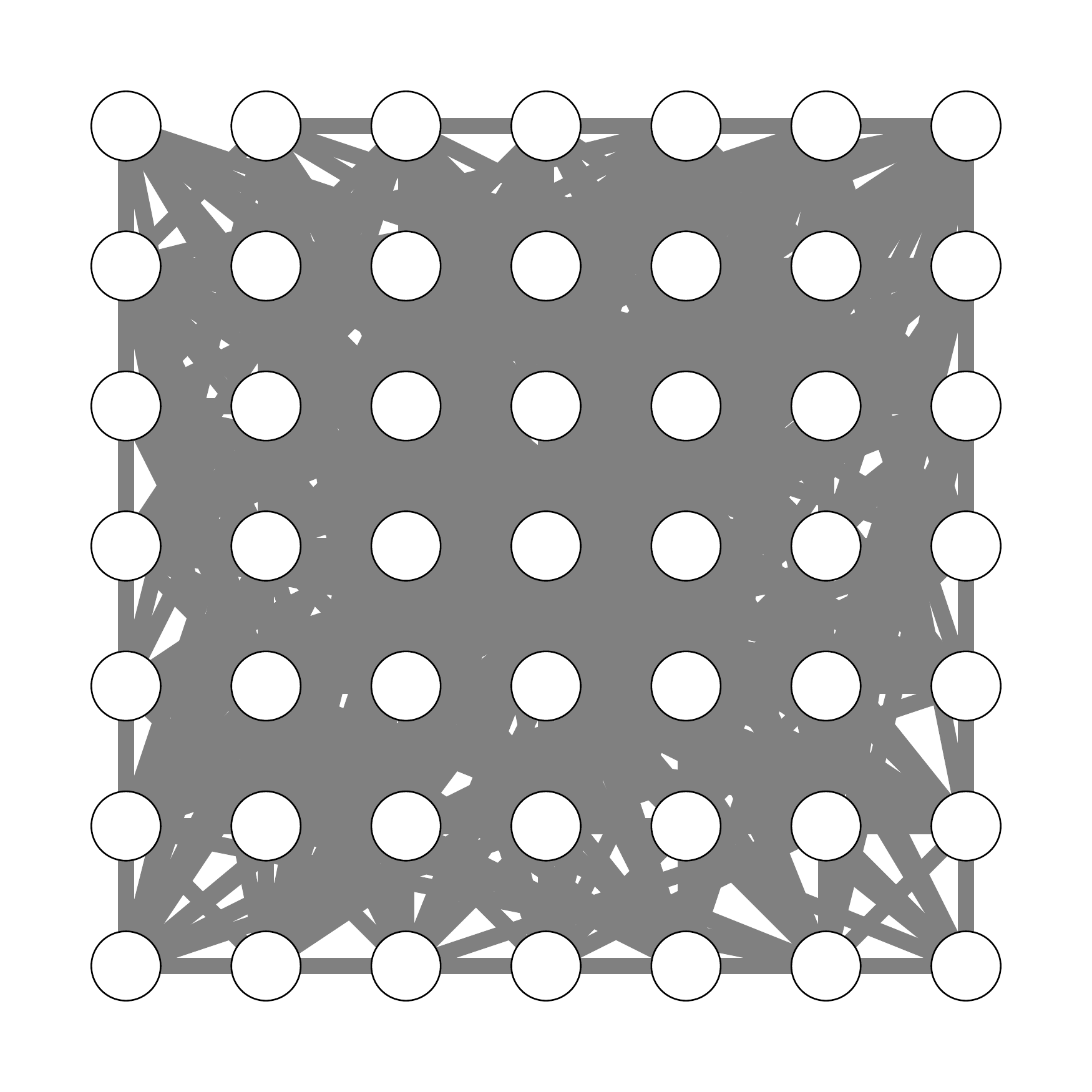} 
\includegraphics[width=0.2\textwidth]{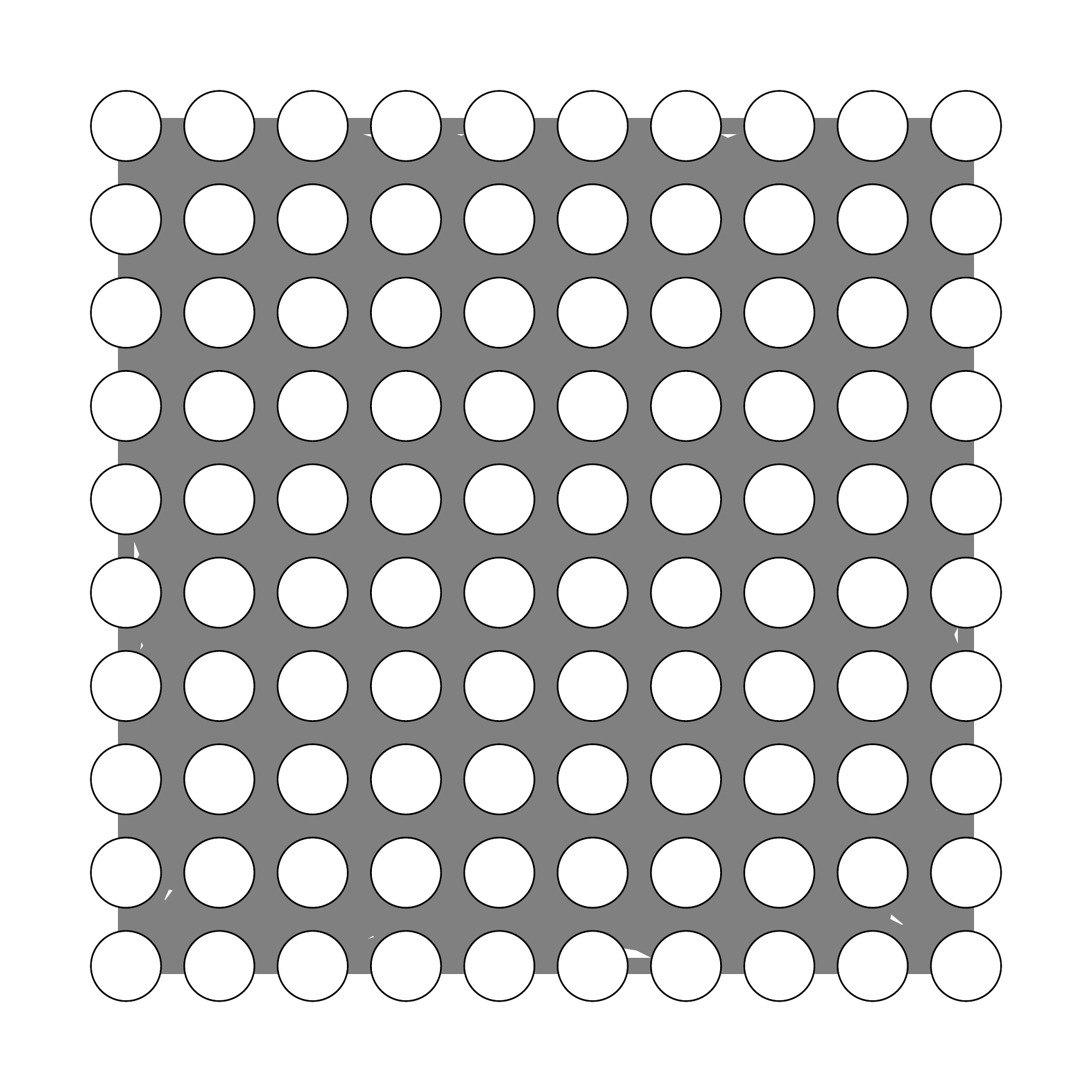} \\

\includegraphics[width=0.2\textwidth]{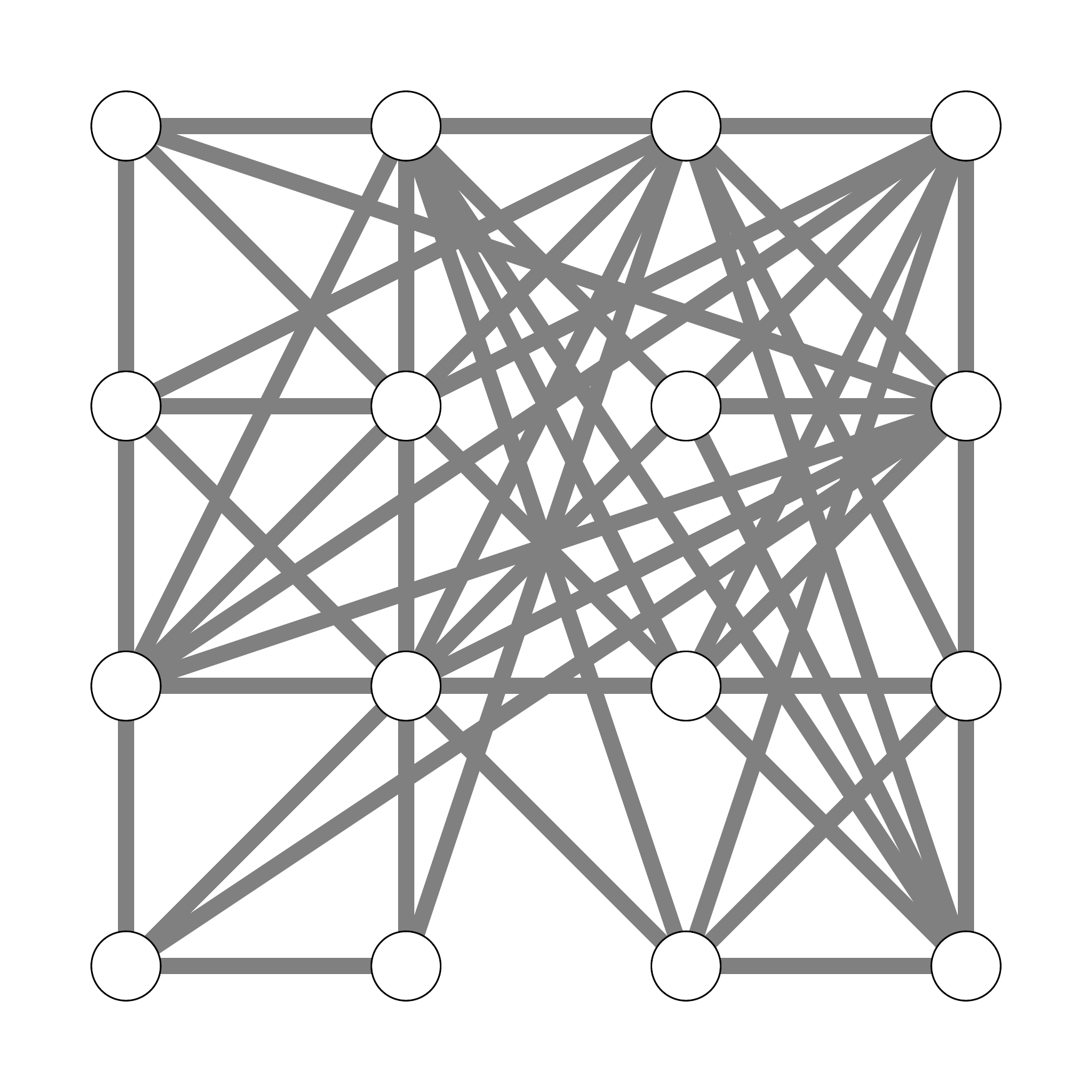}
\put(-120, 40){$p_e = 0.4$}  
\includegraphics[width=0.2\textwidth]{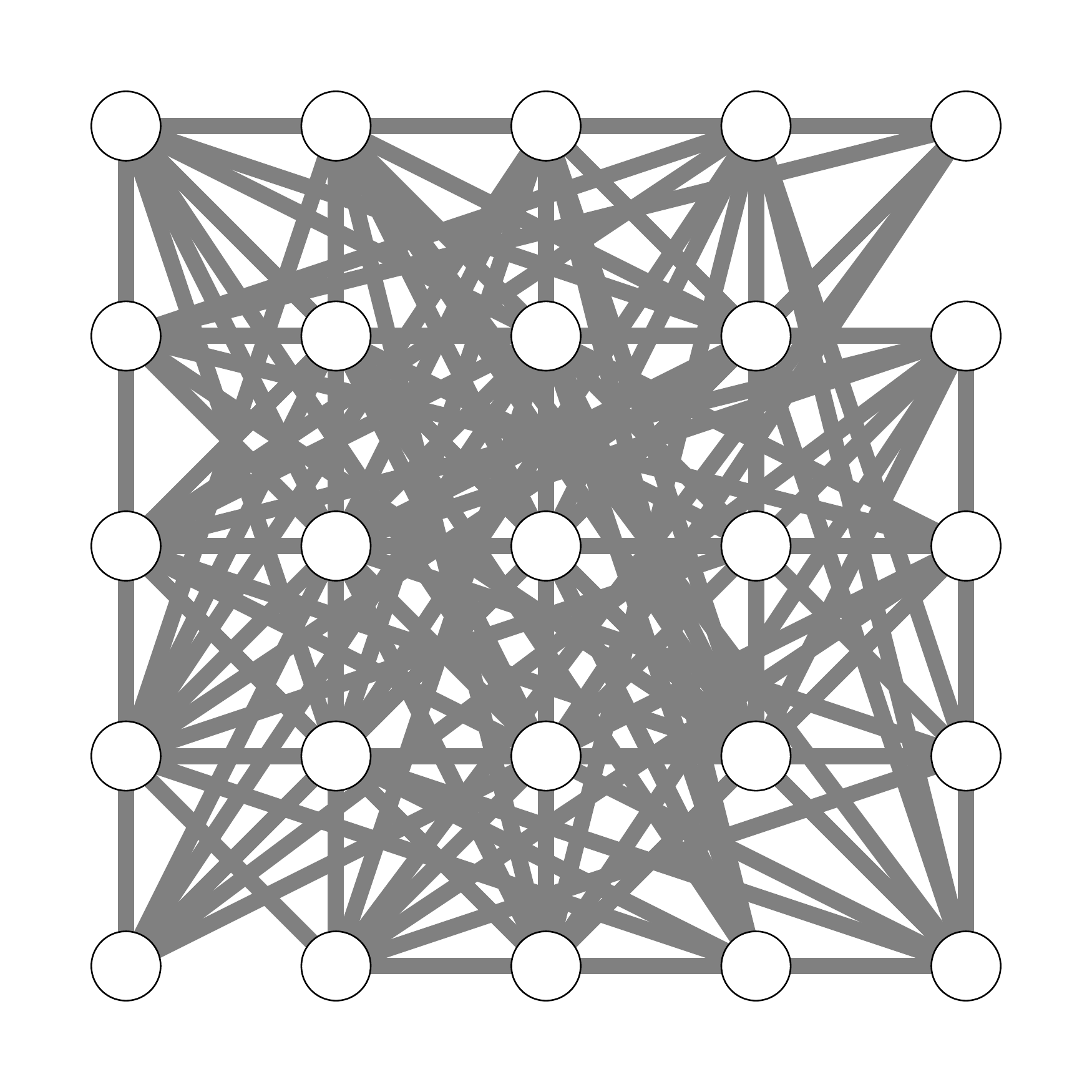} 
\includegraphics[width=0.2\textwidth]{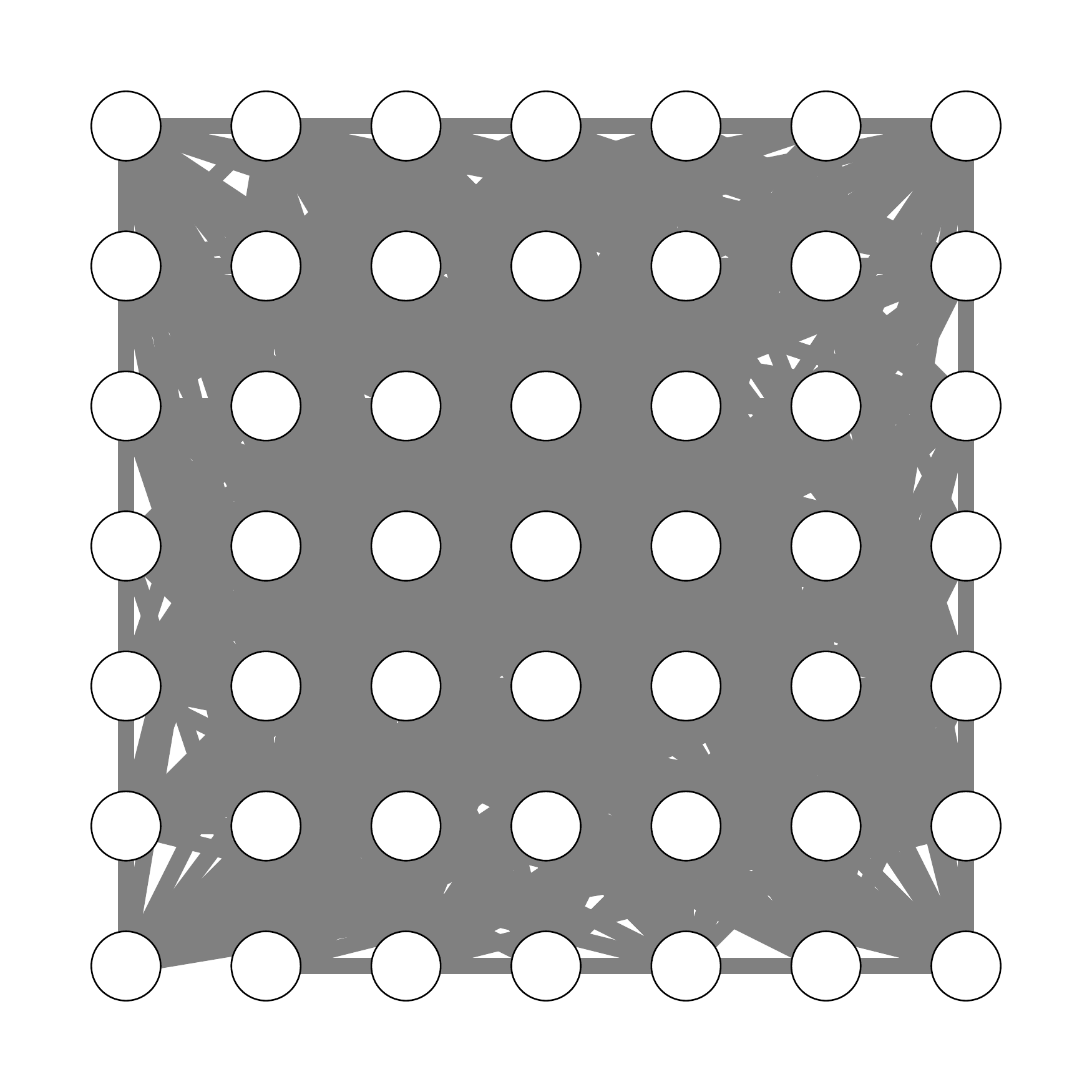} 
\includegraphics[width=0.2\textwidth]{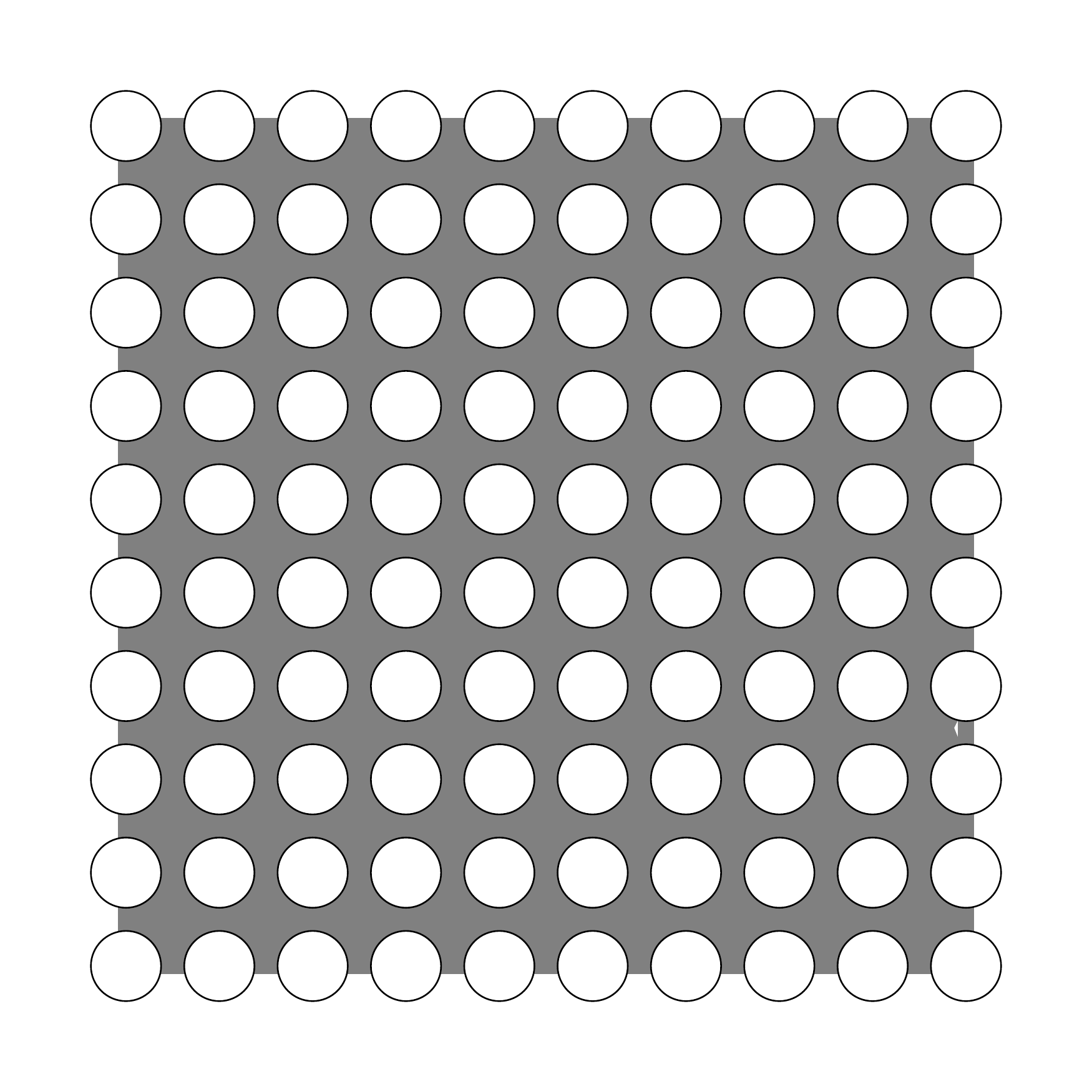} \\
 
\includegraphics[width=0.2\textwidth]{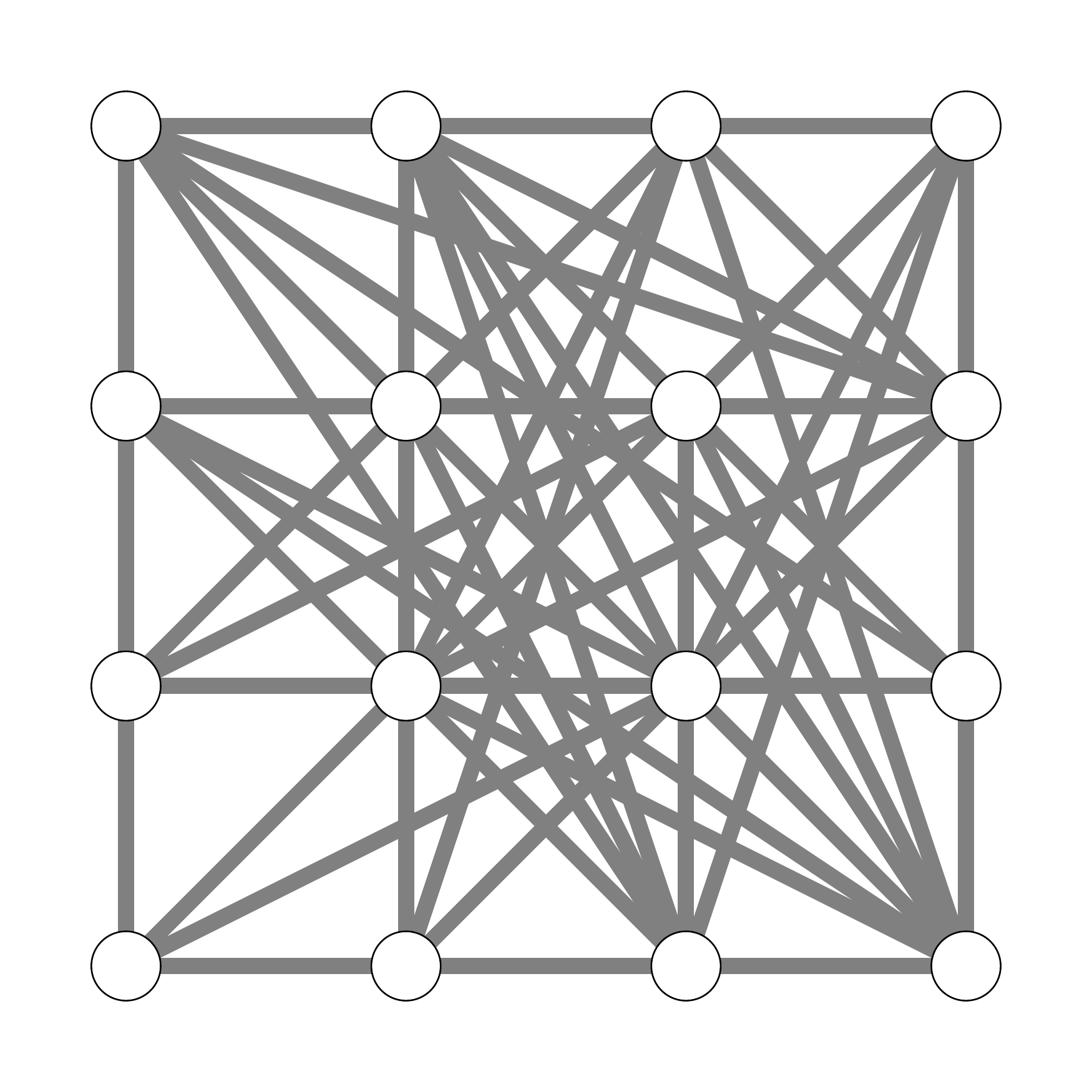}
\put(-120, 40){$p_e = 0.5$}  
\includegraphics[width=0.2\textwidth]{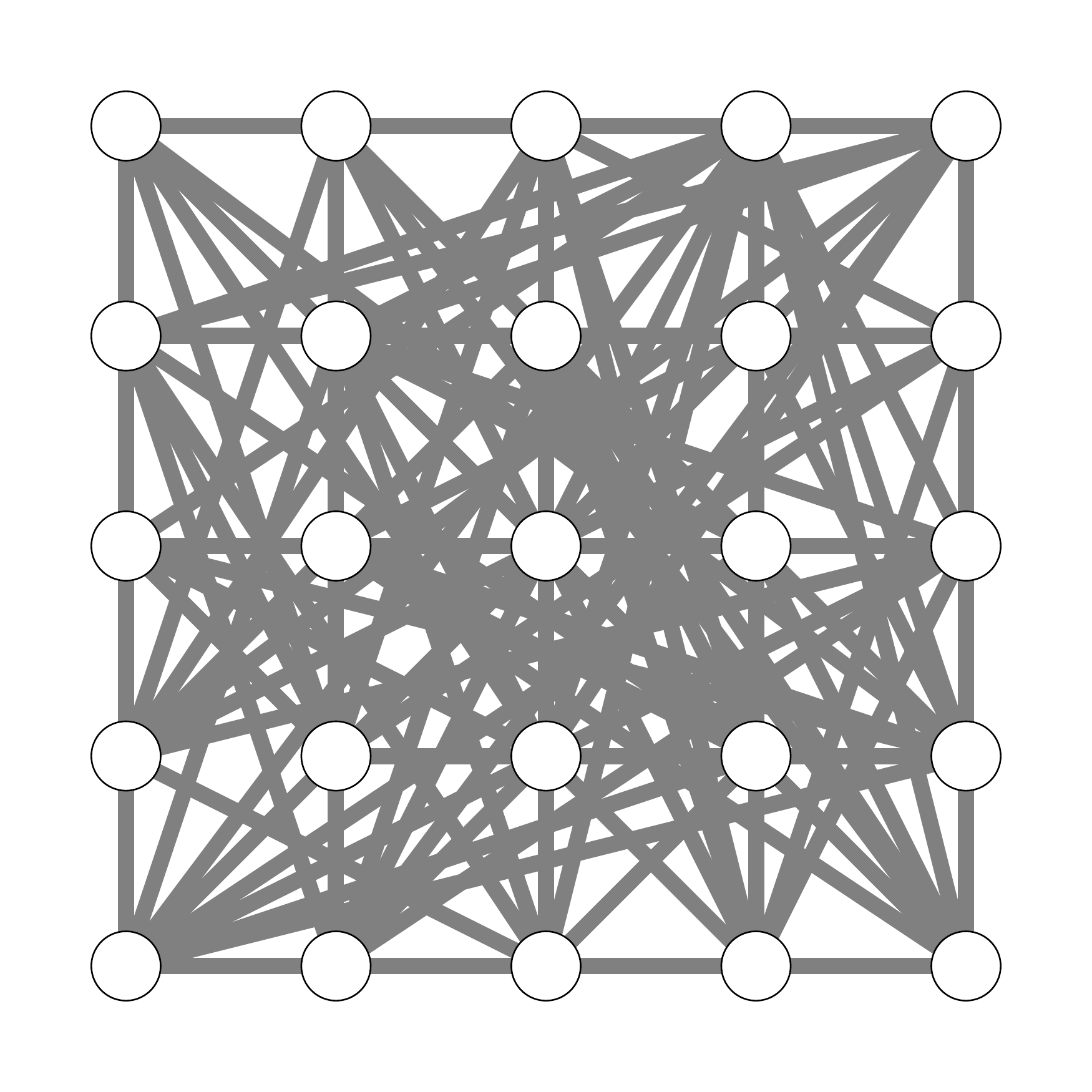} 
\includegraphics[width=0.2\textwidth]{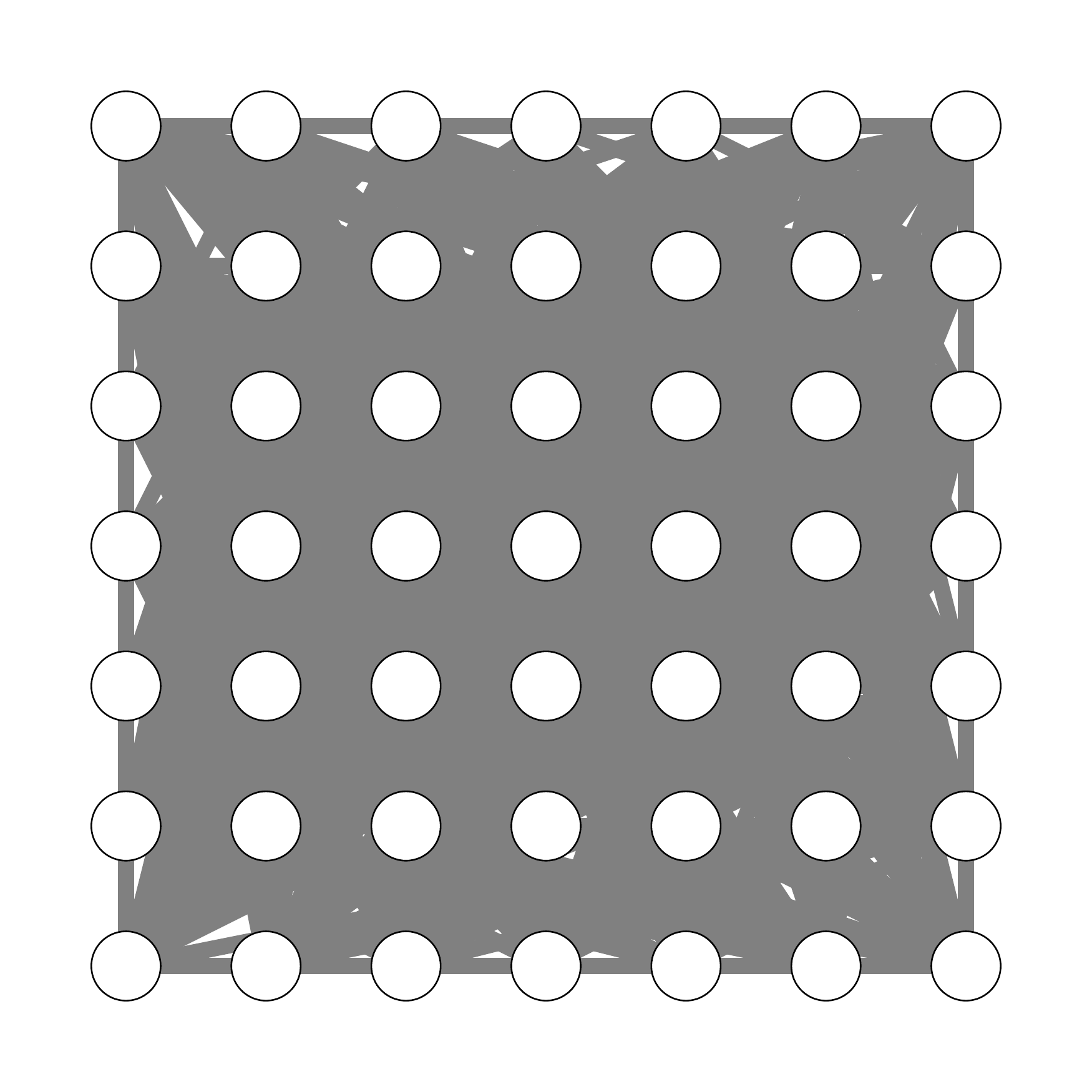} 
\includegraphics[width=0.2\textwidth]{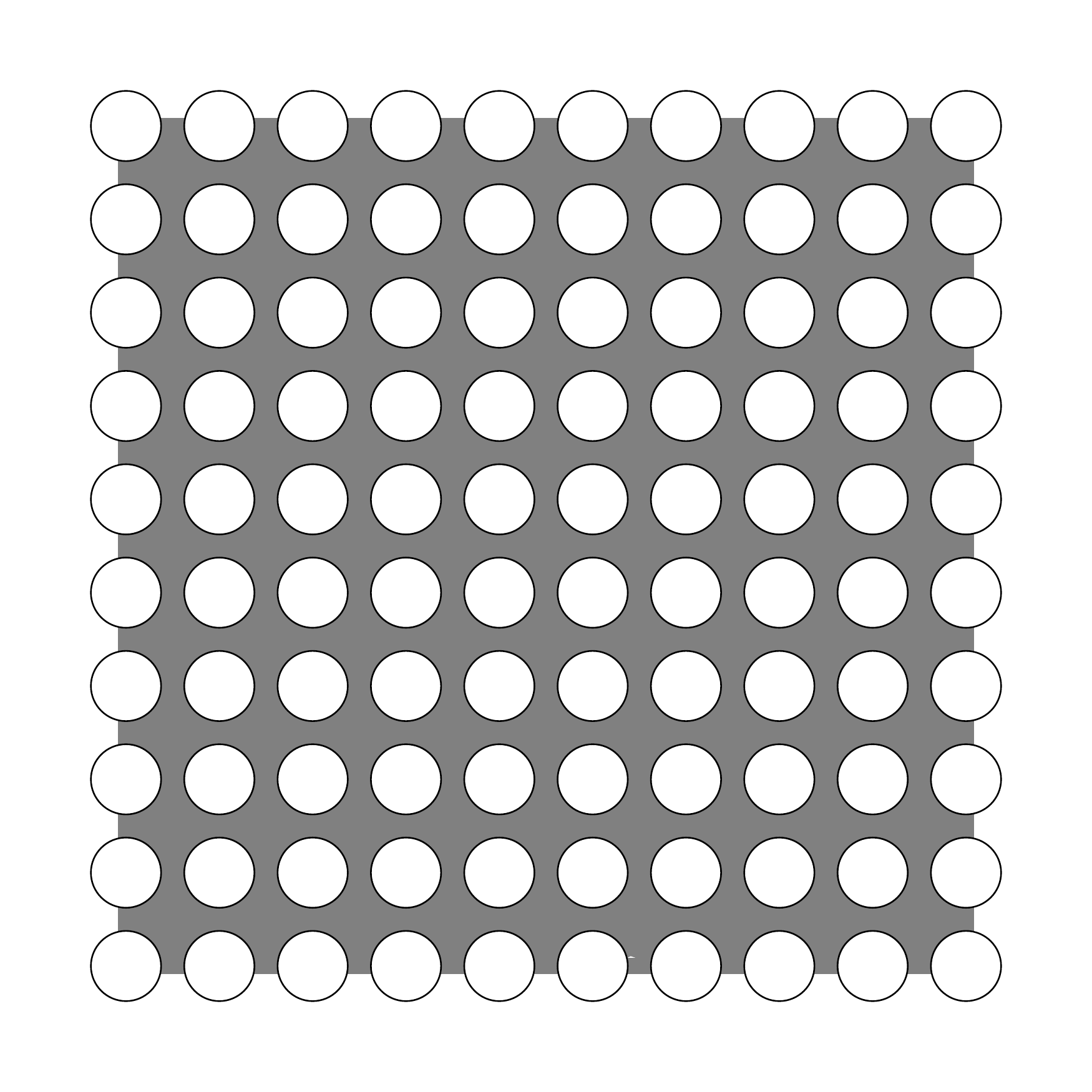} \\
\caption{Examples of simulated random graphs with network size $n \in \left\{16, 25, 49, 100\right\}$ and $p_e \in \left\{0.1, 0.2, 0.3, 0.4, 0.5\right\}$.}
\label{fig:RGNetworks}
\end{figure*}

\begin{figure*}
\includegraphics[width=0.2\textwidth]{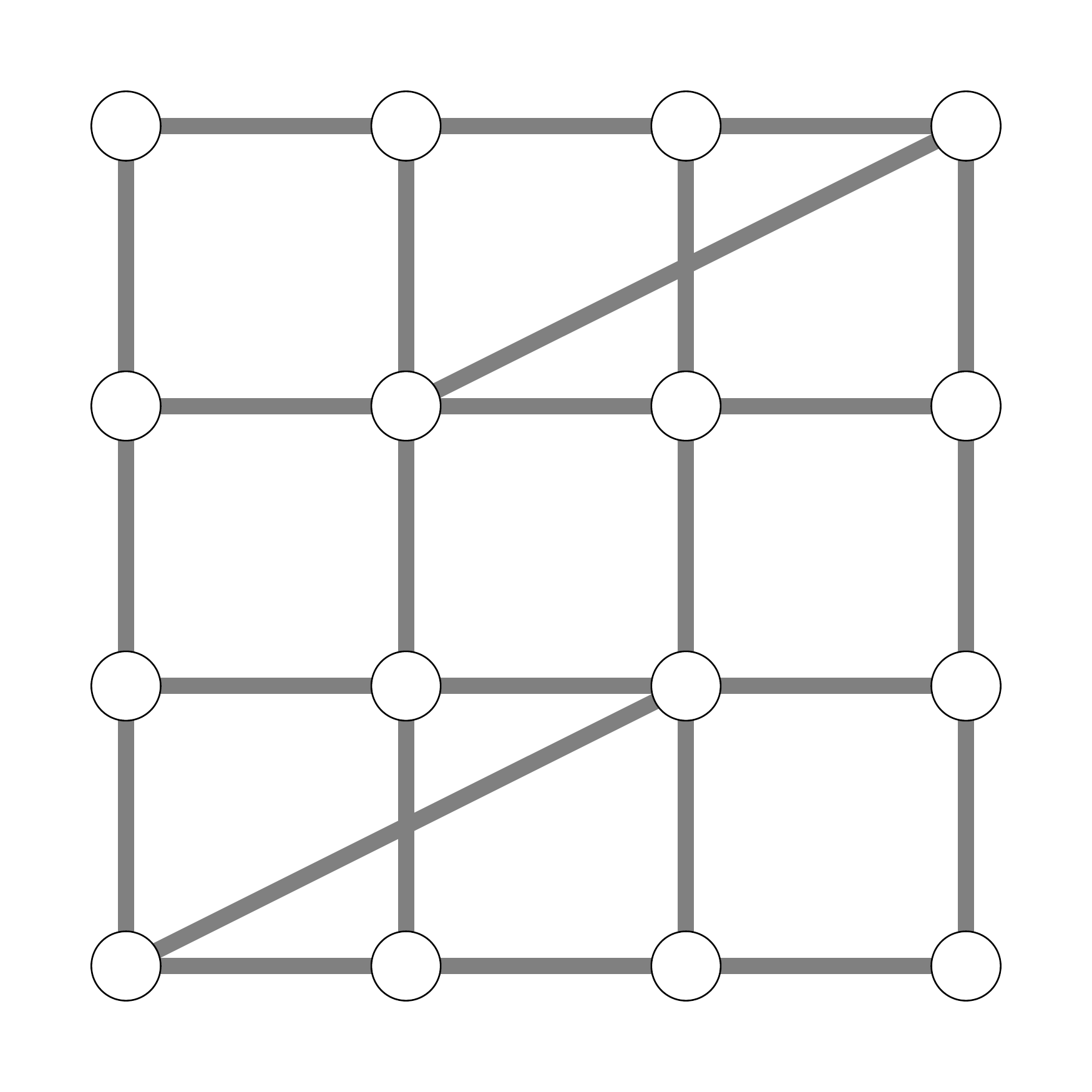}
\put(-55, 90){$n = 16$} 
\put(-120, 40){$p_w = 0.1$}
\includegraphics[width=0.2\textwidth]{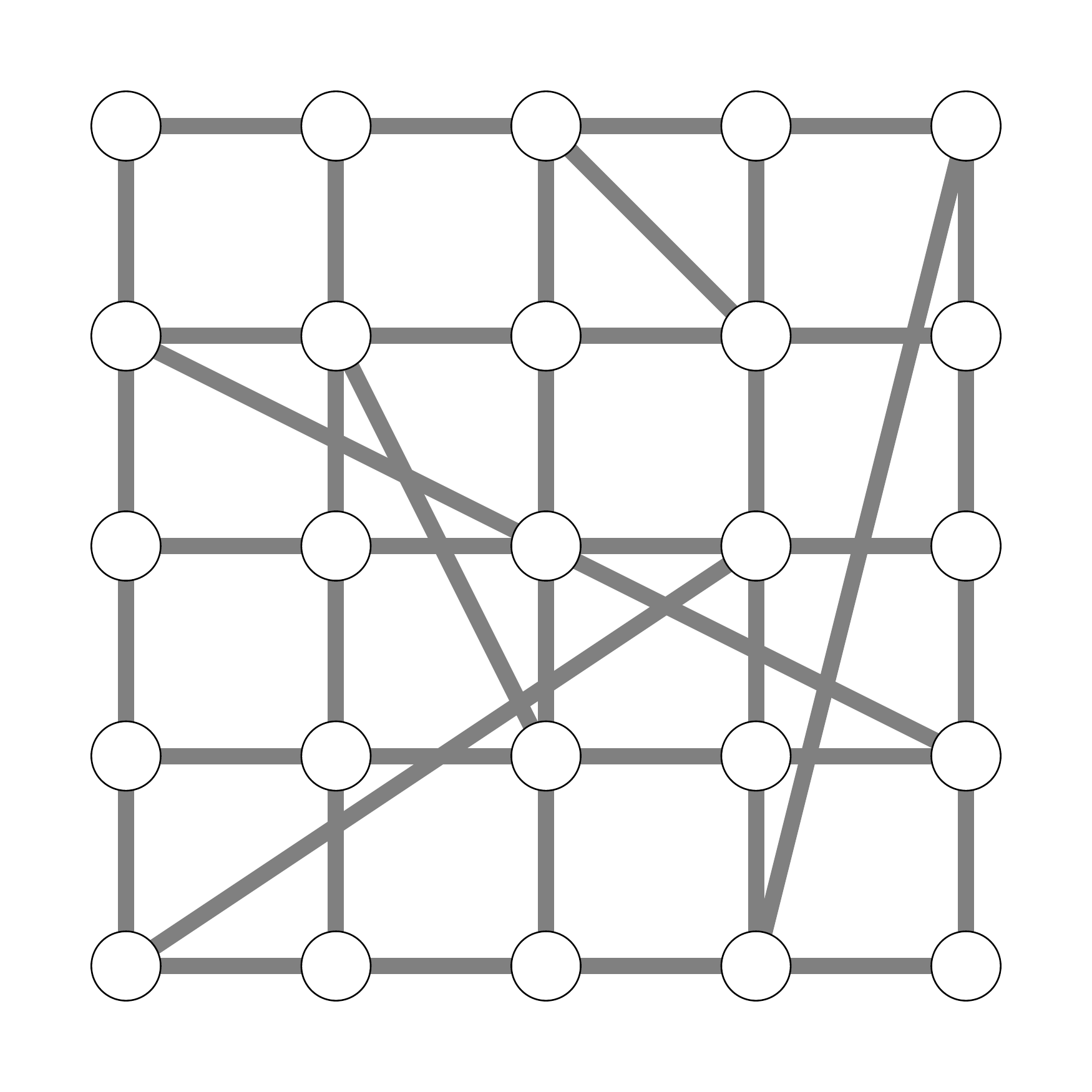}
\put(-55, 90){$n = 25$}   
\includegraphics[width=0.2\textwidth]{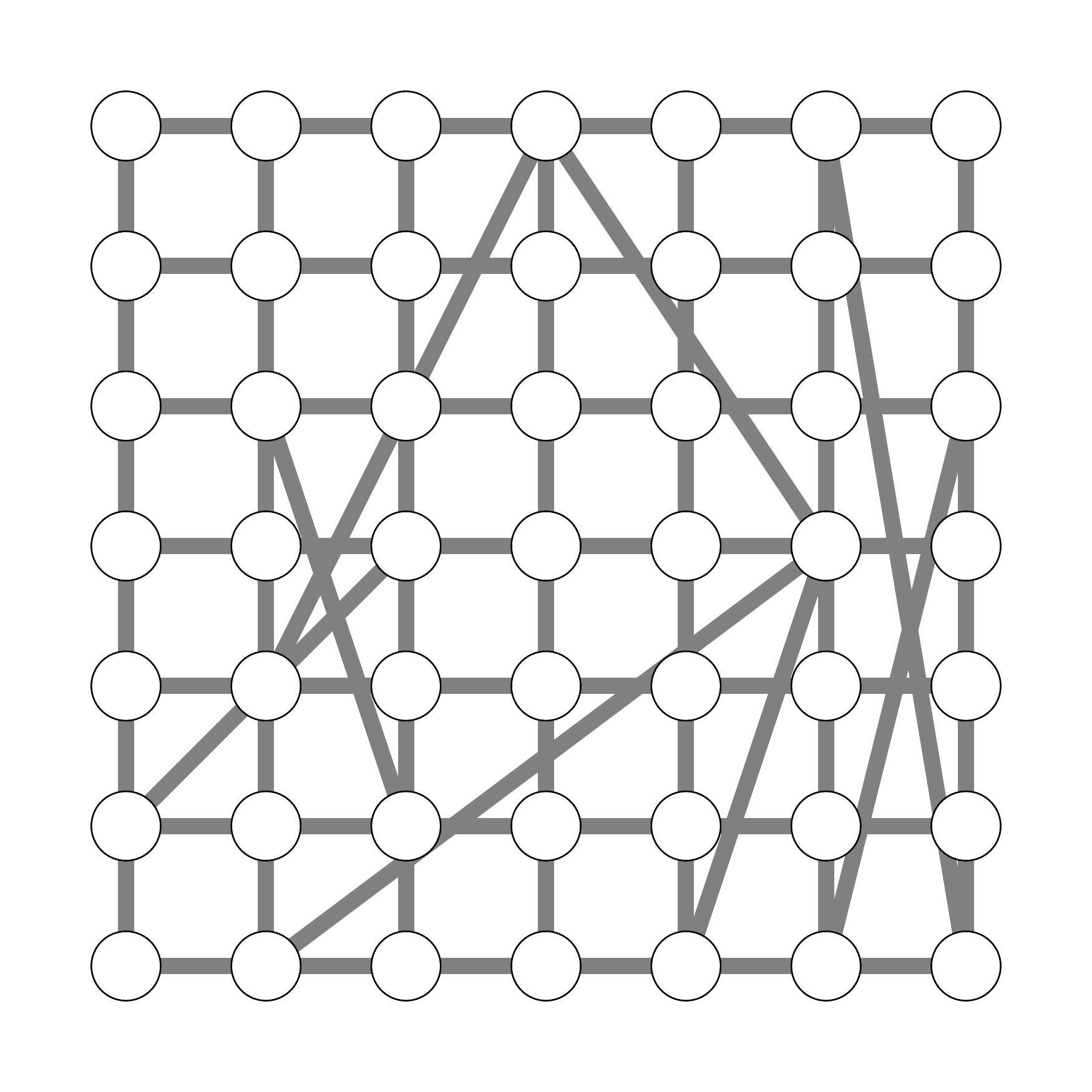}
\put(-55, 90){$n = 49$}   
\includegraphics[width=0.2\textwidth]{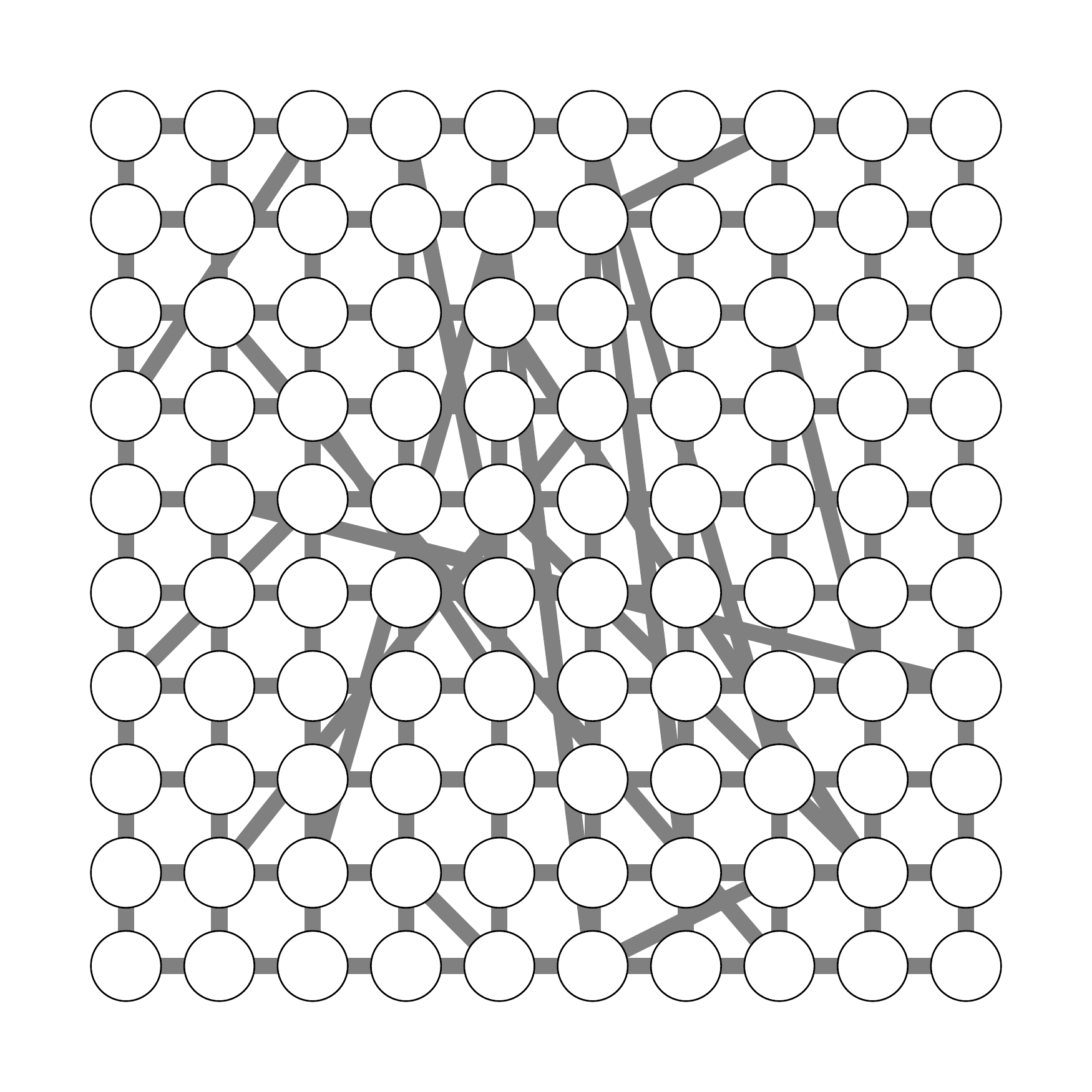}
\put(-55, 90){$n = 100$} \\

\includegraphics[width=0.2\textwidth]{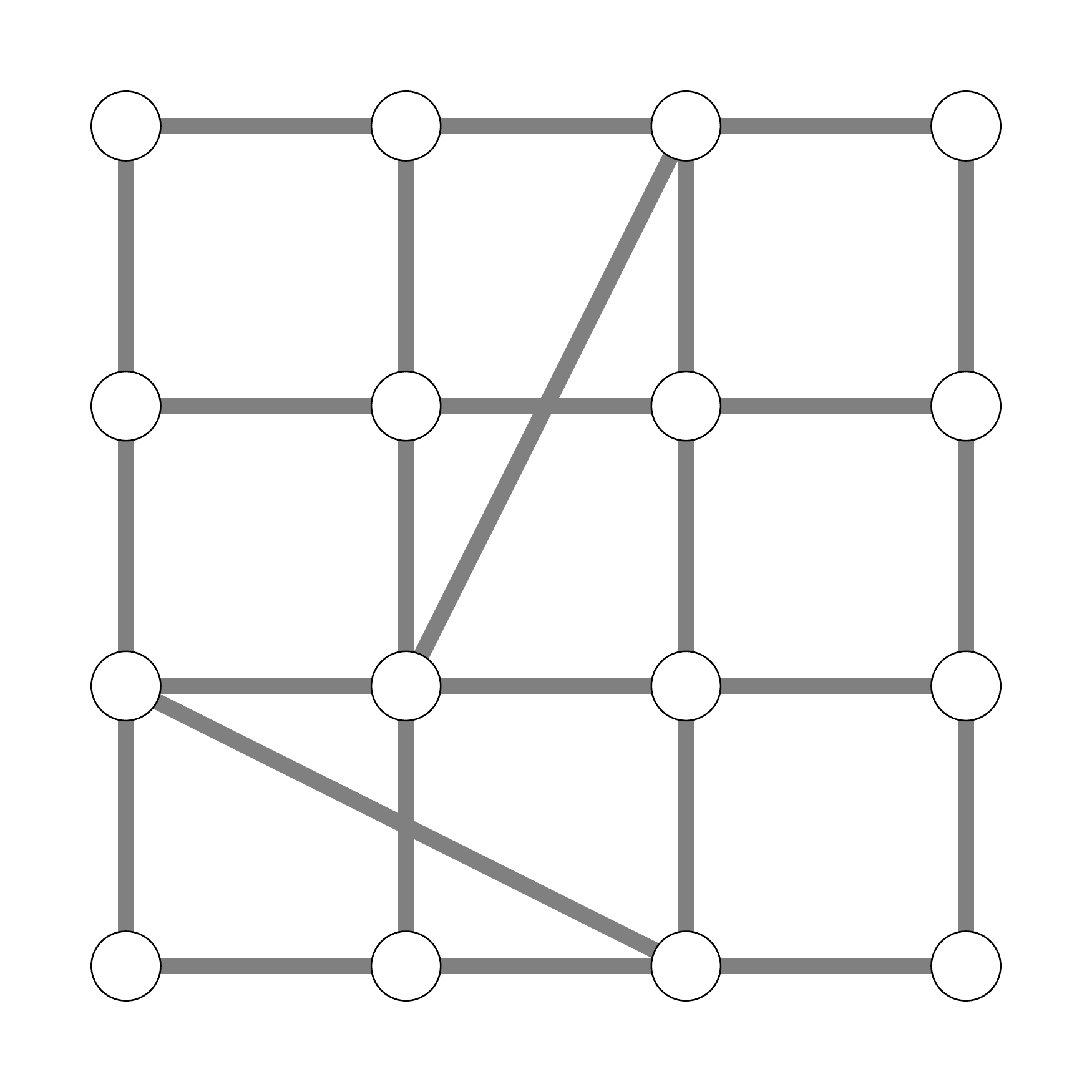}
\put(-120, 40){$p_w = 0.2$} 
\includegraphics[width=0.2\textwidth]{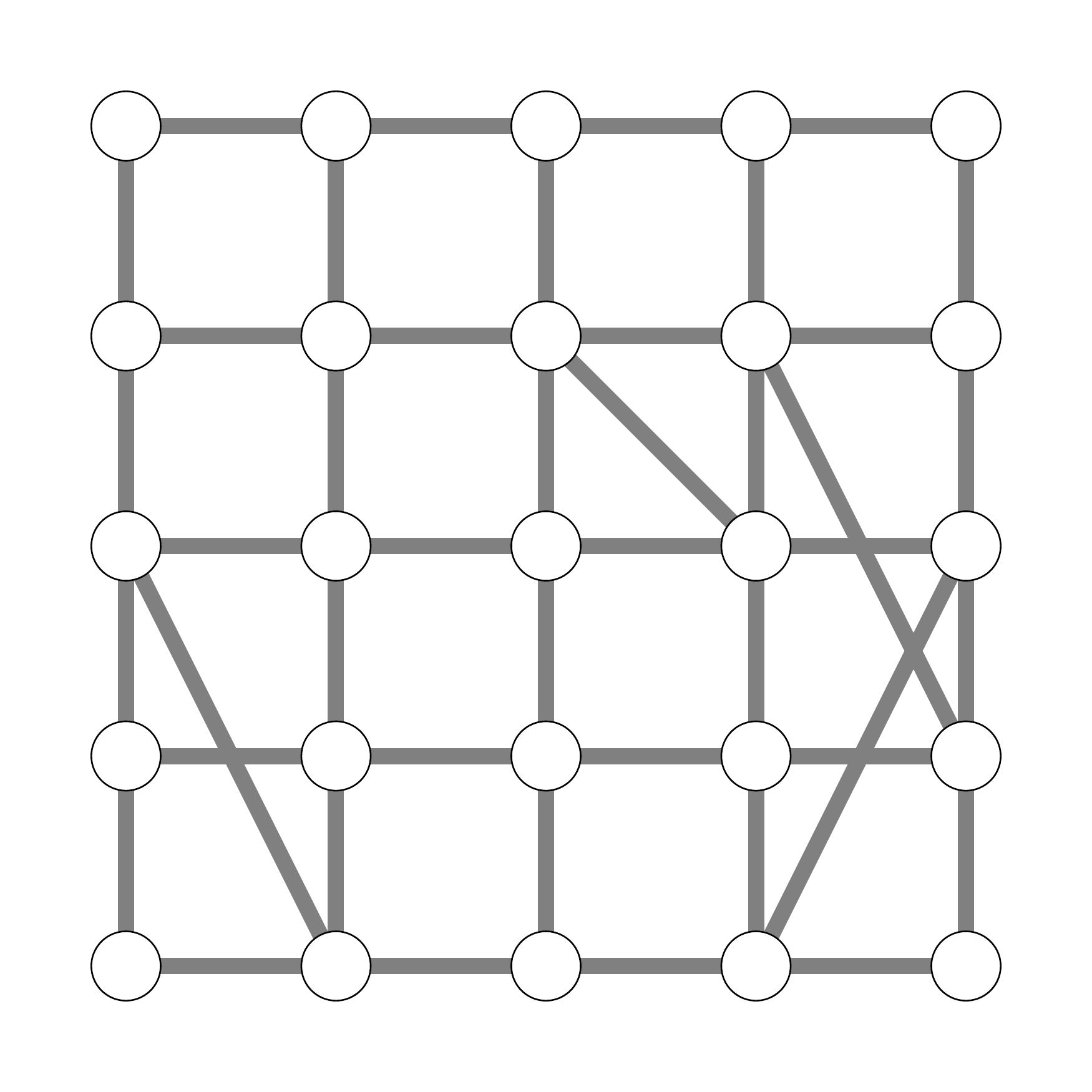}   
\includegraphics[width=0.2\textwidth]{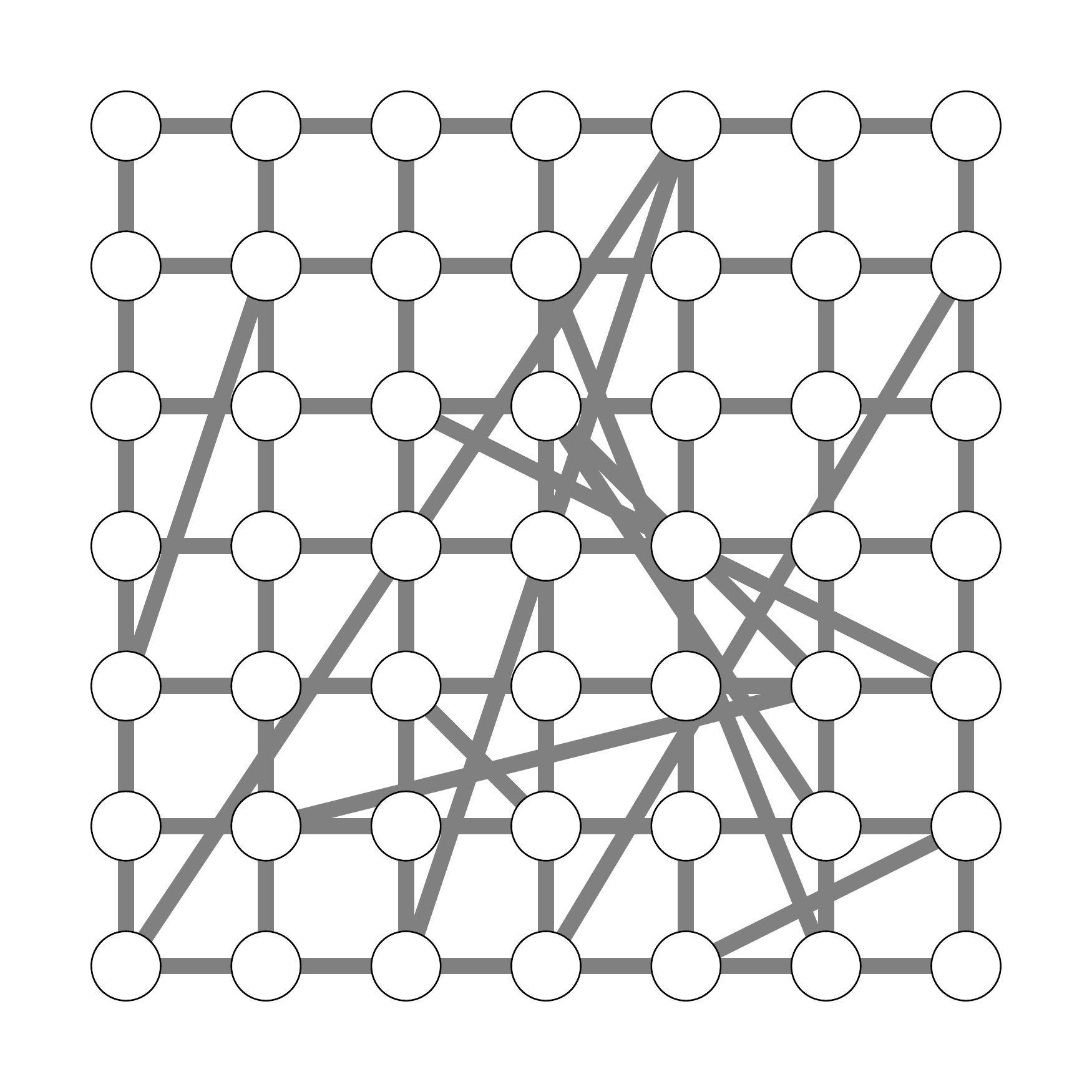} 
\includegraphics[width=0.2\textwidth]{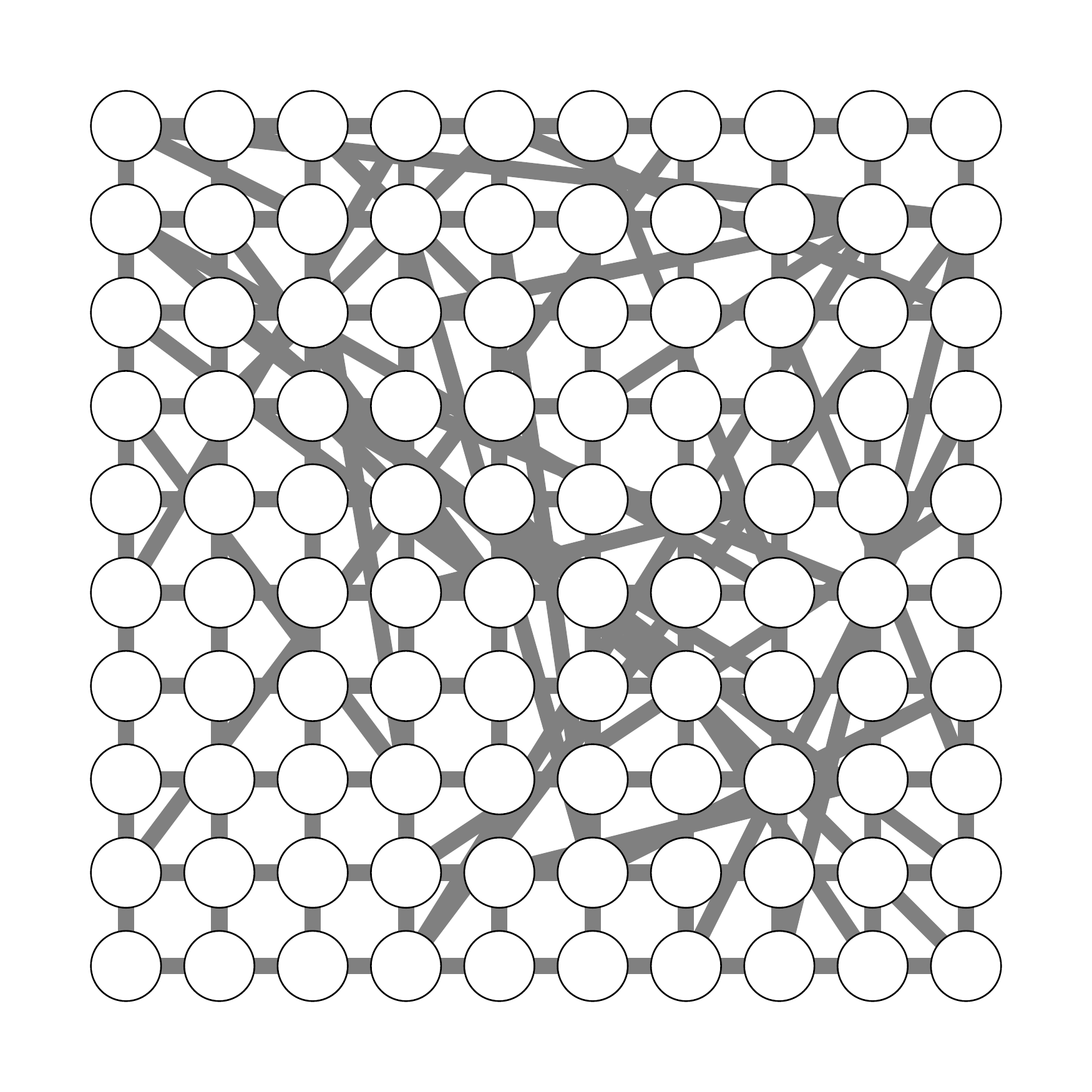} \\ 

\includegraphics[width=0.2\textwidth]{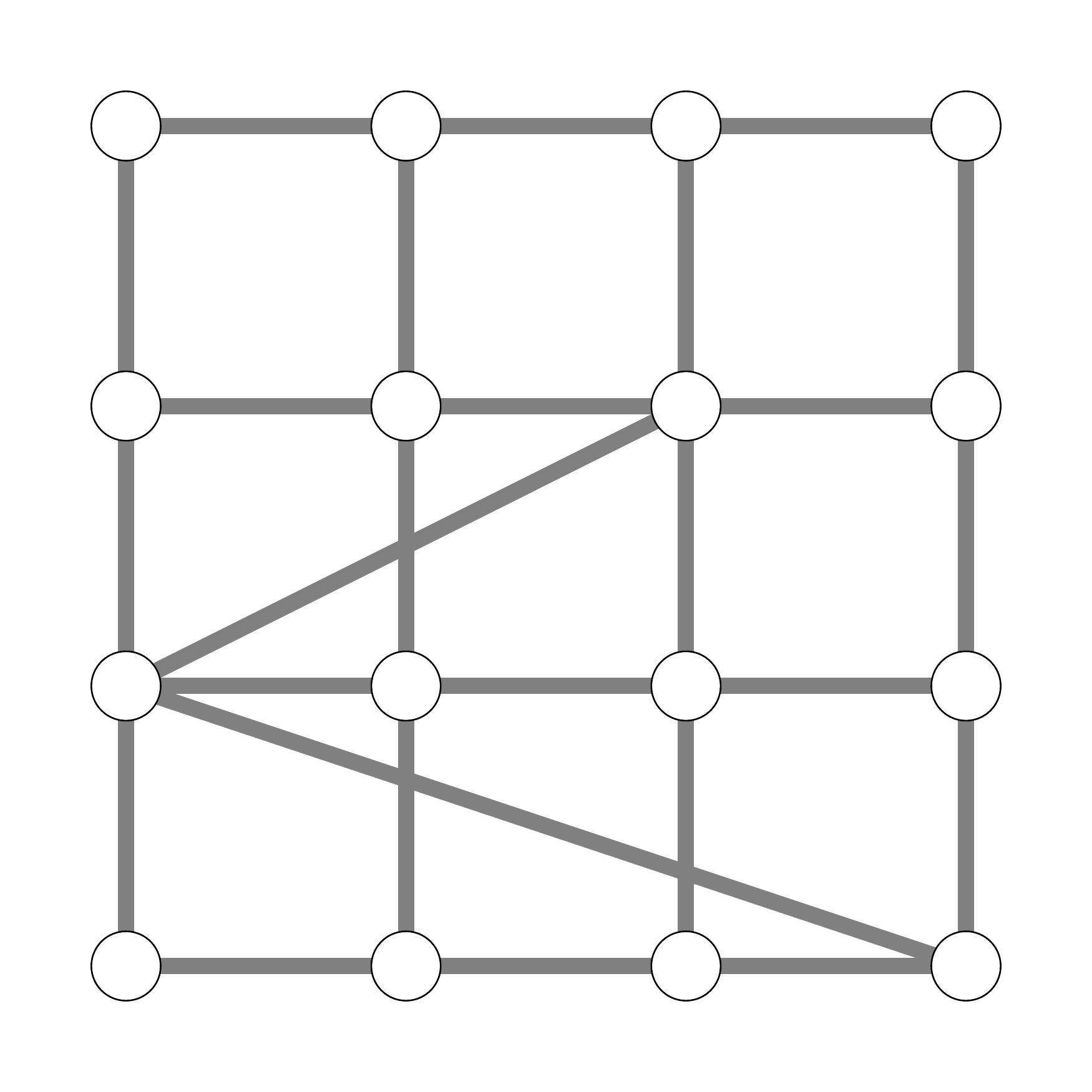}
\put(-120, 40){$p_w = 0.3$}  
\includegraphics[width=0.2\textwidth]{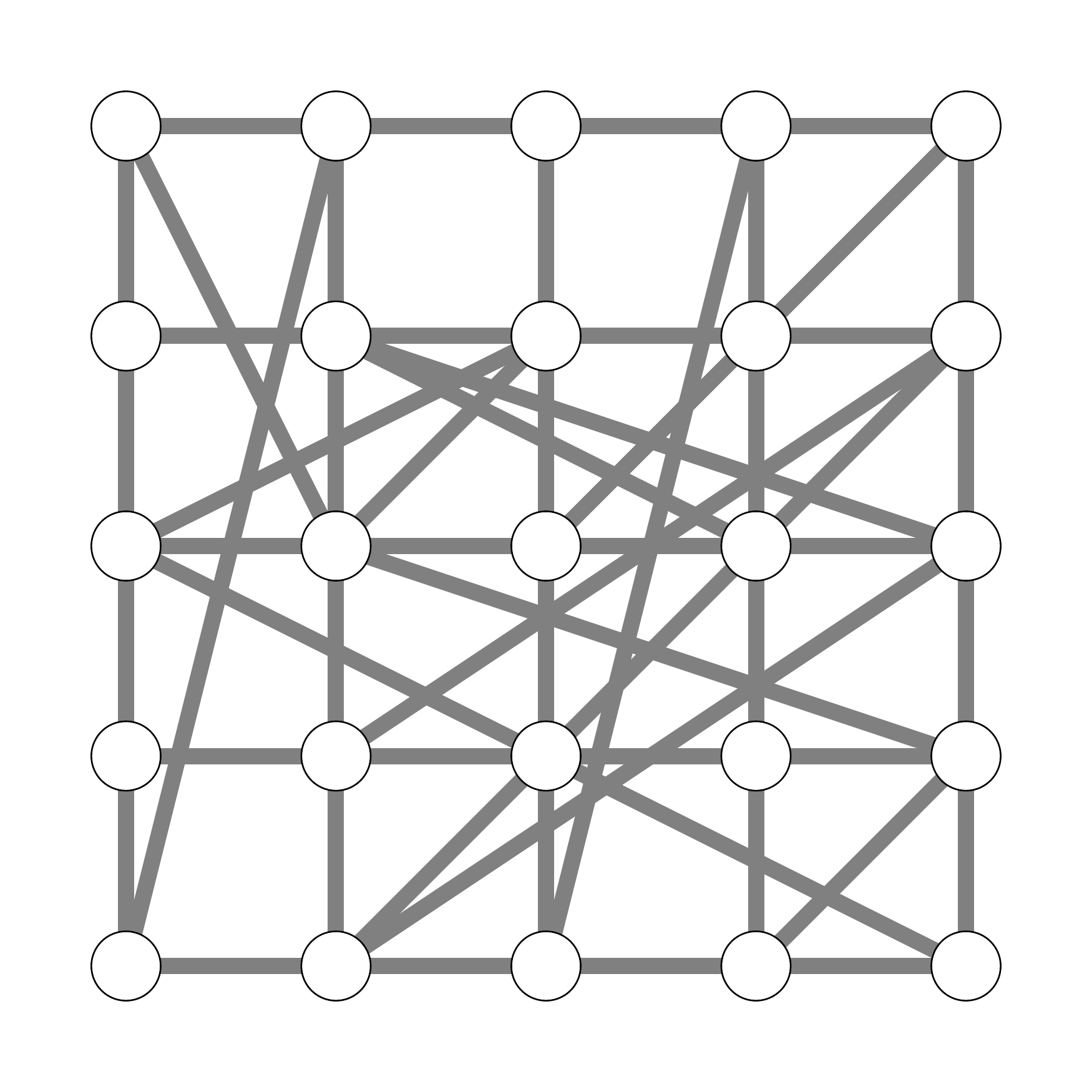}  
\includegraphics[width=0.2\textwidth]{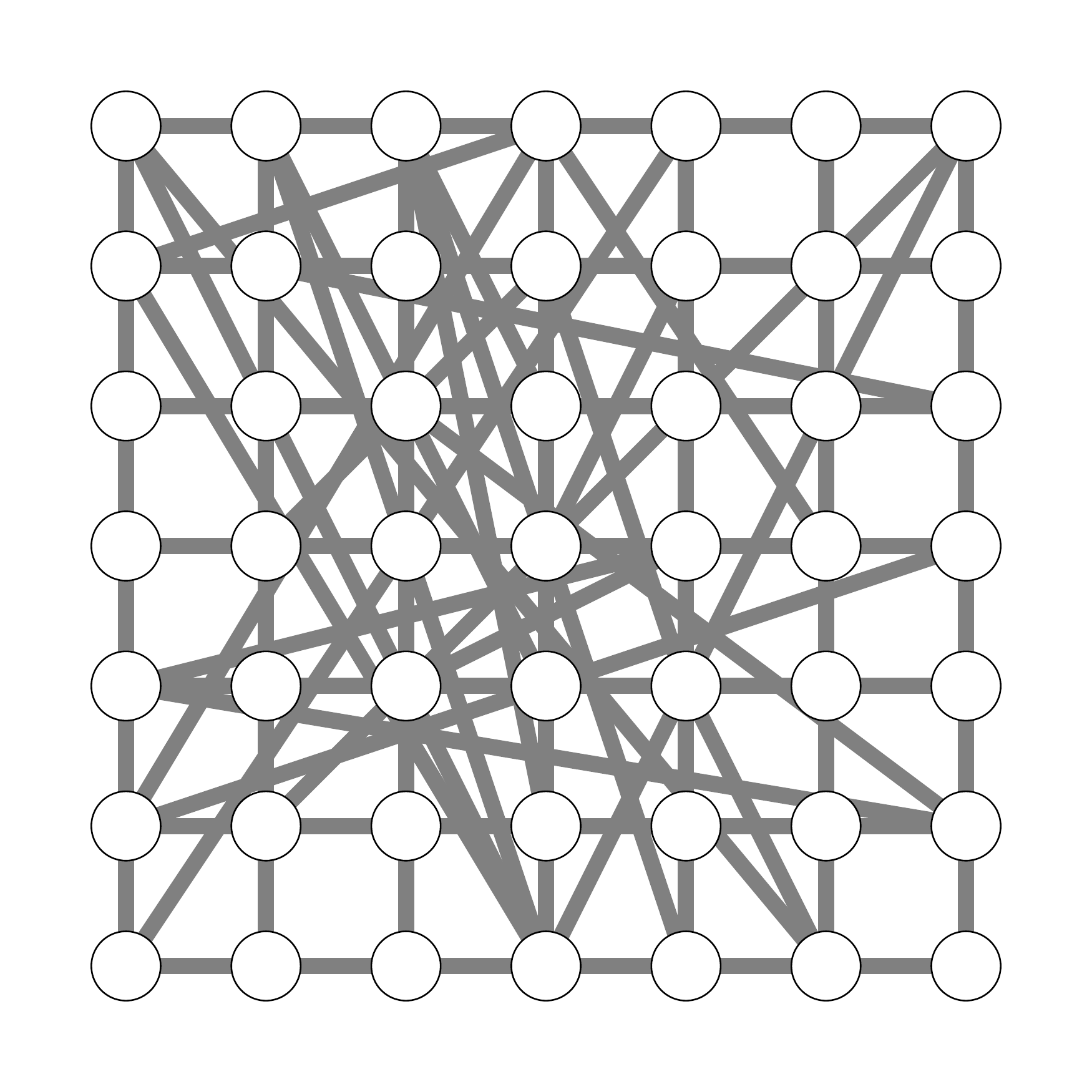}  
\includegraphics[width=0.2\textwidth]{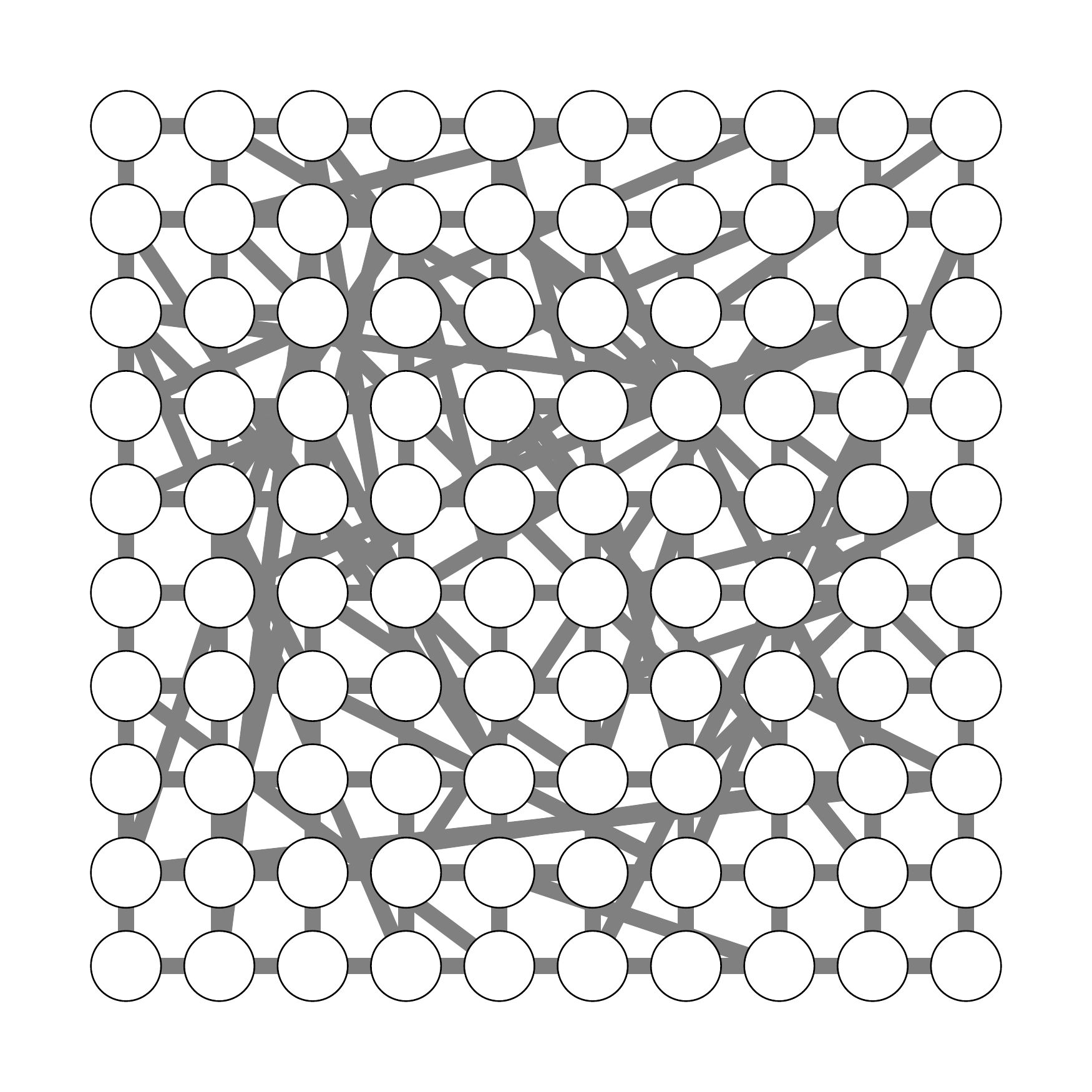} \\

\includegraphics[width=0.2\textwidth]{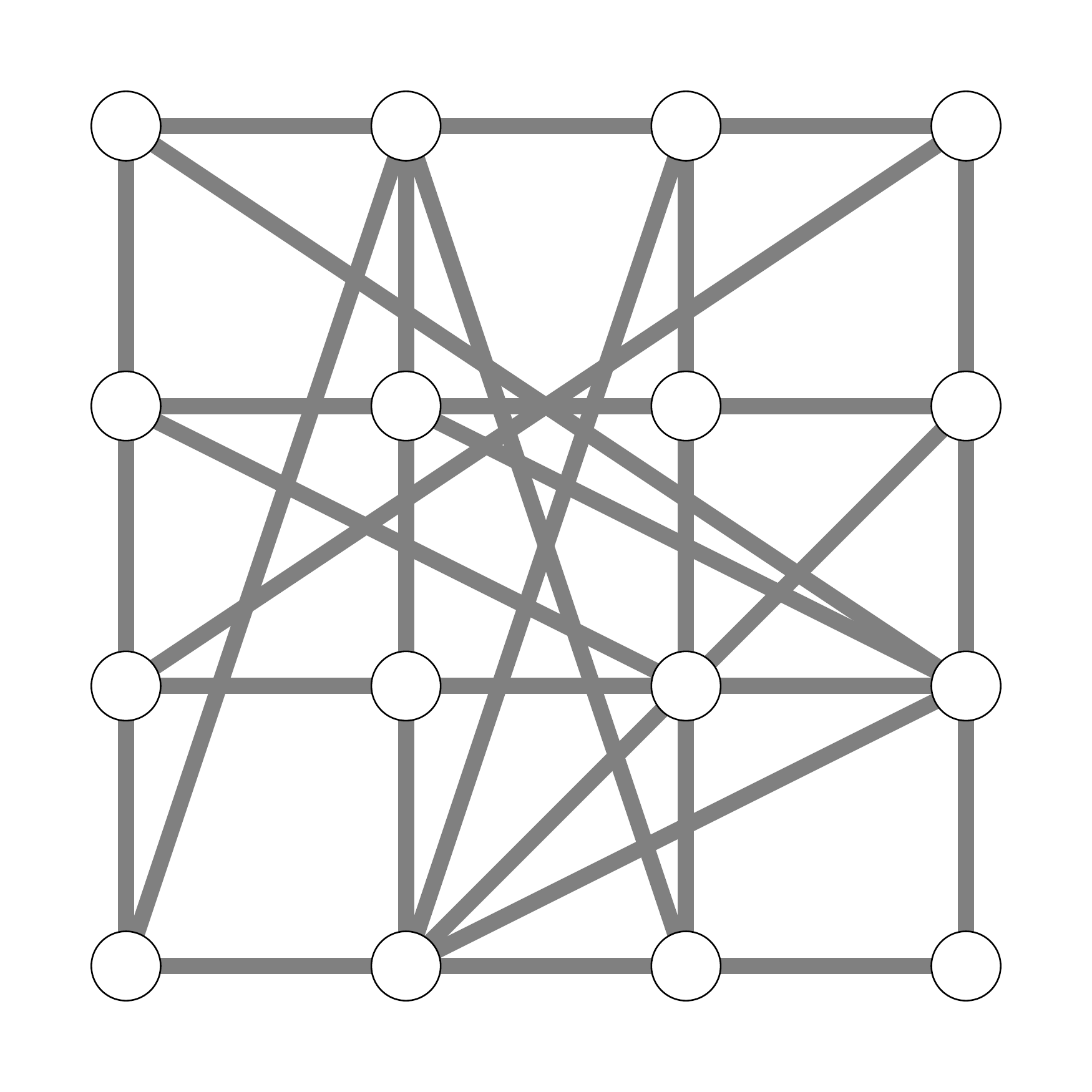} 
\put(-120, 40){$p_w = 0.4$} 
\includegraphics[width=0.2\textwidth]{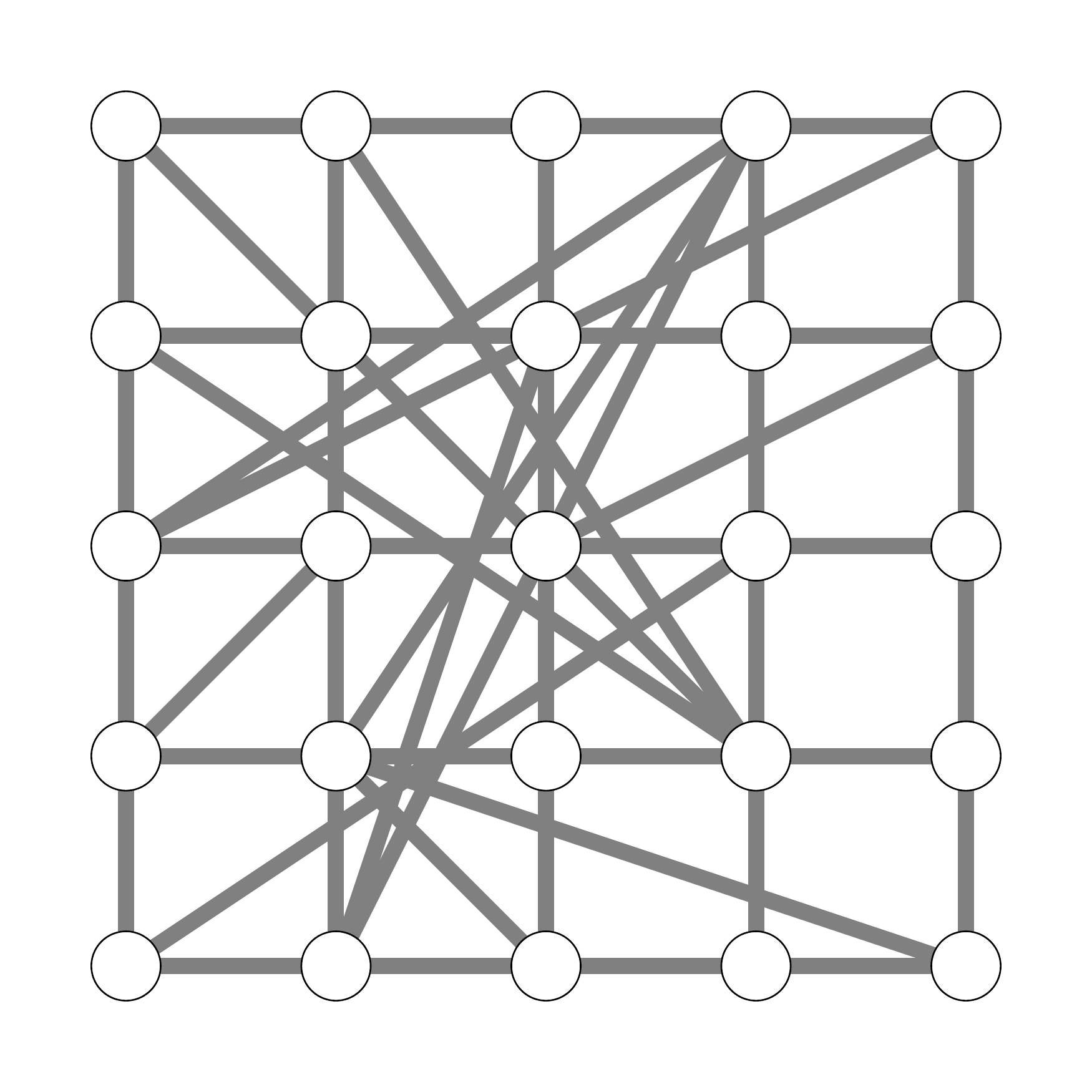}  
\includegraphics[width=0.2\textwidth]{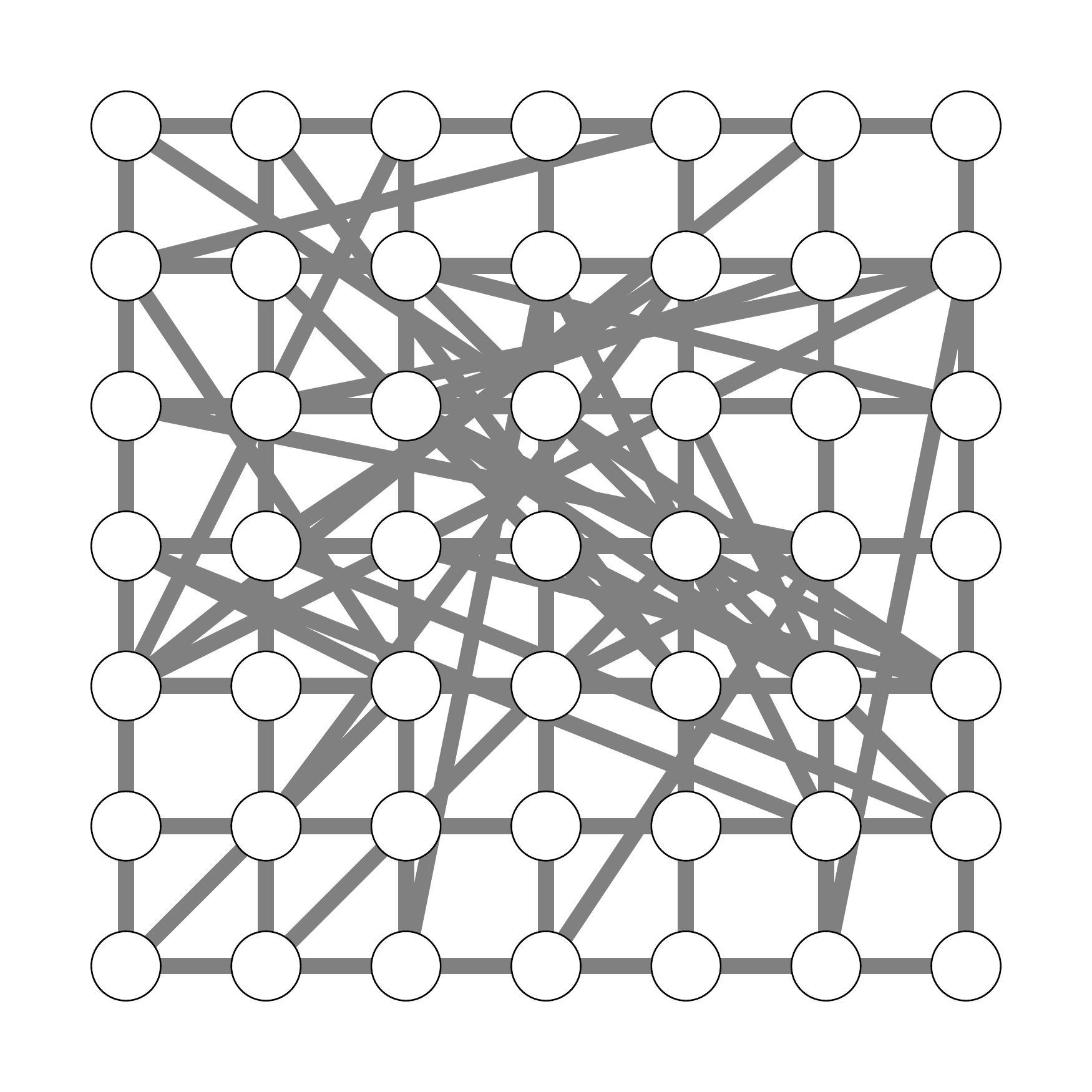}  
\includegraphics[width=0.2\textwidth]{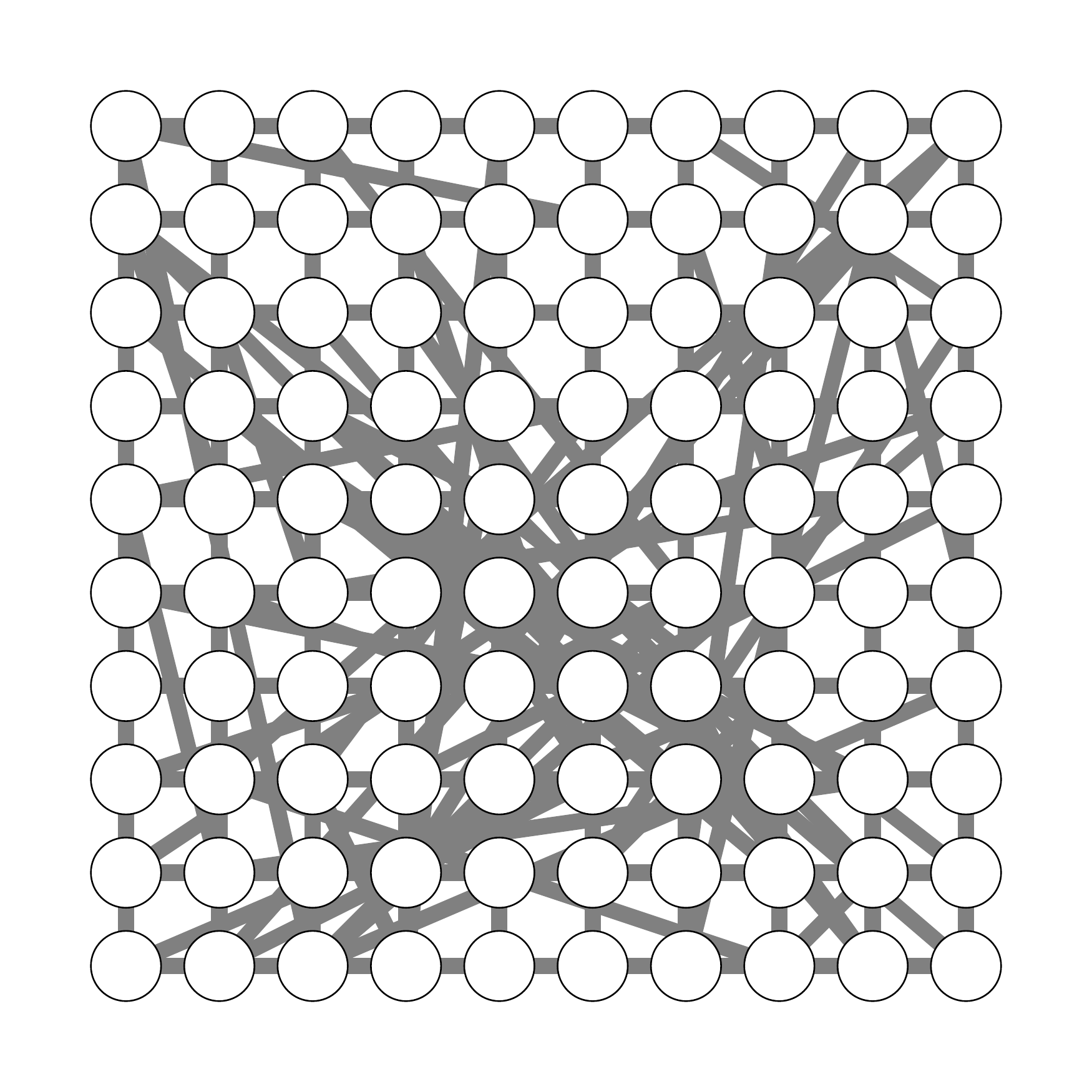} \\

\includegraphics[width=0.2\textwidth]{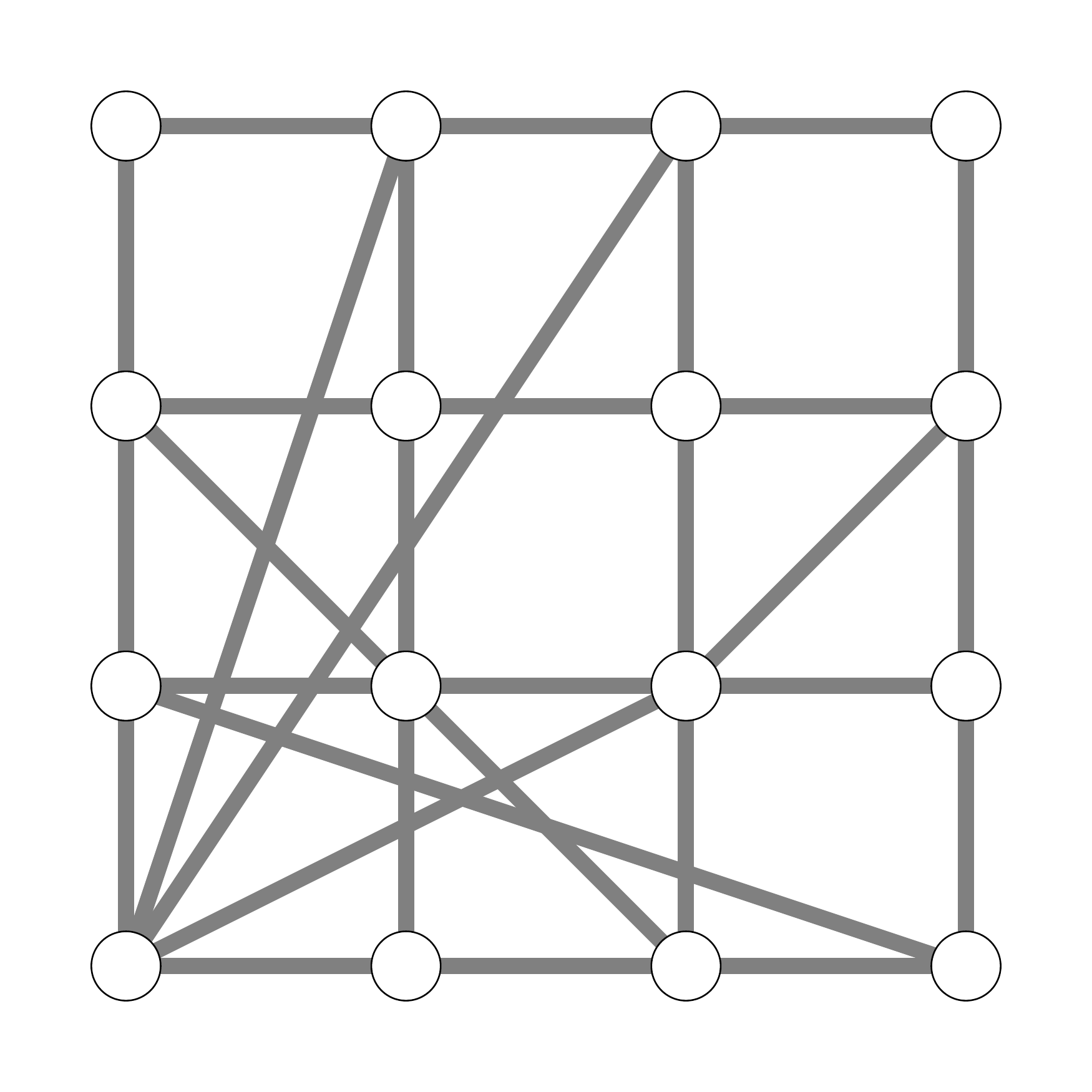}
\put(-120, 40){$p_w = 0.5$} 
\includegraphics[width=0.2\textwidth]{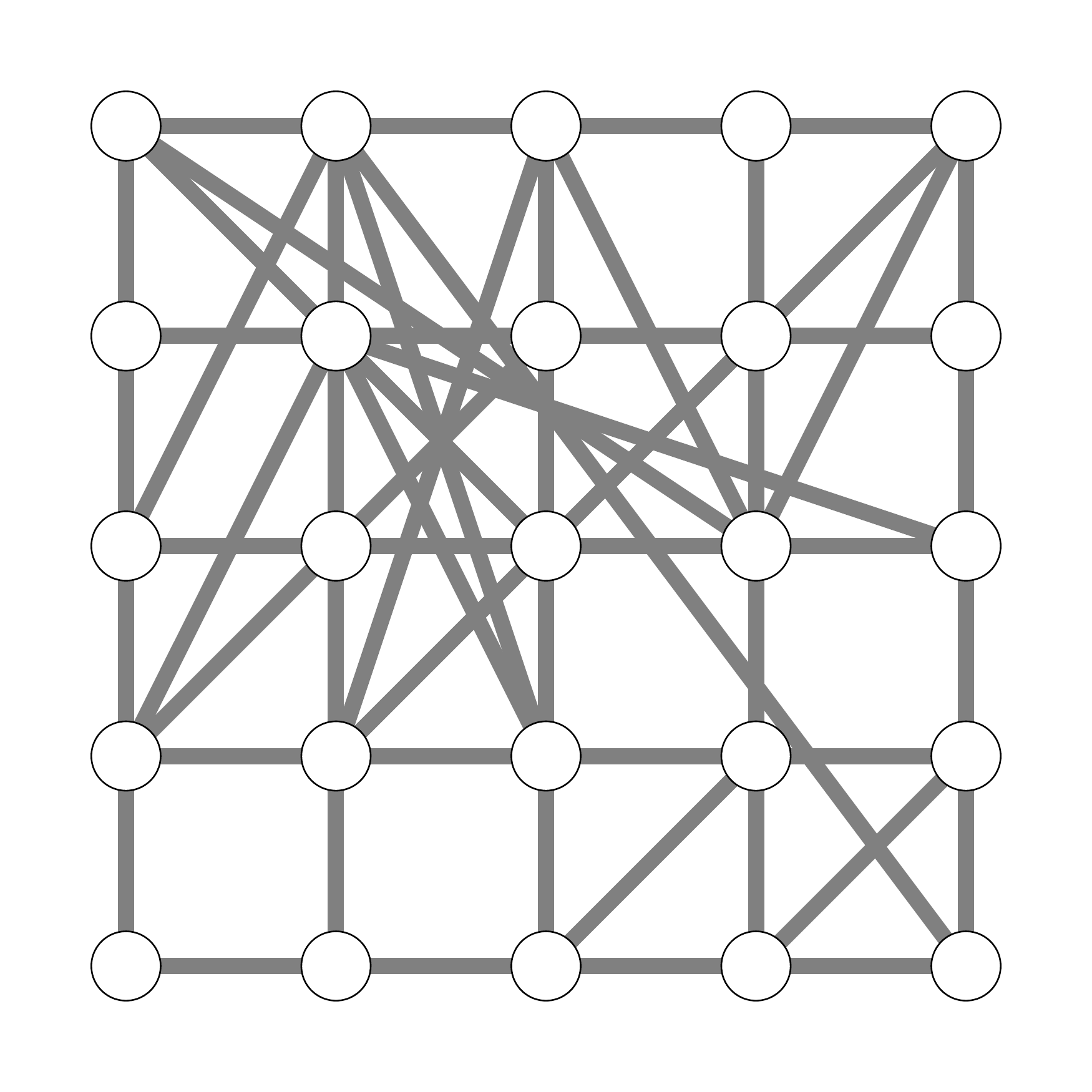} 
\includegraphics[width=0.2\textwidth]{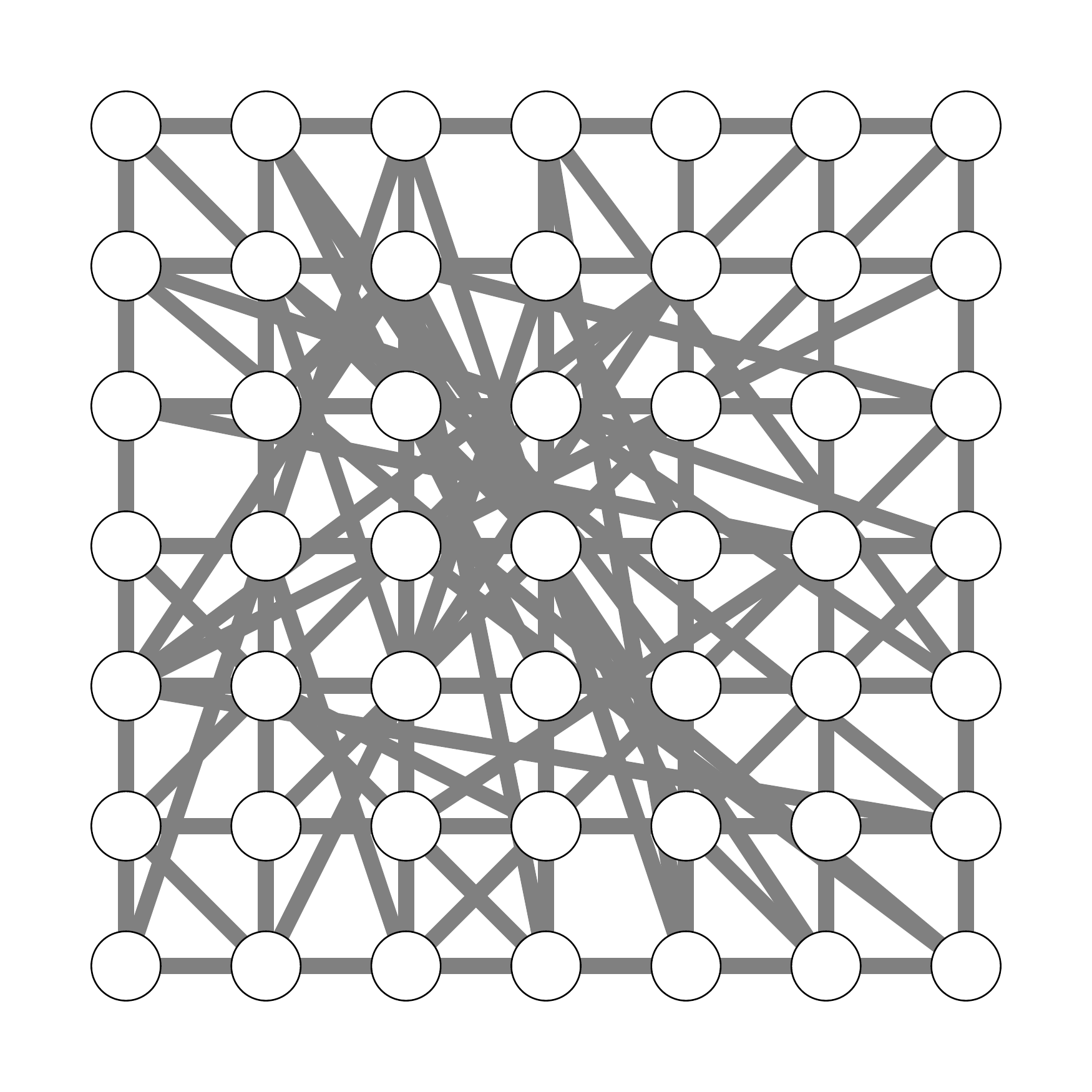} 
\includegraphics[width=0.2\textwidth]{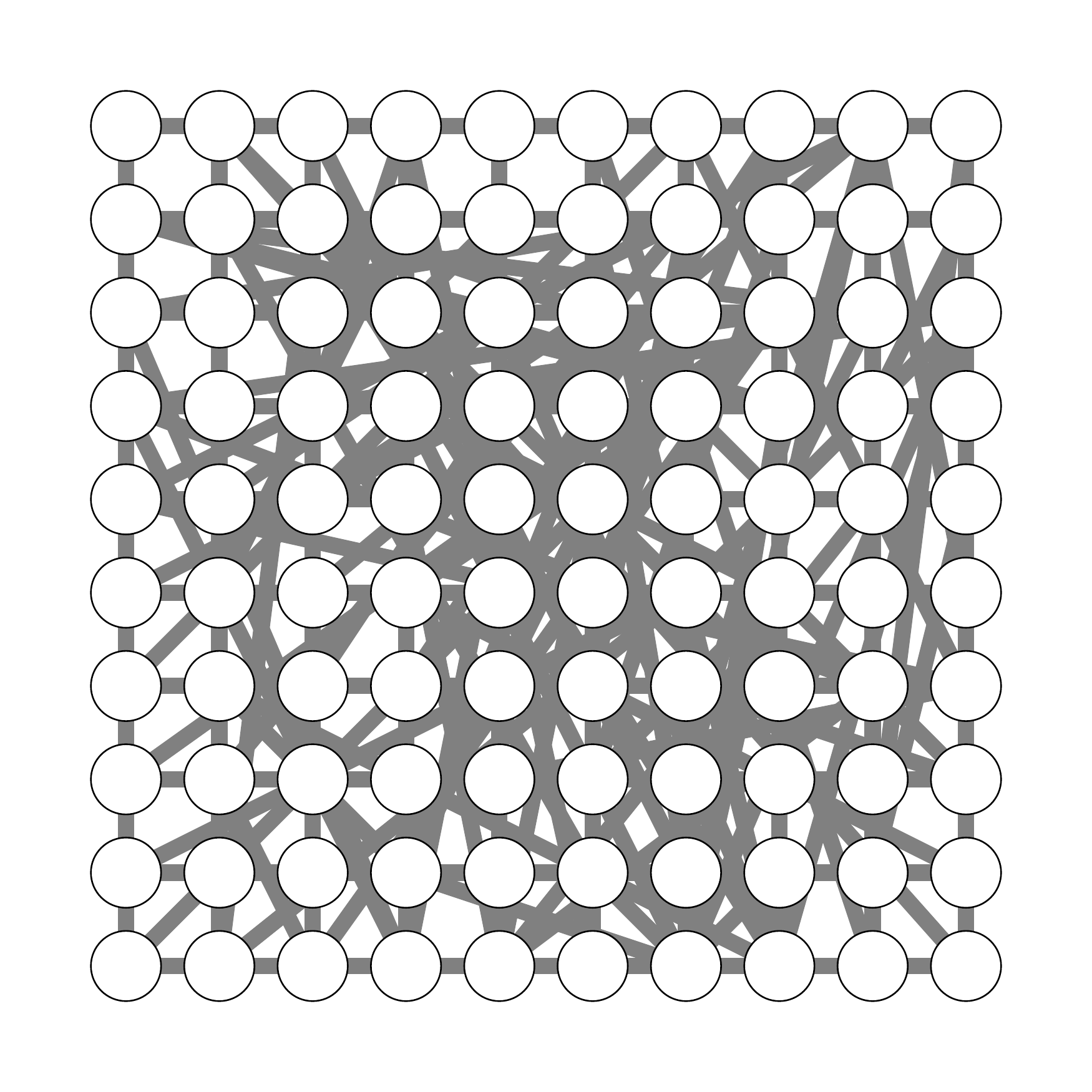} \\
\caption{Examples of simulated small world graphs with network size $n \in \left\{16, 25, 49, 100\right\}$ and $p_w \in \left\{0.1, 0.2, 0.3, 0.4, 0.5\right\}$.}
\label{fig:SWGNetworks}
\end{figure*}

\begin{figure*}
\includegraphics[width=0.3\textwidth]{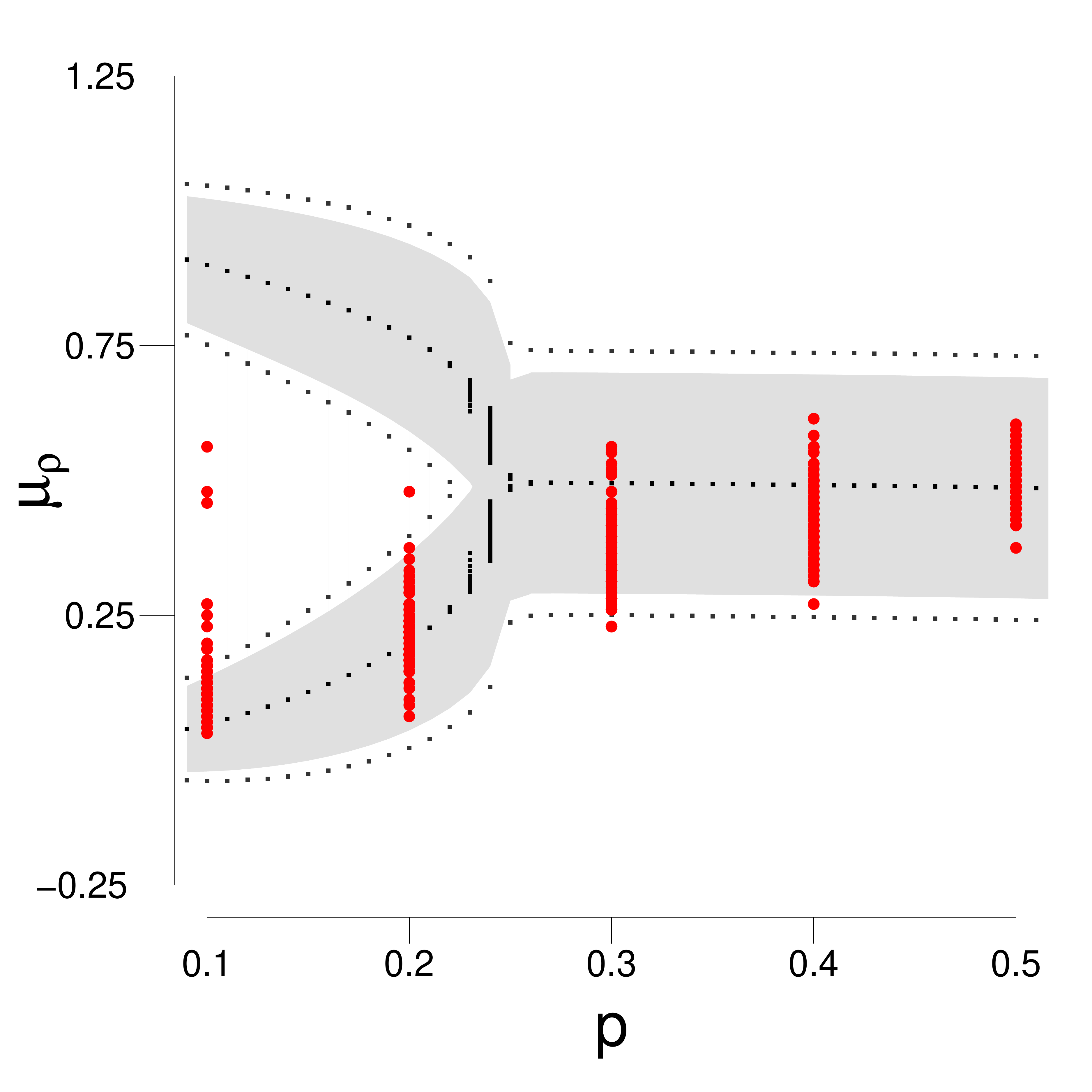}
\put(-65,150){$t = 50$} 
\put(-65,140){$n = 16$}
\put(-160,65){Torus}
\includegraphics[width=0.3\textwidth]{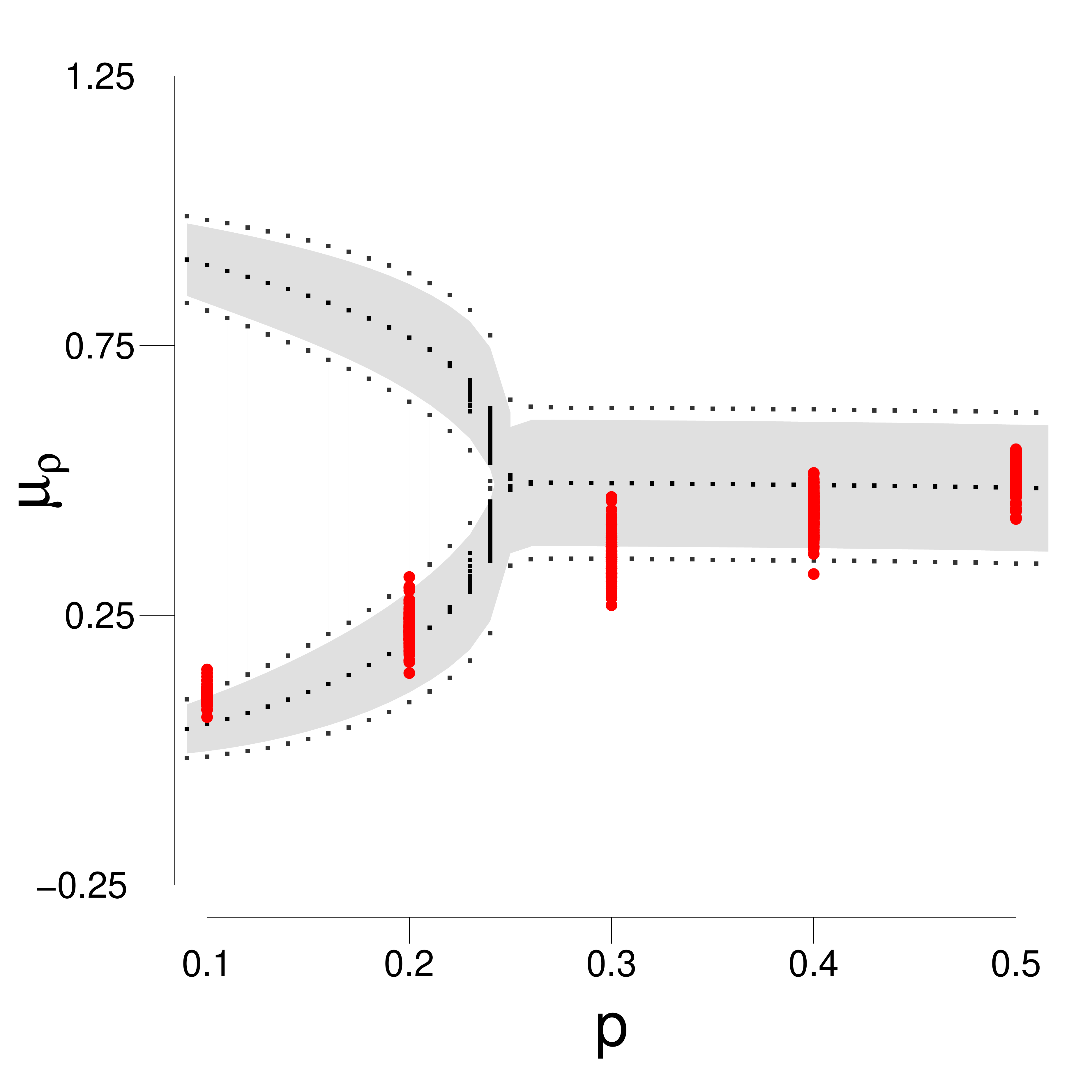}
\put(-65,150){$t = 500$} 
\put(-65,140){$n = 49$} 
\includegraphics[width=0.3\textwidth]{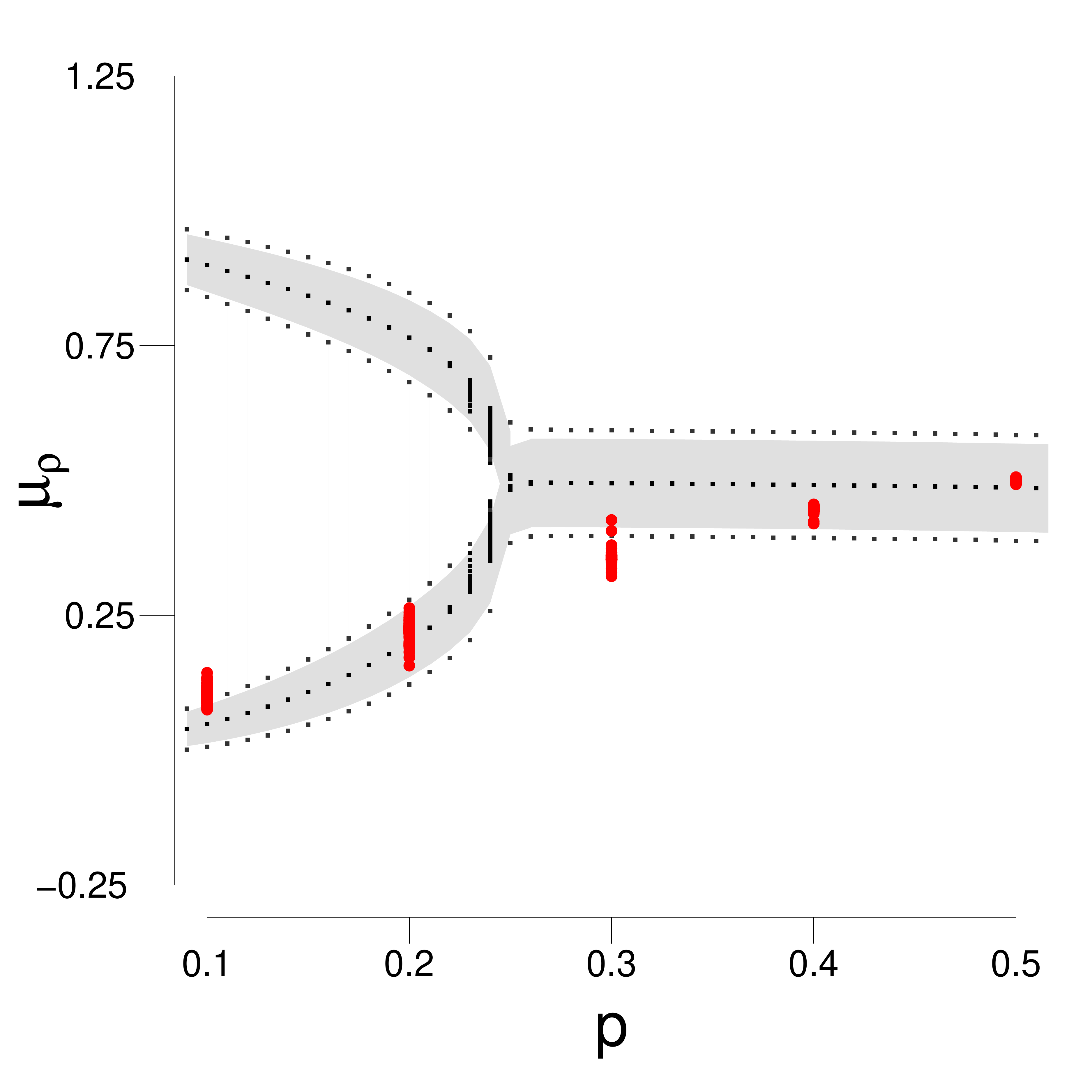}
\put(-65,150){$t = 5000$} 
\put(-65,140){$n = 100$}\\

\includegraphics[width=0.3\textwidth]{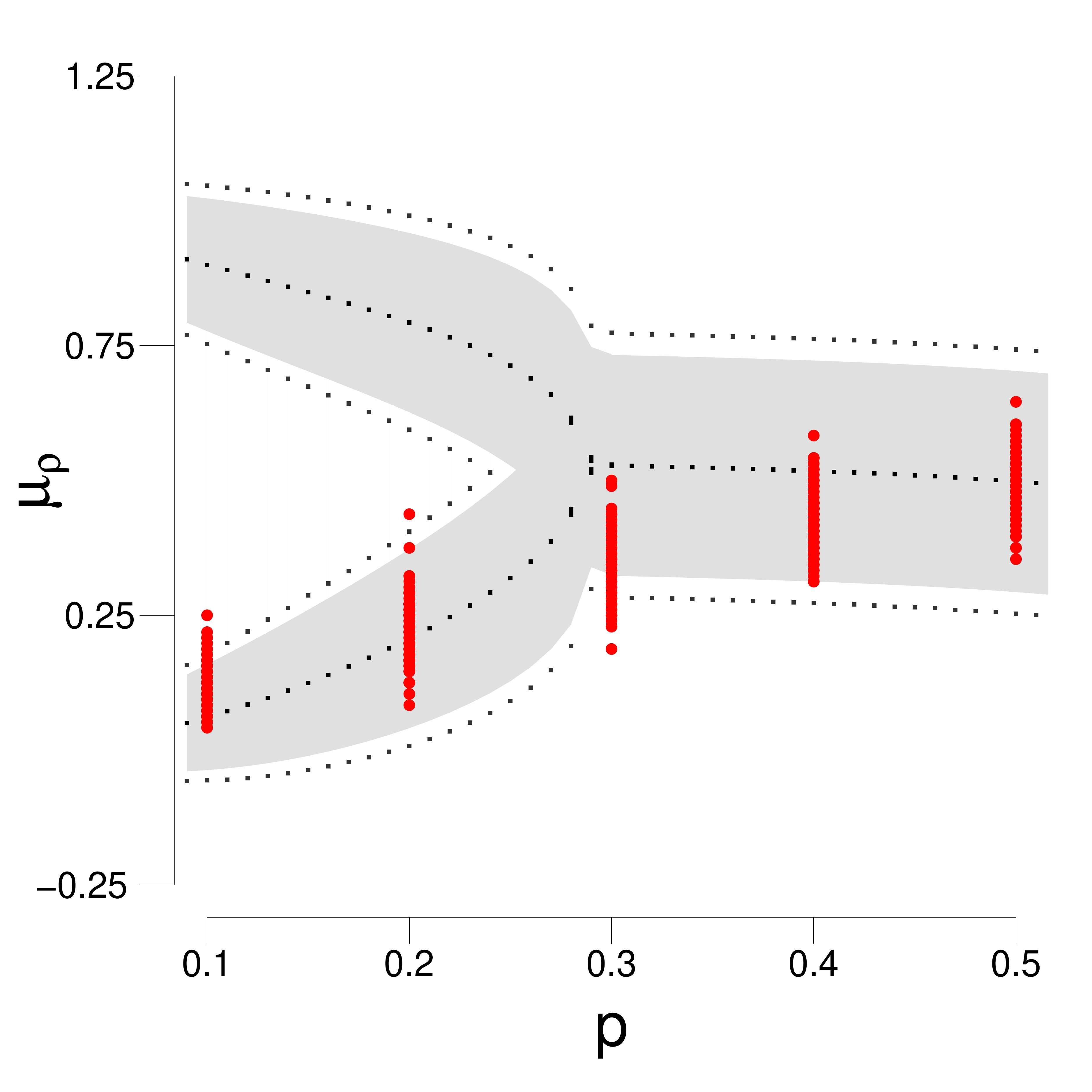} 
\put(-160,70){Random}
\put(-155,60){graph}
\includegraphics[width=0.3\textwidth]{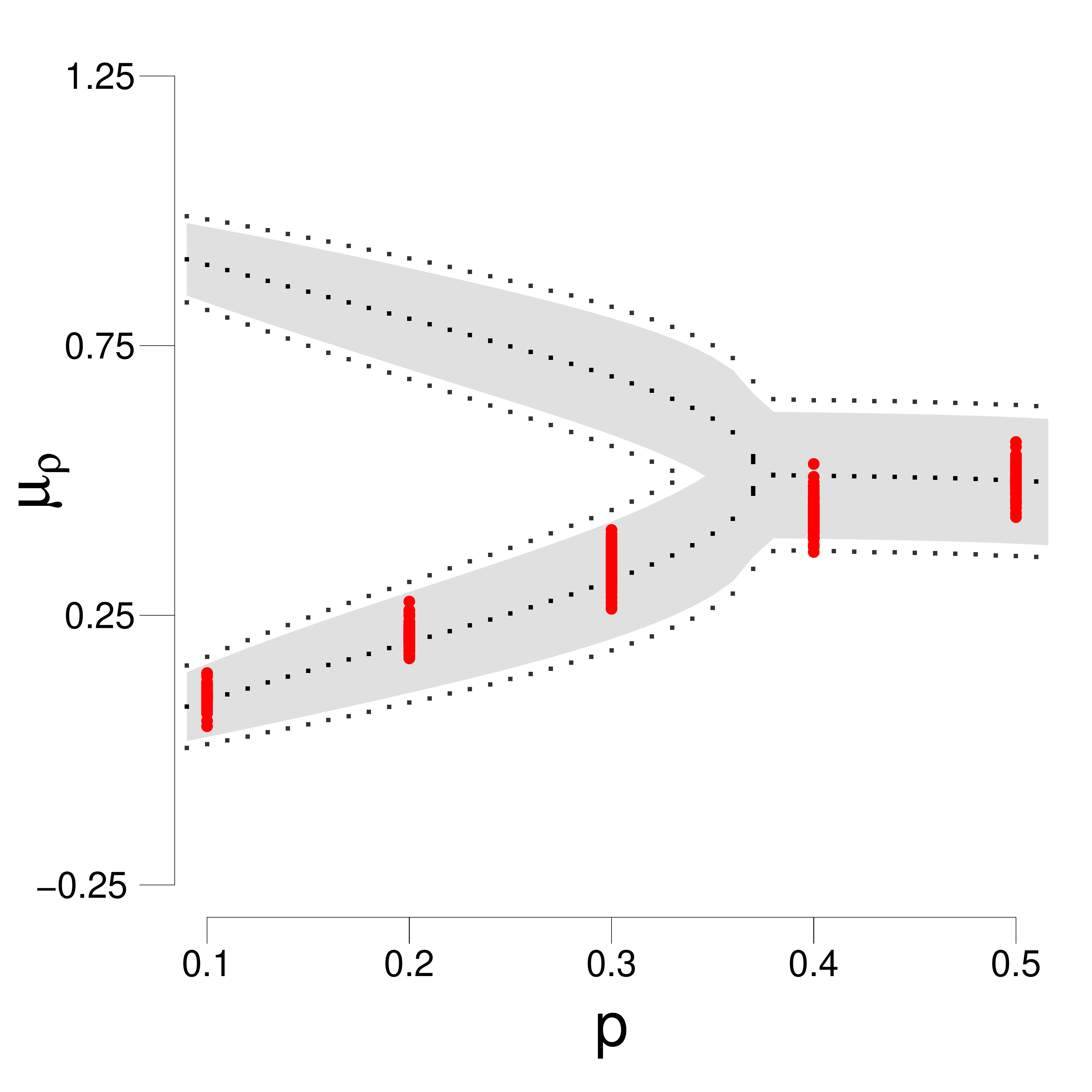} 
\includegraphics[width=0.3\textwidth]{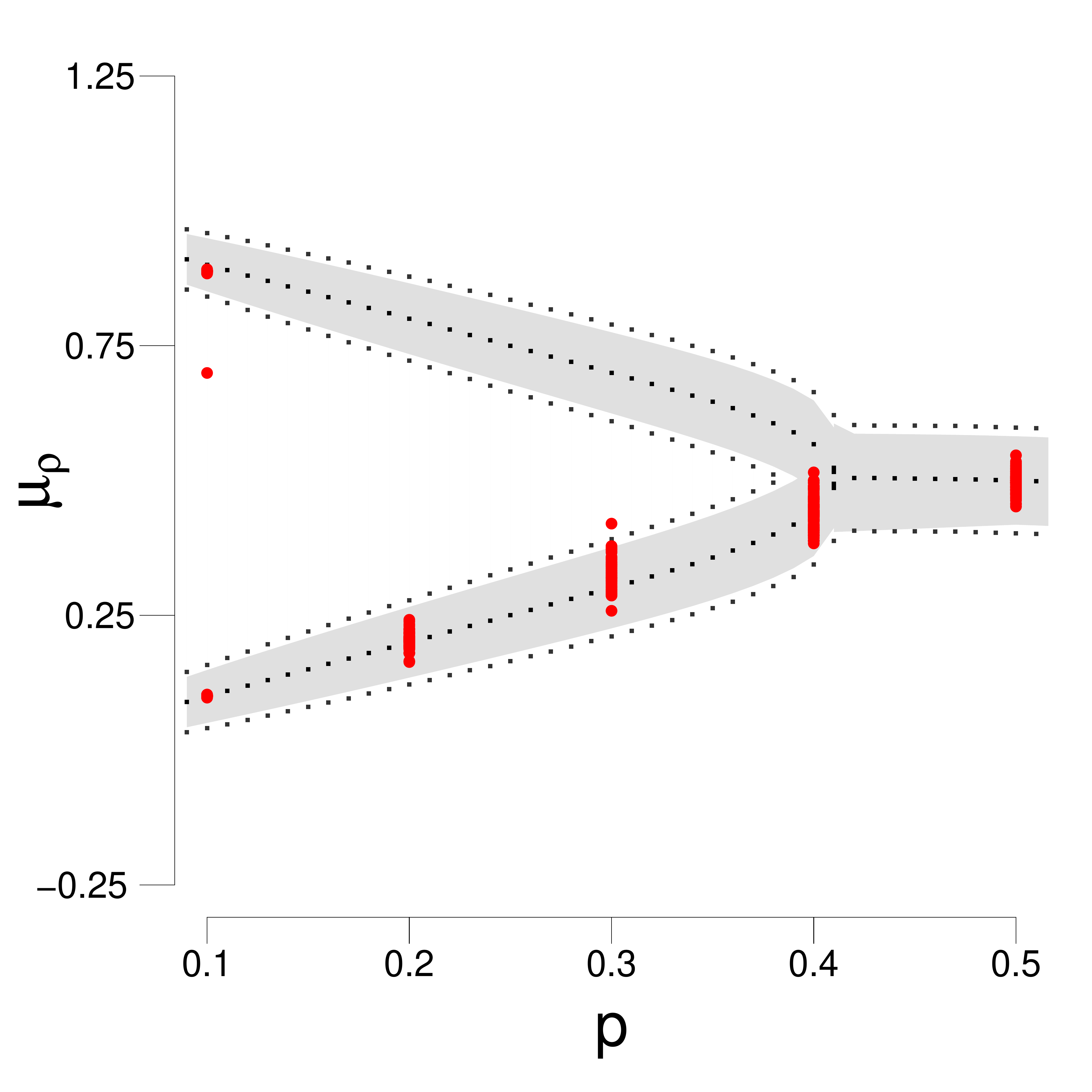}\\

\includegraphics[width=0.3\textwidth]{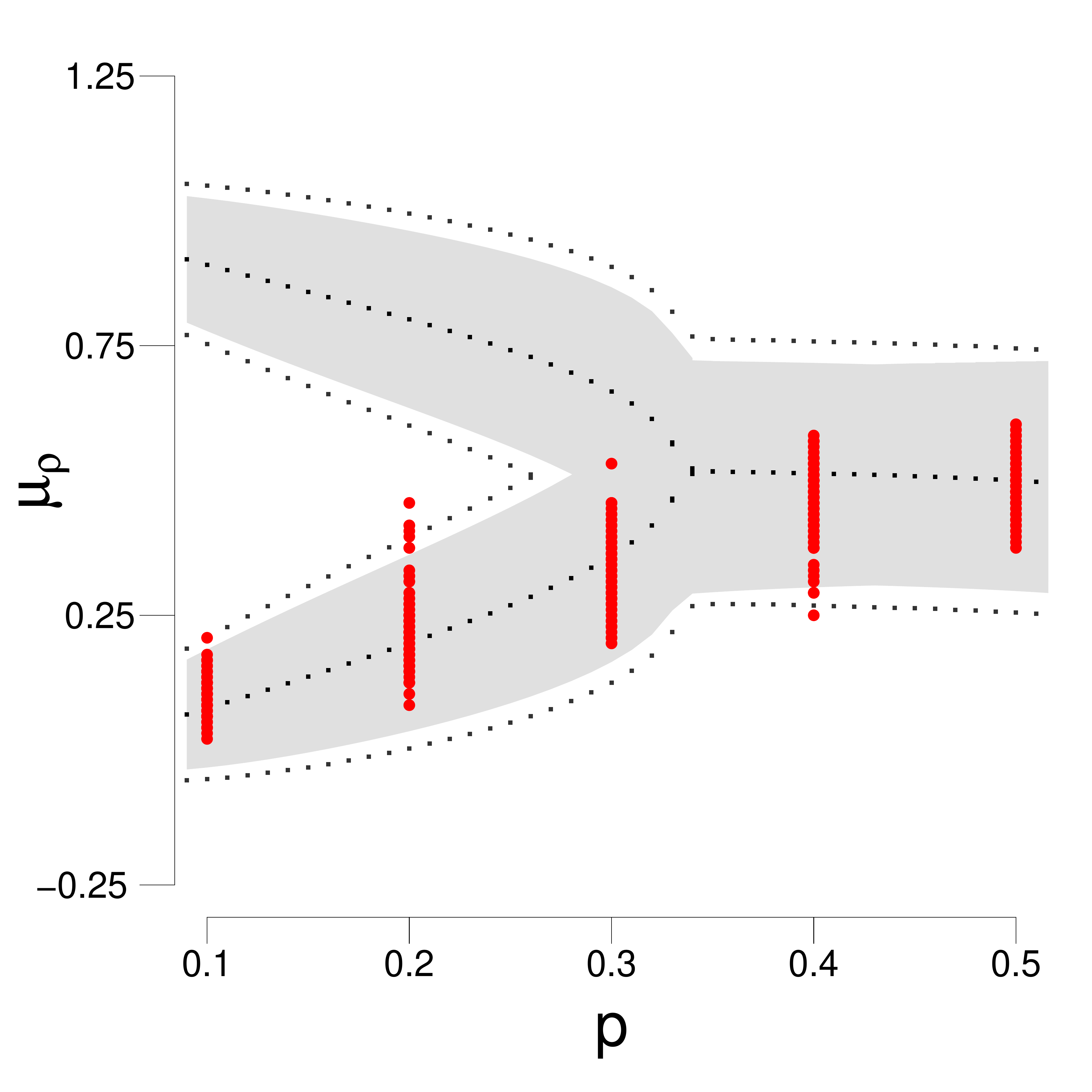}
\put(-160,75){Small-}
\put(-158,65){world}
\put(-158,55){graph}
\includegraphics[width=0.3\textwidth]{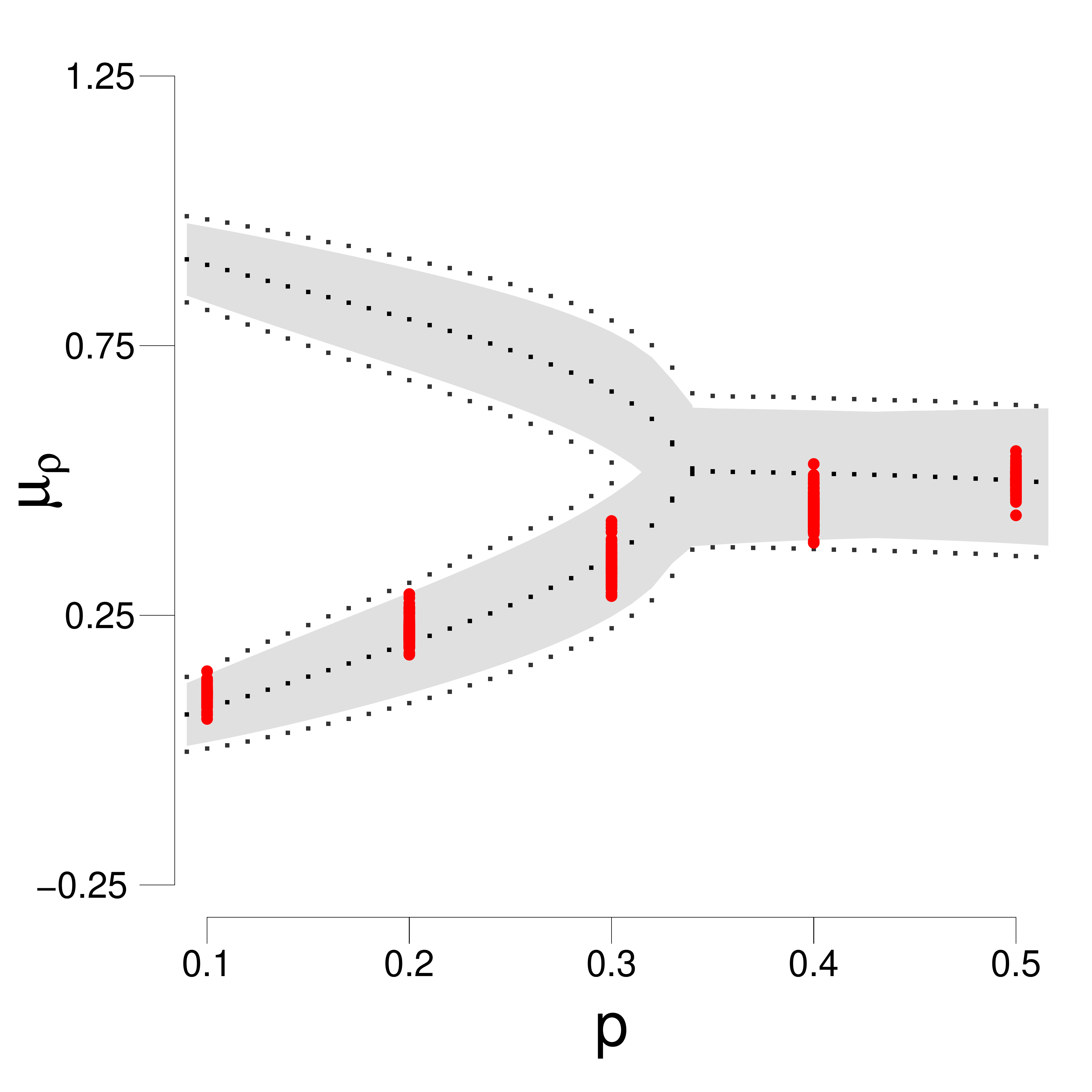} 
\includegraphics[width=0.3\textwidth]{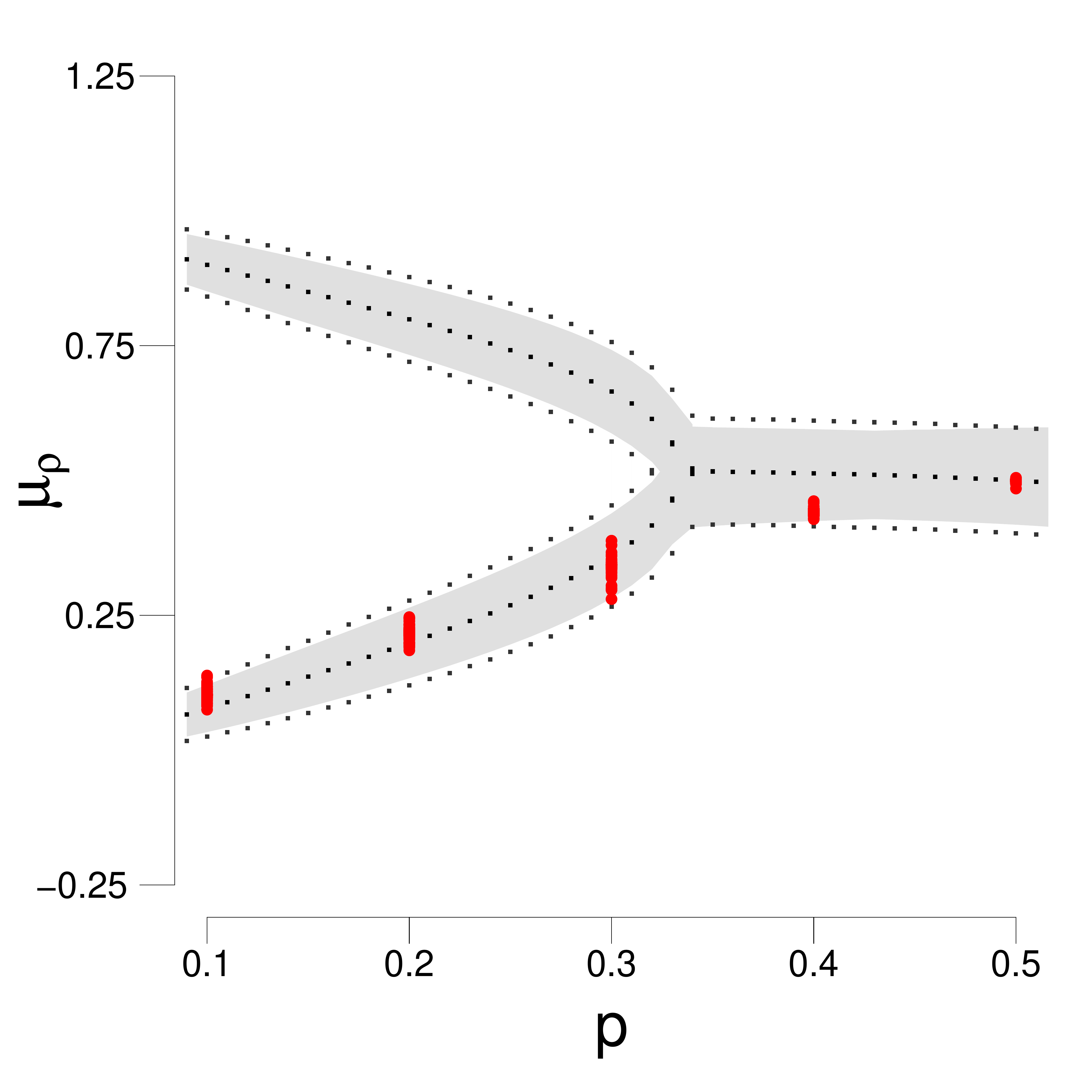}\\
\caption{Bifurcation diagrams of a torus (upper panel), a random graph (middle panel; $p_e = 0.5$) and a small world graph (lower panel; $p_w = 0.5$). Grey solid area = 90\% confidence interval around bifurcation. Dashed grey lines = 95\% confidence interval around bifurcation. Red dots = mean density estimates at different values of $p$.}
\label{fig:bifurcation-diagrams}
\end{figure*}

\begin{figure*}
\includegraphics[width=0.2\textwidth]{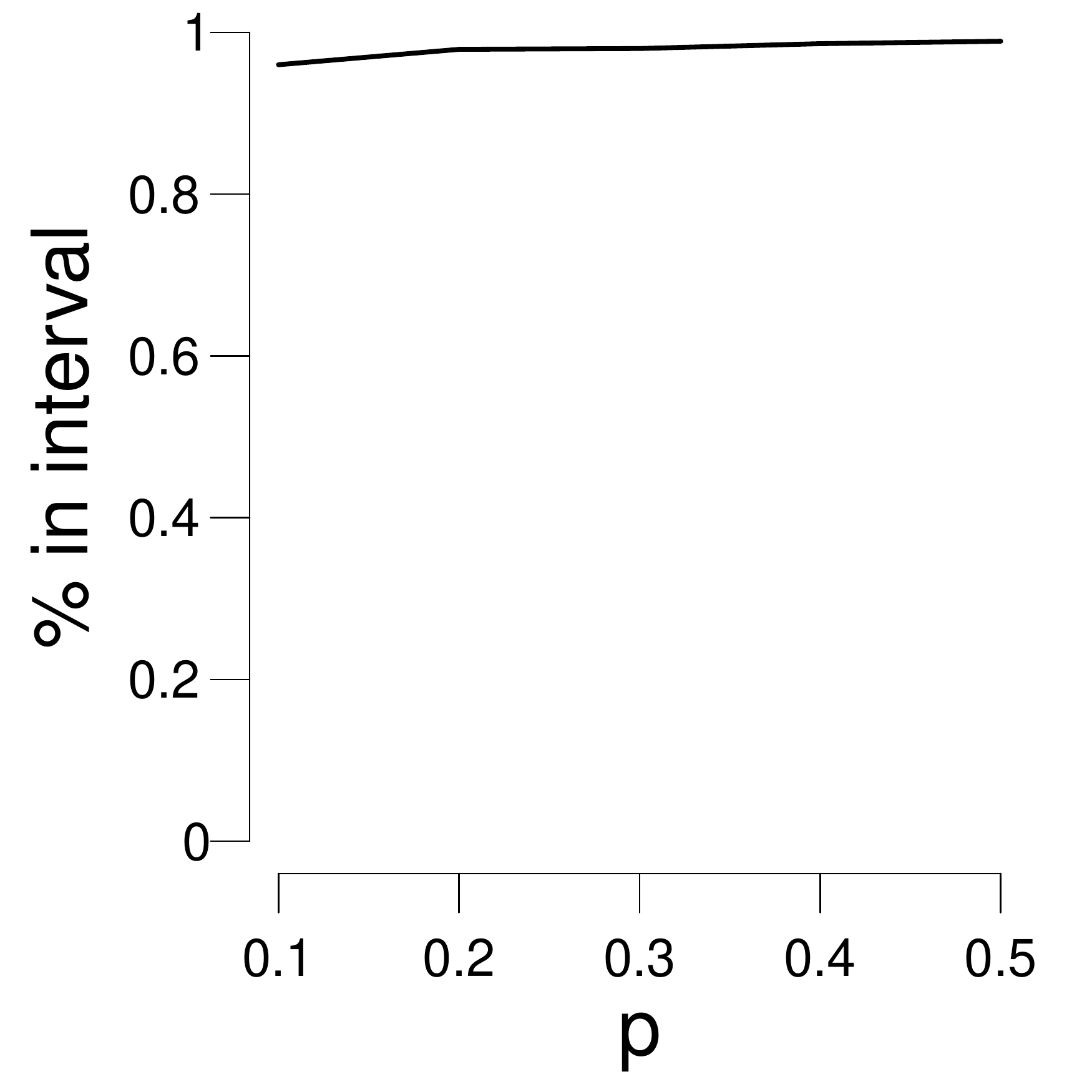}
\put(-50,90){$n = 16$} 
\put(-120,45){Torus}
\includegraphics[width=0.2\textwidth]{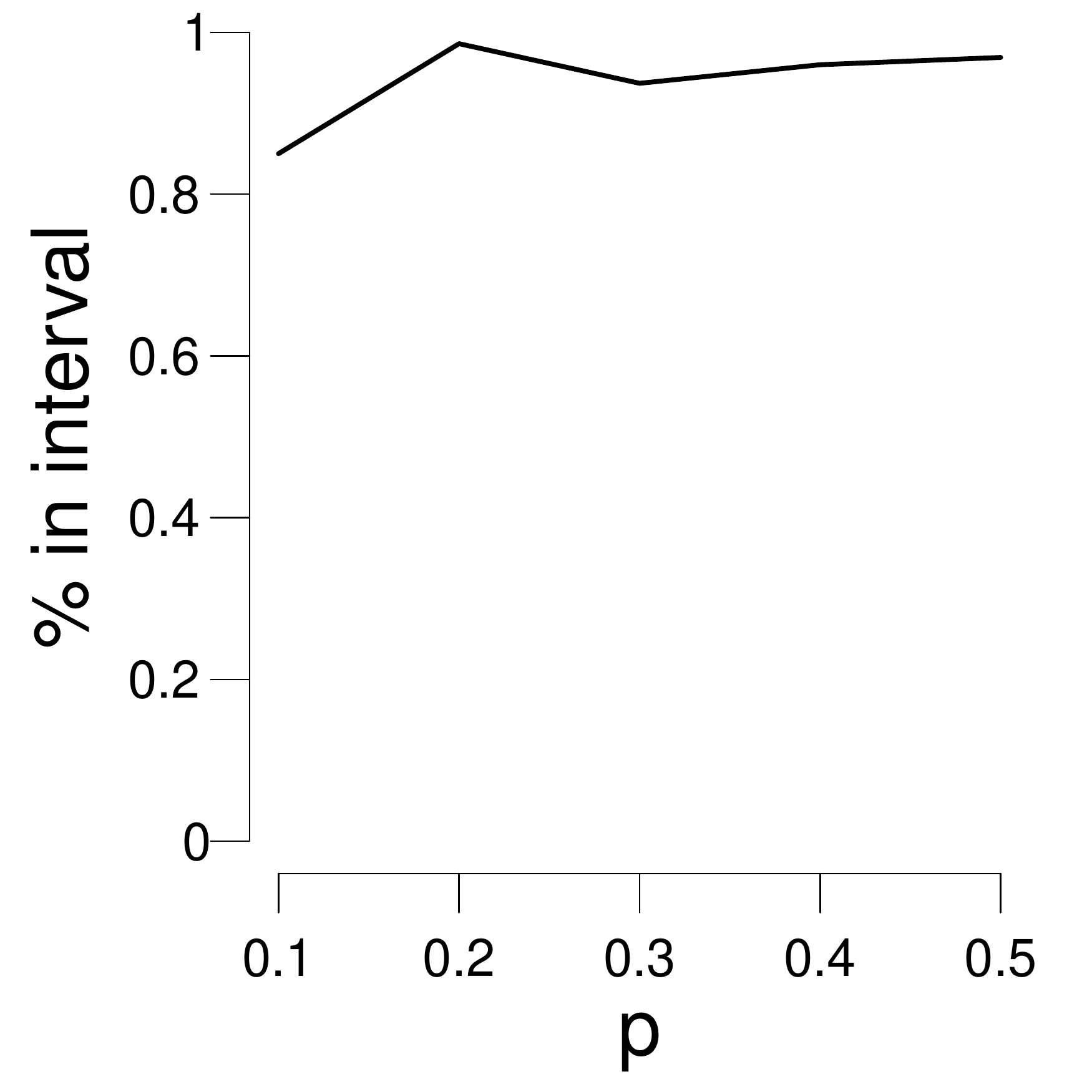}
\put(-50,90){$n = 25$} 
\includegraphics[width=0.2\textwidth]{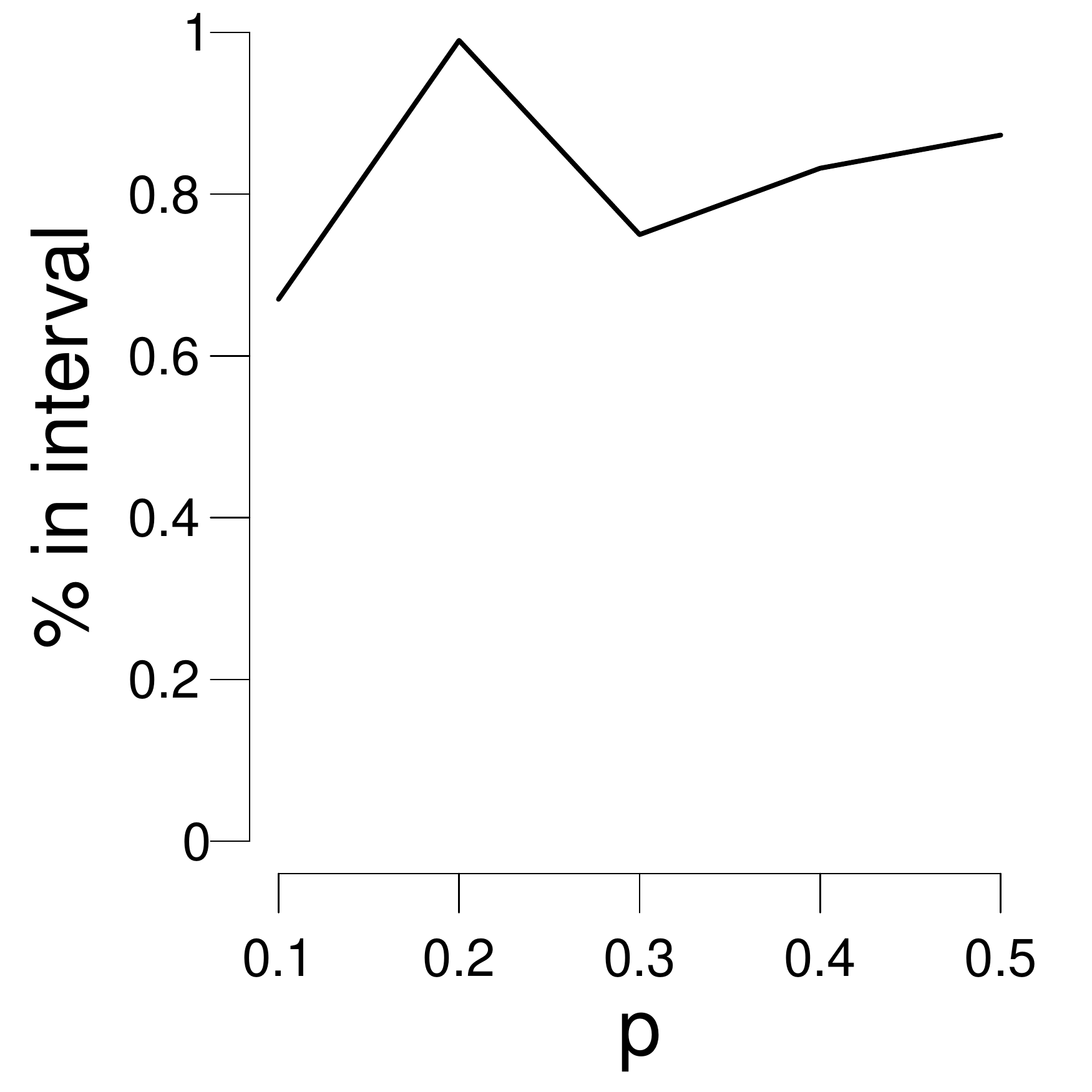}
\put(-50,90){$n = 49$} 
\includegraphics[width=0.2\textwidth]{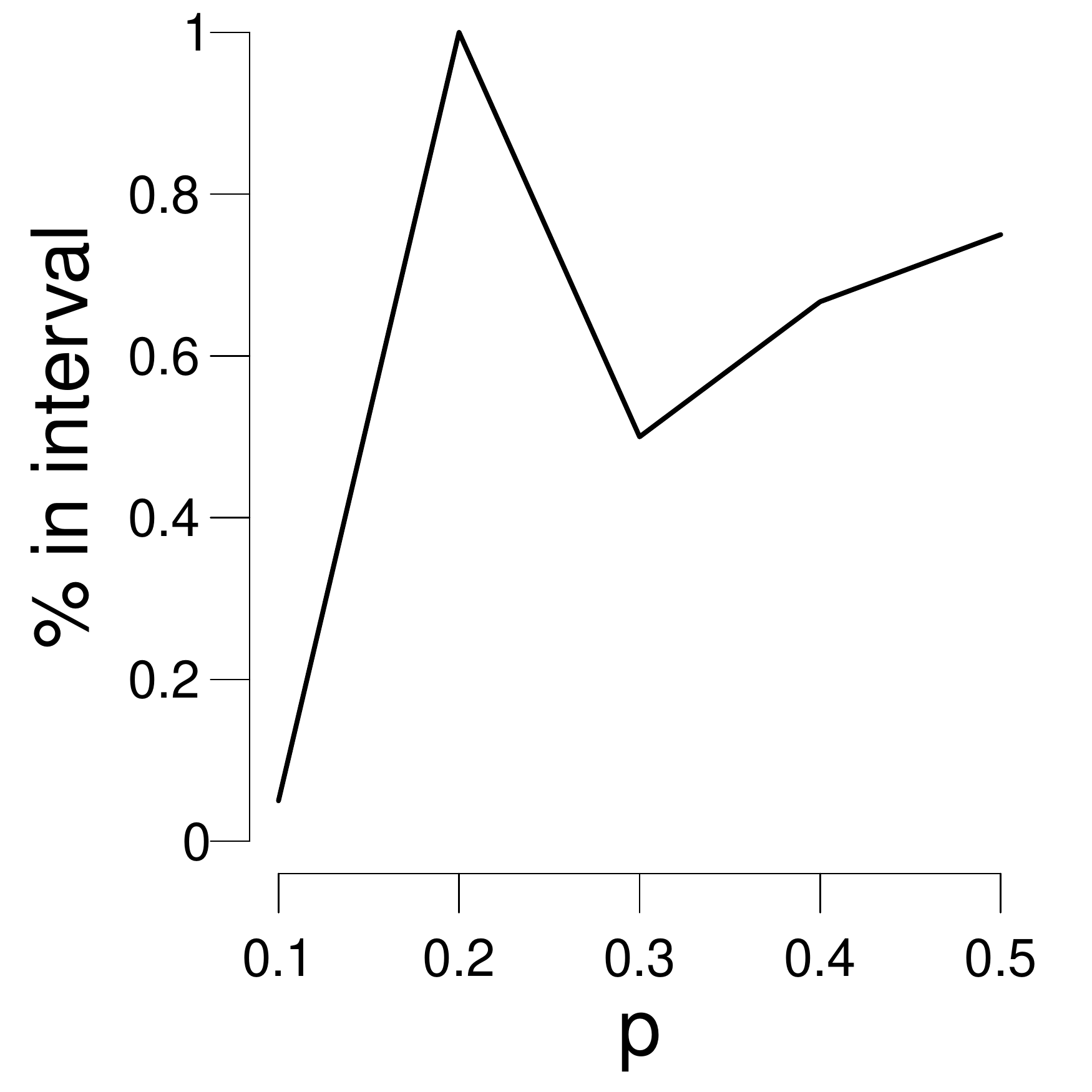}
\put(-50,90){$n = 100$}\\

\includegraphics[width=0.2\textwidth]{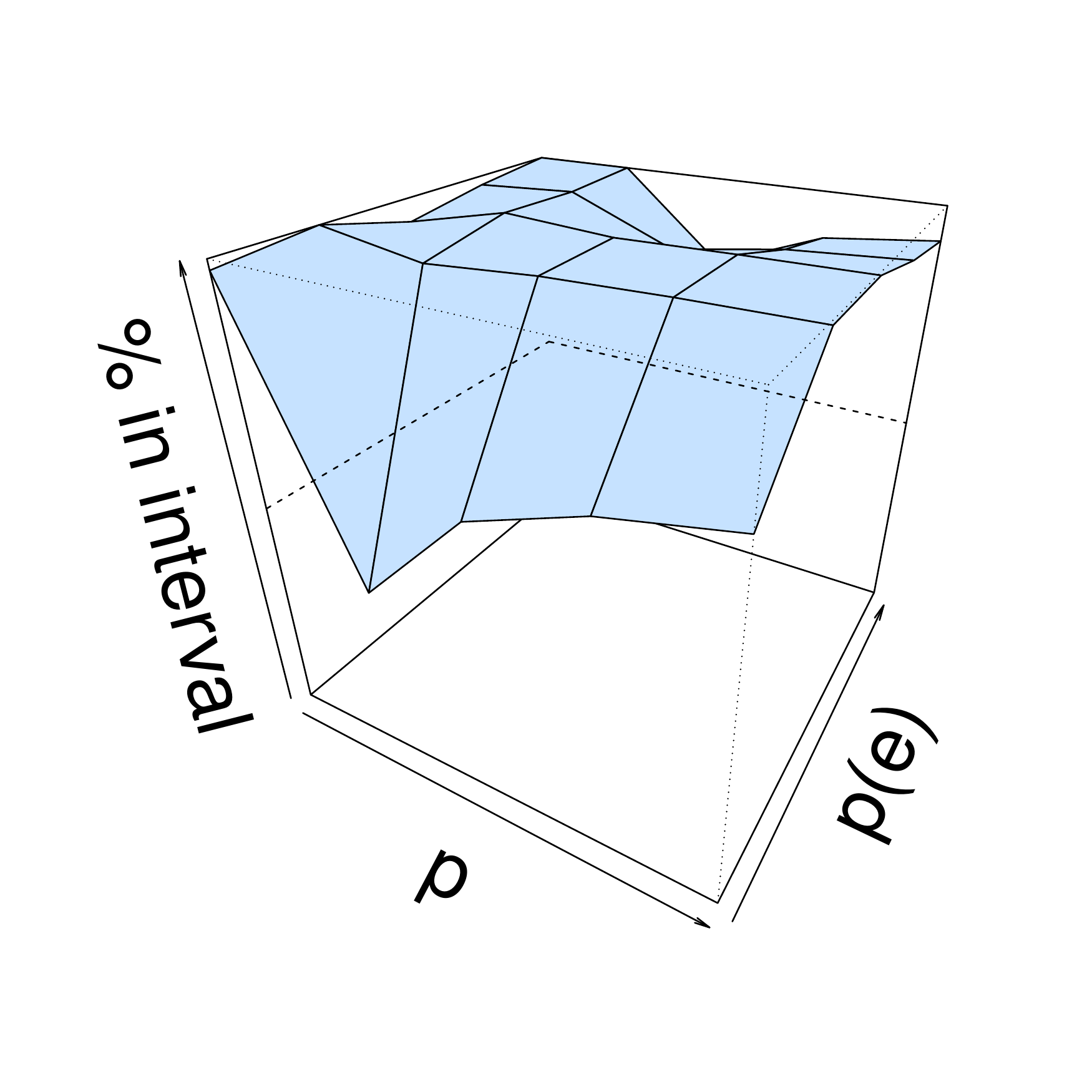} 
\put(-120,50){Random}
\put(-115,40){graph}
\includegraphics[width=0.2\textwidth]{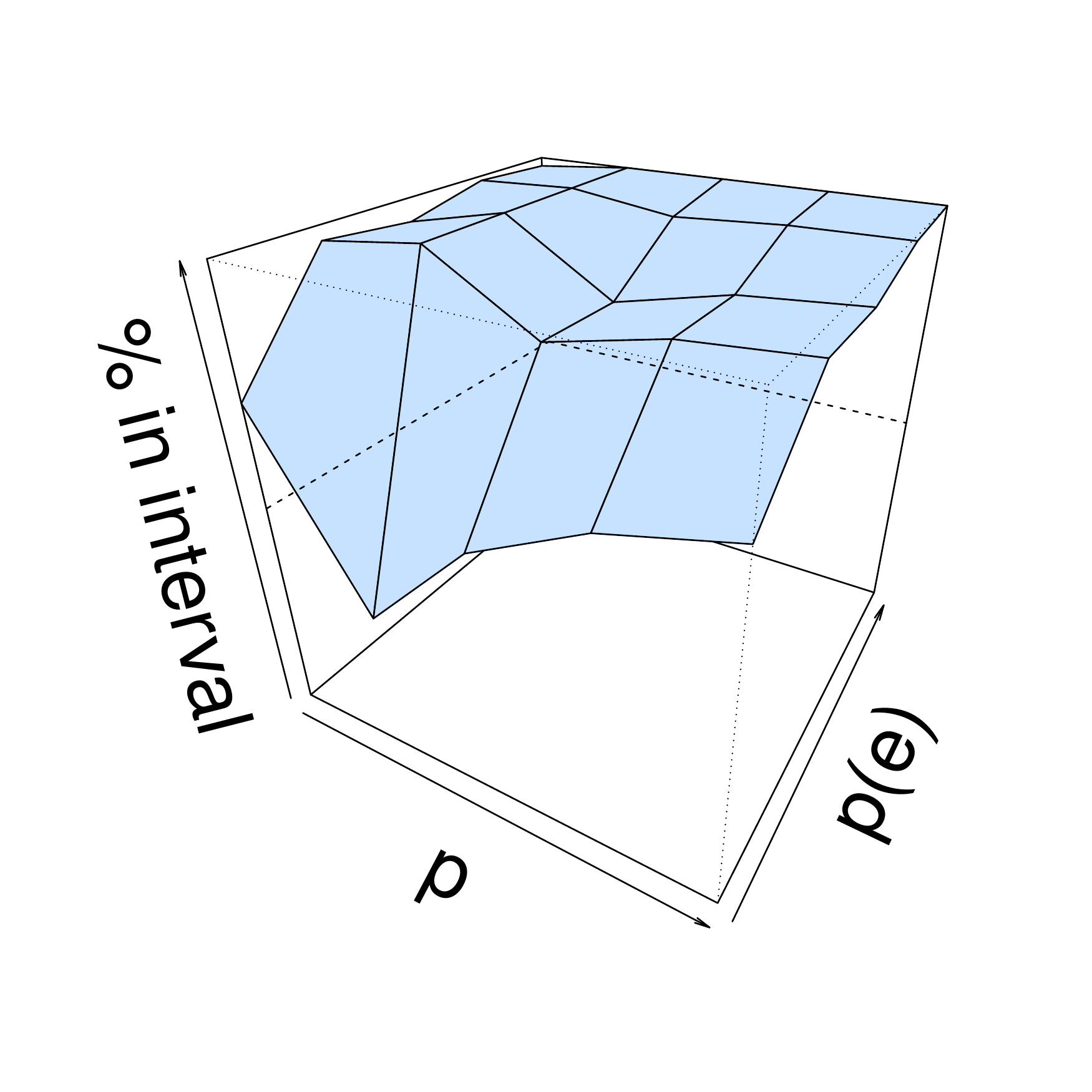} 
\includegraphics[width=0.2\textwidth]{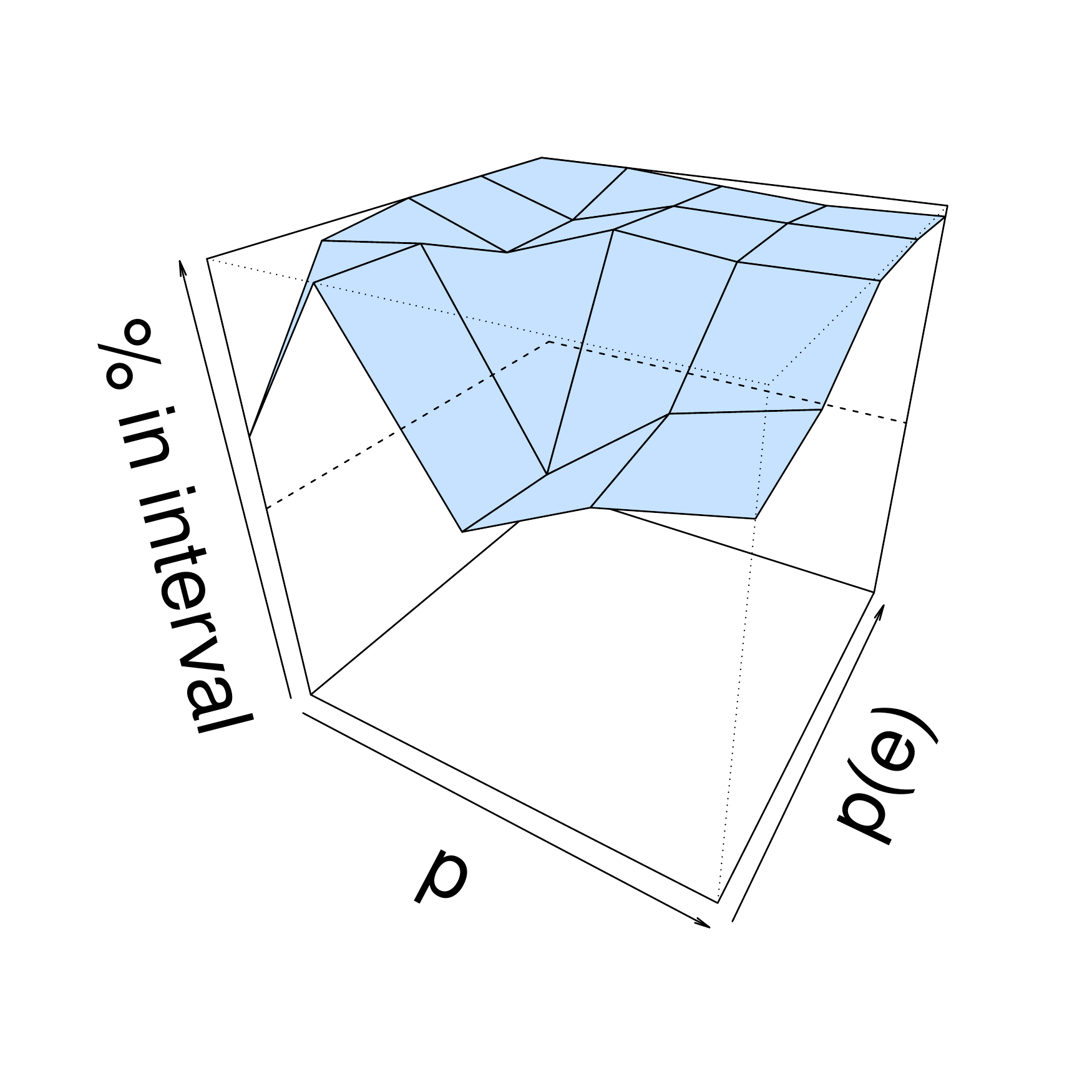} 
\includegraphics[width=0.2\textwidth]{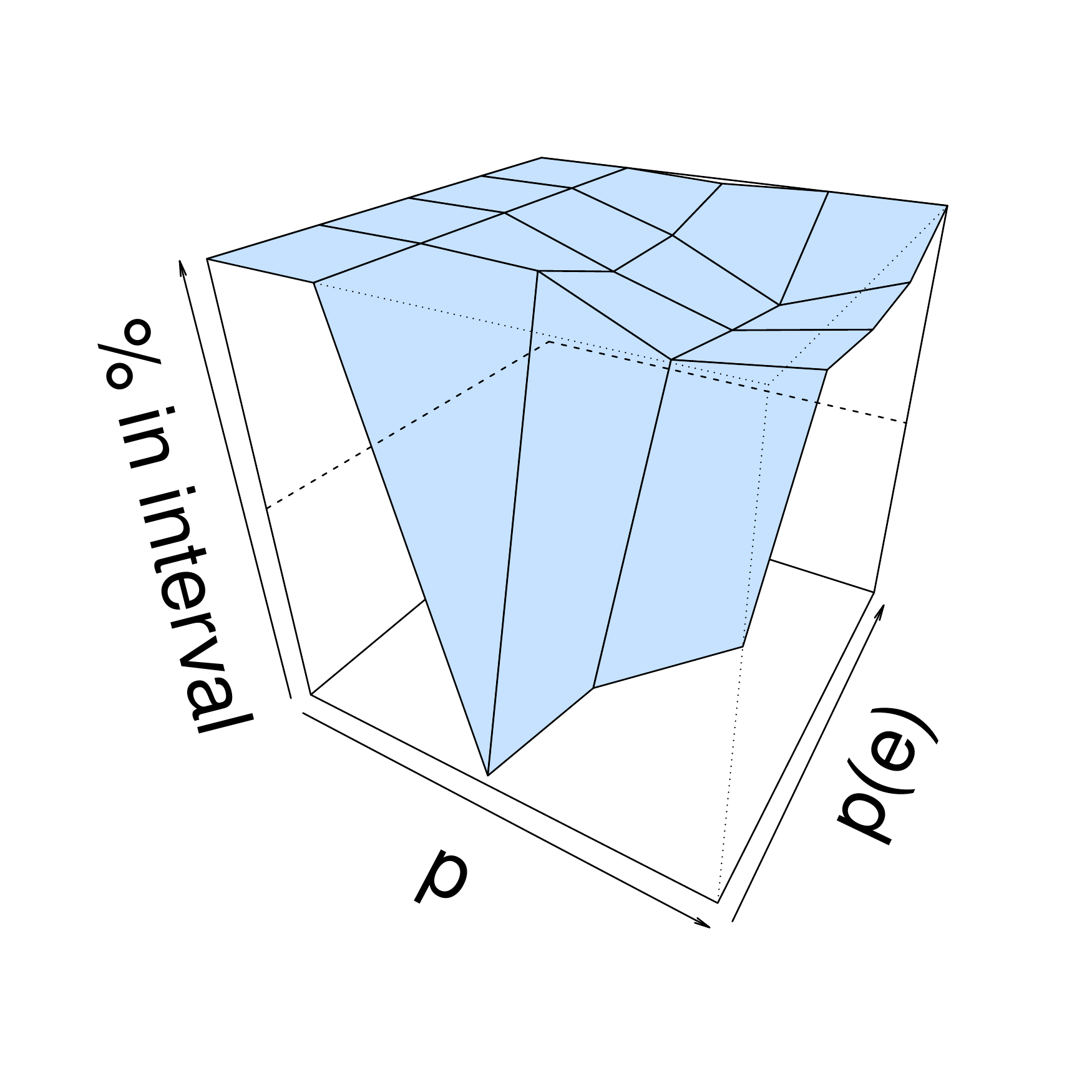}\\

\includegraphics[width=0.2\textwidth]{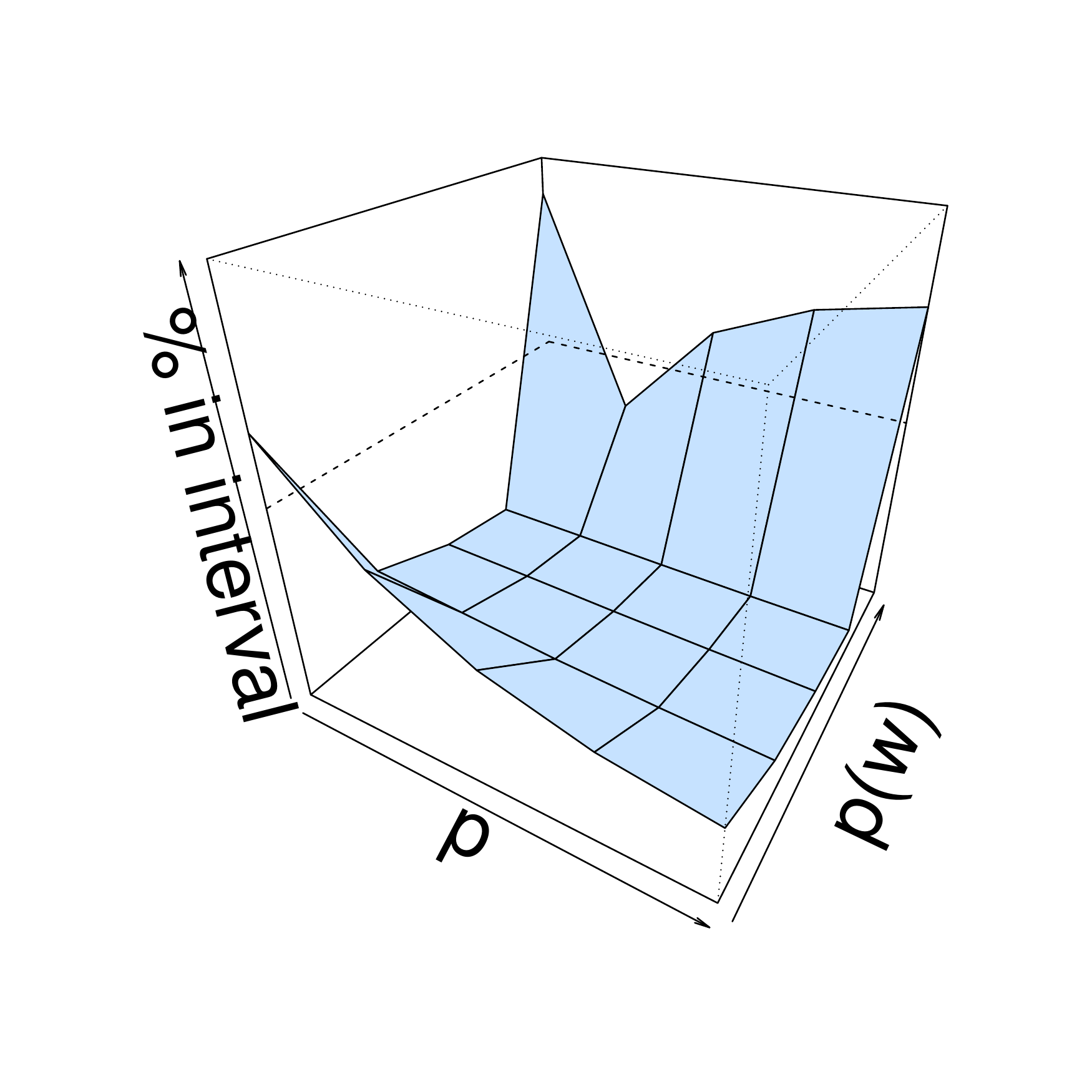}
\put(-120,55){Small-}
\put(-120,45){world}
\put(-120,35){graph} 
\includegraphics[width=0.2\textwidth]{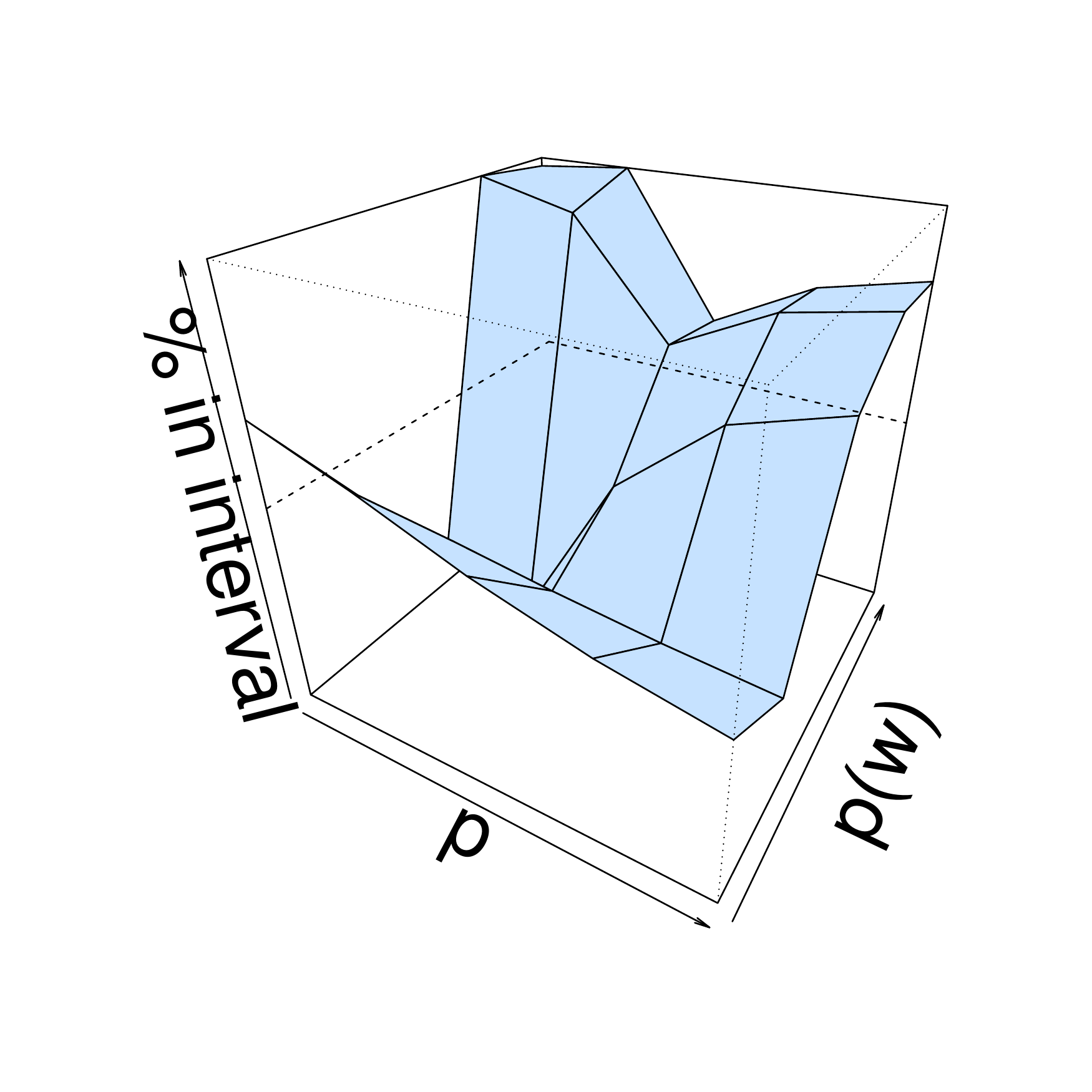} 
\includegraphics[width=0.2\textwidth]{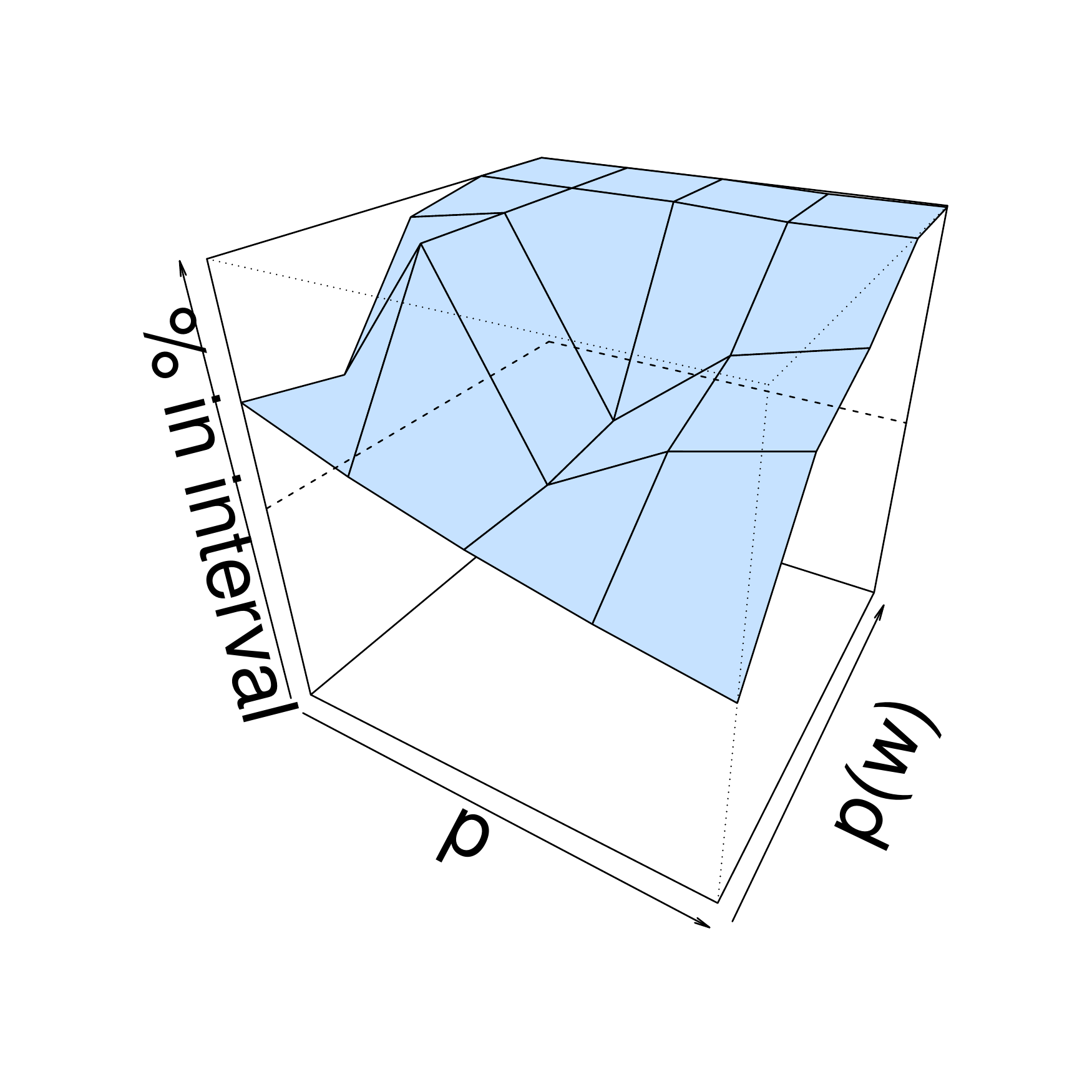} 
\includegraphics[width=0.2\textwidth]{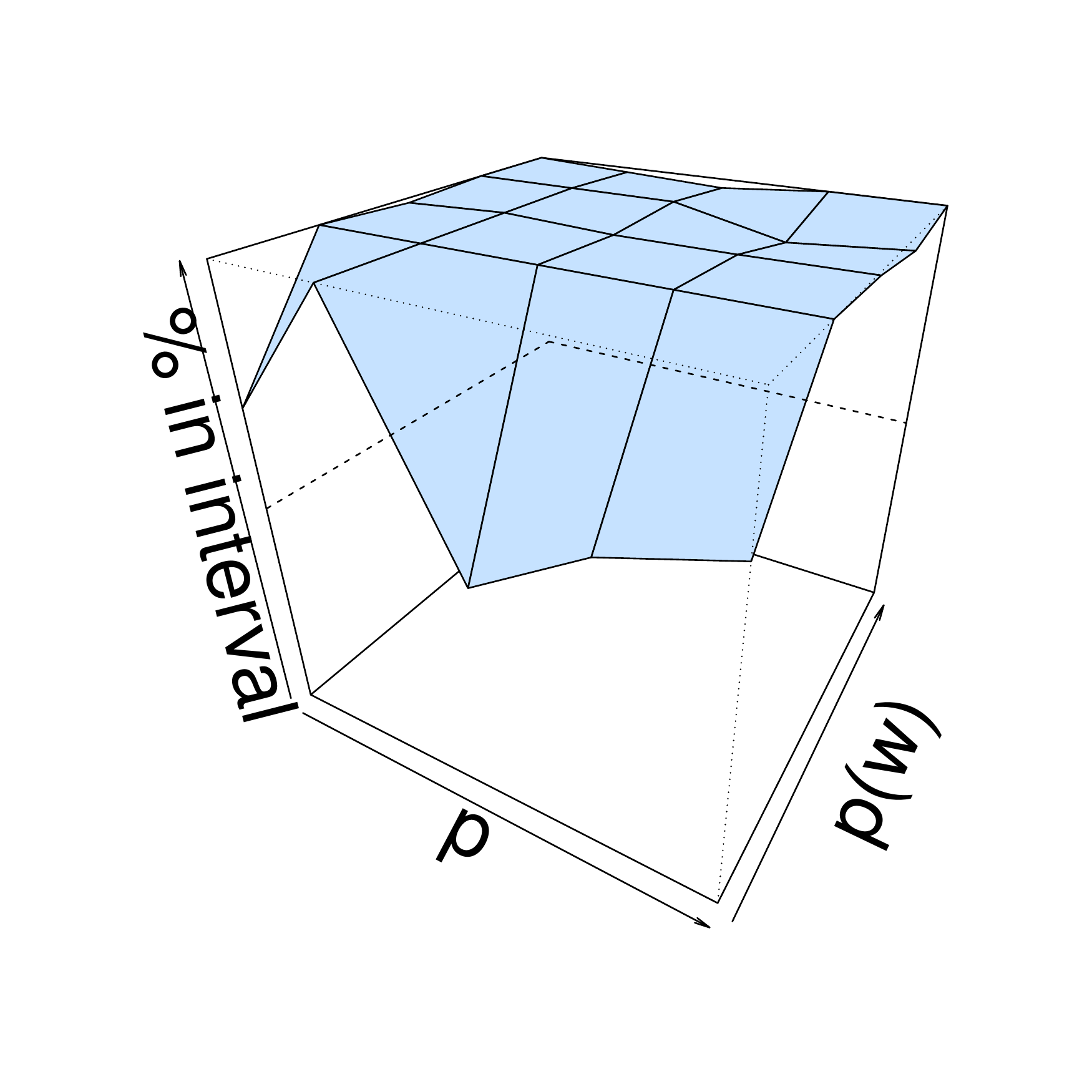}\\
\caption{Visualization of the percentage of mean density estimates that fall within a 95\% confidence interval in 2D (torus; left column) and 3D (random graph; middle column, and small world graph; right column) at $t = 500$.}
\label{fig:3D-accuracy}
\end{figure*}

For each combination of parameters, 100 graphs were simulated in the topology of an unweighted grid, a random graph, and a small-world graph. We varied the size of the network $n\in \{16, 25,49,100\}$, the number of time points $T\in \{50,100,200,500,5000\}$, and the probability of an active node in the majority rule $p\in\{0.1,0.2,0.3,0.4,0.5\}$, see (\ref{eq:psym}). We also varied the probability of an edge in the random graph $p_e\in \{0.1,0.2,\ldots ,0.9\}$, and the probability of wiring in the small-world graph $p_w\in\{0.1,0.2,\ldots ,0.9\}$. For $t = 0$, a random number of nodes was set to active by using the {\tt\small R} package \emph{IsingSampler} version 0.2 \cite{Epskamp:2013}. Figures \ref{fig:RGNetworks} and \ref{fig:SWGNetworks} show the topology of some simulated random graphs (\ref{fig:RGNetworks}) and small-world graphs (\ref{fig:SWGNetworks}) arranged in a square.
All simulated data, figures, as well as the used R-code are publicly available at the OSF \cite{OSF}. 
Results in bifurcation diagrams with 90\% and 95\% confidence intervals are shown in Figure \ref{fig:bifurcation-diagrams}. Mean density estimates, visualized with red dots, were calculated by dividing each simulation into snippets in which the distance between density estimates ($\delta$) did not exceed $0.4$. Other values for $\delta$ were considered, but it was found that changing $\delta$ to $0.3$ or $0.5$ resulted in similar percentages. We calculated for each snippet the percentage of mean density estimates that fell within a 90\% and 95\% confidence interval. Table \ref{tab:percentages} shows the mean percentages, marginalized over the number of time points $T$, the variation in $p_e$, $p_w$ and the number of nodes $n$ obtained from the central limit theorem with bounds $\hat{\rho} \pm 1.96$ for the 95\% confidence interval, and $\hat{\rho} \pm 1.64$ for the 90\% confidence interval. Figure \ref{fig:3D-accuracy} gives a 3-dimensional representation of the percentage of mean density estimates that fall within a 95\% confidence interval. Results from the 90\% confidence interval are not presented, as these were similar to the results from the 95\% confidence interval. It can be seen that the mean field approximation accurately estimates the density of the network structures across various simulation conditions. 

As seen in Figure \ref{fig:3D-accuracy}, a local dip occurs for all network structures at $p = 0.3$ and becomes more extreme as $n$ increases in size. Also shown in Figure \ref{fig:bifurcation-diagrams}, the mean density estimates at $p = 0.3$ fall less often in the 95\% confidence interval in comparison to the mean density estimates at other values for $p$. This is partly due to the fact that the standard error, a parameter needed for the calculation of the confidence intervals, depends on the network size; as the network size $n$ increases, the standard error becomes smaller as well as the resulting confidence interval. Furthermore, as this phenomenon occurs in all simulated network structures, we believe that this behaviour results from the fact that the mean field approximation has a bit of trouble adjusting to the one-phase stability, after being in an area where phase transitions may occur. All in all, results show that the mean field approximation also performs well when non-regular network structures are under consideration.

\begin{table*}
\centering
\caption{Percentages of density estimates with a 90\% or a 95\% confidence interval. $p_e$ = edge probability (random graph). $p_w$ = rewiring probability (small-world graph).}
\label{tab:percentages}
\begin{tabular}{p{3cm}p{1cm}p{1cm}p{1cm}p{1cm}p{1.5cm}p{1.5cm}}
  \hline
Structure & T & N & $p_e$/$p_w$ & $p$ & 90\%CI & 95\%CI \\ 
  \hline
Torus &  50 &  16 & - & 0.1 & 0.75 & 0.80 \\ 
  &  &  &  & 0.2 & 0.99 & 0.99 \\ 
  &  &  &  & 0.3 & 0.92 & 0.99 \\ 
  &  &  &  & 0.4 & 0.94 & 0.99 \\ 
  &  &  &  & 0.5 & 0.96 & 0.99 \\ 
  & 500 &  49 & - & 0.1 & 0.29 & 0.67 \\ 
  & &  &  & 0.2 & 0.97 & 0.99 \\ 
  & &  &  & 0.3 & 0.63 & 0.75 \\ 
  & &  &  & 0.4 & 0.75 & 0.83 \\ 
  & &  &  & 0.5 & 0.81 & 0.87 \\ 
  & 5000 & 100 & - & 0.1 & 0.12 & 0.55 \\ 
  & & &  & 0.2 & 1.00 & 1.00 \\ 
  & & &  & 0.3 & 0.50 & 0.50 \\ 
  & & &  & 0.4 & 0.75 & 0.83 \\ 
  & & &  & 0.5 & 0.75 & 0.75 \\ 
  Random graph &  50 &  16 & 0.5 & 0.1 & 0.97 & 0.97 \\ 
  &  &  & & 0.2 & 1.00 & 1.00 \\ 
  &  &  & & 0.3 & 0.55 & 0.76 \\ 
  &  &  & & 0.4 & 0.78 & 0.88 \\ 
  &  &  & & 0.5 & 0.85 & 0.92 \\ 
  & 500 &  49 & 0.5 & 0.1 & 1.00 & 1.00 \\ 
  & &  & & 0.2 & 1.00 & 1.00 \\ 
  & &  & & 0.3 & 0.98 & 0.98 \\ 
  & &  & & 0.4 & 0.92 & 0.97 \\ 
  & &  & & 0.5 & 0.94 & 0.98 \\ 
  & 5000 & 100 & 0.5 & 0.1 & 1.00 & 1.00 \\ 
  & & & & 0.2 & 1.00 & 1.00 \\ 
  & & & & 0.3 & 0.87 & 0.91 \\ 
  & & & & 0.4 & 0.98 & 1.00 \\ 
  & & & & 0.5 & 0.99 & 1.00 \\ 
  Small-world graph & 50 & 16 & 0.5 & 0.1 & 0.34 & 0.76 \\ 
  &  &  & & 0.2 & 0.20 & 0.43 \\ 
  &  &  & & 0.3 & 0.38 & 0.59 \\ 
  &  &  & & 0.4 & 0.54 & 0.69 \\ 
  &  &  & & 0.5 & 0.63 & 0.75 \\ 
  & 500 &  49 & 0.5 & 0.1 & 1.00 & 1.00 \\ 
  & &  & & 0.2 & 1.00 & 1.00 \\ 
  & &  & & 0.3 & 1.00 & 1.00 \\ 
  & &  & & 0.4 & 0.98 & 0.99 \\ 
  & &  & & 0.5 & 0.99 & 1.00 \\ 
  & 5000 & 100 & 0.5 & 0.1 & 1.00 & 1.00 \\ 
  & & & & 0.2 & 1.00 & 1.00 \\ 
  & & & & 0.3 & 0.99 & 0.99 \\ 
  & & & & 0.4 & 1.00 & 1.00 \\ 
  & & & & 0.5 & 1.00 & 1.00 \\ 
   \hline
\end{tabular}
\end{table*}

\begin{figure*}
\includegraphics[width=1\textwidth]{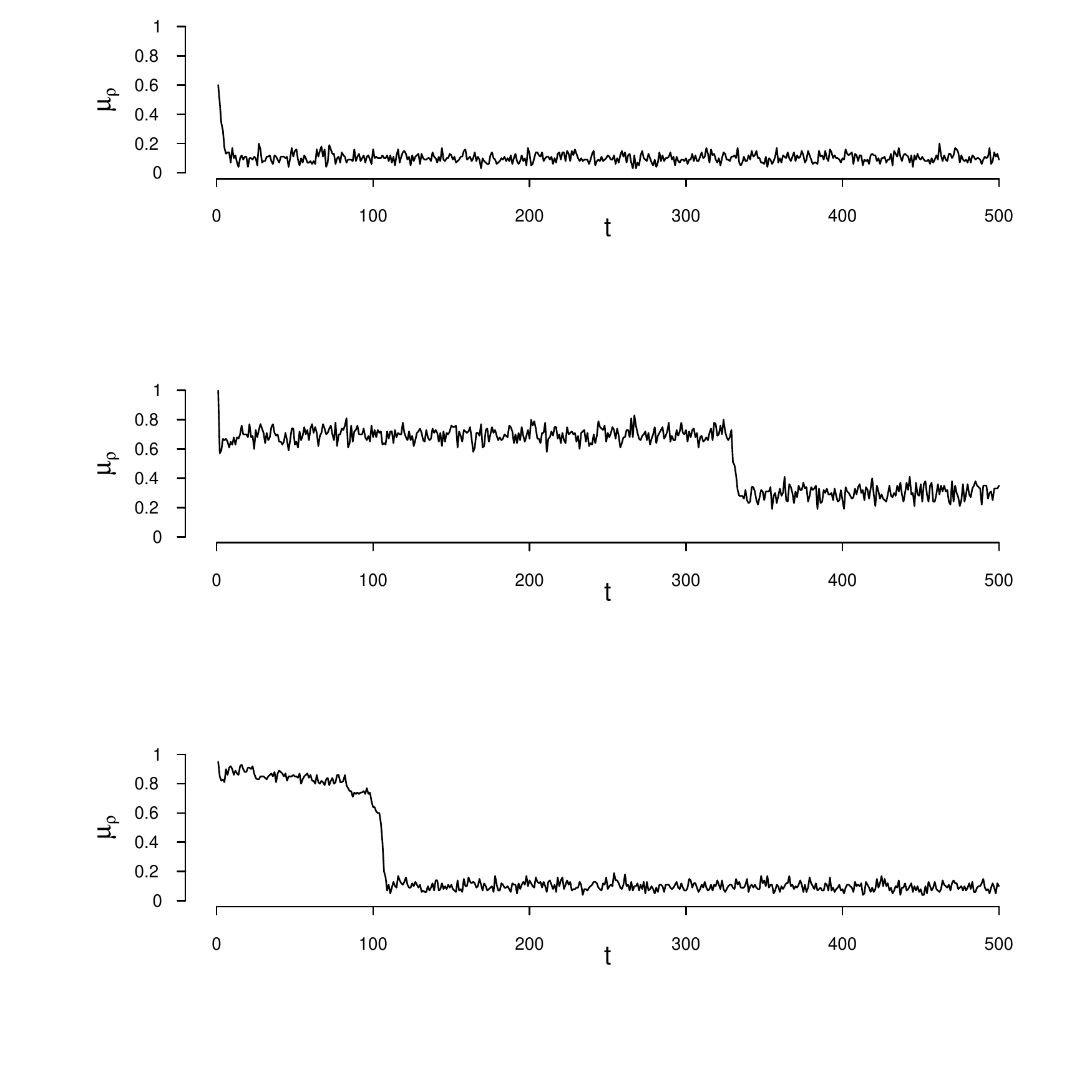}
\caption{Examples of the evolution of a torus (upper panel; $p = 0.1$), a random graph (middle panel; $p = 0.3, p_e = 0.6$) and a small world graph (lower panel; $p = 0.1, p_w = 0.9$).}
\label{fig:evolution-plots}
\end{figure*}

Figure \ref{fig:evolution-plots} visualizes the evolution of selected simulation conditions. Phase transitions were obtained for the random graph, specifically at $p_e = \left\{0.3, 0.4\right\}$ and $p = \left\{0.3, 0.4\right\}$. We also obtained phase transitions for the small world graph at $p_w = \left\{0.9\right\}$ and $p = \left\{0.1\right\}$. For the duration in our simulations, we did not obtain phase transitions in any other simulations. However, as we observed stable chains throughout the simulation study, we expect to obtain phase transitions in all conditions when $t$ approaches infinity.

\section{Conclusions and discussion}
To model the complex dynamics of large-scale networks (graphs) is in general difficult. This is because there are many different 'agents' that operate within the graph. In particular, if the nodes in the graph represent symptoms and the edges represent their mutual influence, then the interacting symptoms show complex behaviour on a macroscopic scale, e.g., at the level of the number of active symptoms. Here we showed that the mean field model for a probabilistic cellular automaton with majority rule, can serve as an accurate approximation to such large-scale graphs, and can simplify analysis of the dynamics of such systems. Specifically, we showed that averaging across the different possible degrees for a random and small-world graph, results in approximations that lie with high probability close to a generalised version of the mean field on a torus. These theoretical results were confirmed by extensive simulations, showing correspondence between the mean field and simulated graph activity for different size graphs.

Our approximation is based on the formulation of the grid (torus) where a relatively simple sum over possible active nodes determines the probability of a randomly selected node in the graph being active. We showed that for large graphs this approximation is accurate. This simplification could serve to obtain a more extensive analysis of the dynamics such as that presented in \citet{Janson:2015}. There the majority rule (the probabilistic element) was removed from the model, to obtain exact fixed points for the model. Here we chose not to remove the probabilistic element since we aim to introduce different rules for updates than the majority rule, like a conditional Ising probability.

Our initial motivation for these results was to obtain a model where we could assess the risk of a single person based on the estimate of the graph and the corresponding probability of an active node $p$, to determine the risk of that person 'jumping' from one state into another. This risk assessment might be useful in a clinical setting where a decision in a particular type of intervention is required. This idea is pursued in the companion paper in this issue.

\section*{Appendix}
\begin{proof}(Lemma \ref{lem:chernov})
Let the Kullback-Leibler divergence between $p+\varepsilon$ and $p$ be defined as a function of $0\le \varepsilon \le p$
\begin{align*}
h_{+}(\varepsilon) &= (p+\varepsilon)\log\frac{p+\varepsilon}{p} \\
&+ (1-p-\varepsilon)\log\frac{1-p-\varepsilon}{1-p}
\end{align*}
and similarly, define $h_{-}(\varepsilon)=h_{+}(-\varepsilon)$. Then Chernov's bound \citep{Lesigne:2005,Venkatesh:2013} for the density $\rho_{t}$ of a grid with $n$ nodes and its mean at time $t$, $p_{{\rm grid}}(\rho_{t})$ defined in (\ref{eq:pr1}), for $0<\varepsilon<\min\{p_{\rm grid},1-p_{\rm grid}\}$ immediately gives 
\begin{align*}
&\Prob(|\rho_{t}-p_{{\rm grid}}(\rho_{t})|>\varepsilon)\le\\
& \exp(-nh_{+}(\varepsilon)) + \exp(-nh_{-}(\varepsilon))
\end{align*}
The Kullback-Leibler divergence can be approximated quadratically by 
\begin{align*}
h_{+}(\varepsilon) = \frac{\varepsilon^{2}}{2p(1-p)}+O(\varepsilon^{3})\quad \text{as } \varepsilon\to 0
\end{align*}
Similarly for $h_{-}(\varepsilon)$ gives
\begin{align}
\Prob(|\rho_{t}-p_{{\rm grid}}(\rho_{t})|>\varepsilon) \le 2\exp(-\varepsilon^{2}/2\sigma^{2}_{{\rm grid}}(\rho_{t}))
\end{align}
where $\sigma^{2}_{\rm grid}=p_{\rm grid}(1-p_{\rm grid})/n$. Let $\delta=2\exp(-\varepsilon^{2}/2\sigma^{2}_{{\rm grid}}(\rho_{t}))$ such that $\varepsilon = \sqrt{2\sigma^{2}_{\rm grid}\log(2/\delta)}$. Then we obtain the result with probability at least $1-\delta$.
\end{proof}
\begin{proof}(Lemma \ref{lem:prg})
By Theorem 2.1 in \citet{Balister:2006} we have that the probability of state 1 at time $t+1$ on the event $\{|\Gamma|=k\}$ and given $\rho_{t}$ is
\begin{align*}
&\Prob(\Phi_{t}(x)=1\mid |\Gamma|=k,\rho_{t}) =\\
& \sum_{r=0}^{k}\xi_{k}(r)\binom{k}{r}\rho_{t}^{r}(1-\rho_{t})^{k-r}
\end{align*}
and in the random graph of Erd\"{o}s-Renyi the probability of $k$ neighbours for any node is 
\begin{align*}
\Prob_{p_{e}}(|\Gamma|=k) = \binom{n-1}{k}p_{e}^{k}(1-p_{e})^{n-k-1}
\end{align*}
It follows that the marginal $\sum_{k}\Prob(\Phi_{t}(x)=1\mid |\Gamma|=k,\rho_{t})\Prob(|\Gamma|=k)$ is
\begin{align*}
&p_{\rm rg}(\rho_{t}) =\\ 
&\sum_{k=0}^{n-1}\sum_{r=0}^{k}\xi_{k}(r)\binom{k}{r}\rho_{t}^{r}(1-\rho_{t})^{k-r}\Prob_{p_{e}}(|\Gamma|=k)
\end{align*}
as claimed.
\end{proof}

\begin{proof}(Equation \ref{eq:pr-grid-pe}) To obtain (\ref{eq:pr-grid-pe}) assume a fixed value $\nu$ for all $k$, $\xi_{k}=\xi_{\nu}$ for all $k$. Then $\xi_{\nu}(r)$ only depends on $r$. 
First note that 
\begin{align*}
\binom{k}{r}\binom{n-1}{k}=\binom{n-1}{r}\binom{n-r-1}{k-r}
\end{align*}

Second, observe that the order of the two sums in $p_{\Phi,G}$ can be switched since the first sum is for the sequence $(k=0,1,\ldots, n-1)$ and the second is $(r=0,1,\ldots, k)$, which can be seen as the lower triangular of the two-dimensional array for $(k=0,1,\ldots, n-1)$ and $(r=0,1,\ldots, n-1)$.  Upon switching to $(r=0,1,\ldots,n-1)$ and $(k=r,r+1,\ldots, n-1)$, we obtain the upper triangle of the array. 
Second, by changing the order of summation and reordering the sums, we get
\begin{align*}
&\sum_{r=0}^{n-1}\xi_{\nu}(r)\binom{n-1}{r}\rho_{t}^{r}p_{e}^{r} \sum_{k=r}^{n-r-1}\binom{n-r-1}{k-r} \\
&\times (p_{e}(1-\rho_{t}))^{k-r}(1-p_{e})^{n-k-1}
\end{align*}
In the sum on the right we can use the binomial theorem with $m=k-r$ and $N=n-r-1$, which gives
\begin{align*}
&\sum_{m=0}^{N}\binom{N}{m}(p_{e}(1-\rho_{t}))^{m}(1-p_{e})^{N-m}=\\
&\quad (p_{e}(1-\rho_{t})+1-p_{e})^{N}
\end{align*}
which leads to (\ref{eq:pr-grid-pe}). 

For the approximation error, write $\xi_{k}(r)=p\mathbbm{1}\{r\le k/2\}+(1-p)\mathbbm{1}\{r>k/2\}$ and recall that $\nu$ is fixed. Then
\begin{align*}
 &p_{\rm rg}(\rho_{t})-p_{\rm rand}(\rho_{t})=\\
 &\quad\sum_{k=0}^{n-1}\sum_{r=0}^{k}\binom{k}{r}\rho_{t}^{r}(1-\rho_{t})^{k-r}\binom{n-1}{k} \\
 &\quad \times p_{e}^{k}(1-p_{e})^{n-k-1}[\xi_{k}(r)-\xi_{\nu}(r)]
\end{align*}
Using H\"{o}lder's inequality with the $\ell_{\infty}$ and $\ell_{1}$ norms, gives
\begin{align*}
&|p_{\rm rg}(\rho_{t})-p_{\rm rand}(\rho_{t})|\\
\le &\sum_{k=0}^{n-1}\sum_{r=0}^{k}\binom{k}{r}\rho_{t}^{r}(1-\rho_{t})^{k-r}\binom{n-1}{k}\\
&\times p_{e}^{k}(1-p_{e})^{n-k-1}\max_{r,k}|\xi_{k}(r)-\xi_{\nu}(r)|
\end{align*}

The binomial theorem for the first term of the right hand side gives
\begin{align*}
 &\sum_{k=0}^{n-1}\sum_{r=0}^{k}\binom{k}{r}\rho_{t}^{r}(1-\rho_{t})^{k-r}\binom{n-1}{k}p_{e}^{k}(1-p_{e})^{n-k-1}\\
 = &\sum_{r=0}^{n-1}\binom{n-1}{r}(\rho_{t}p_{e})^{r}(1-\rho_{t}p_{e})^{n-r-1}=1.
\end{align*}
For each $r,k$ such that $r\le k$ we have that 
\begin{align*}
&\xi_{k}(r)-\xi_{\nu}(r)=\\
&\quad p(\mathbbm{1}\{r\le k/2\}-\mathbbm{1}\{r\le \nu/2\})+\\
&\quad (1-p)(\mathbbm{1}\{r>k/2\}-\mathbbm{1}\{r>\nu/2\})
\end{align*}
The term $|\xi_{k}(r)-\xi_{\nu}(r)|$ is at most $2p-1$ if $\nu<k$ or $1-2p$ if $\nu \ge k$ for any $r,k$, which gives the size of the error bound. 
\end{proof}
%
%
\begin{proof}(Lemma \ref{lem:prg-ext})
If we fix $\nu=\lfloor p_{e}(n-1)\rfloor$, the expectation of the random variable for each node of the possible number of neighbours $B(n-1,p_{e})$, such that each $k=\nu$ in the part for the density we obtain 
\begin{align*}
&\sum_{k=0}^{n-1}\binom{n-1}{k}p_{e}^{k}(1-p_{e})^{n-k-1}\\
&\times \left( \sum_{r=0}^{\nu}\binom{\nu}{r}\xi_{\nu}(r)\rho_{t}^{r}(1-\rho_{t})^{\nu-r} \right),
\end{align*}
from which we obtain $p_{\rm grid}^{\nu}(\rho_{t})$. The approximation error for the probabilities is then
\begin{align*}
& |p_{\rm rg}(\rho_{t})-p_{\rm rand}^{\nu}(\rho_{t})|=\\
 &\left|\sum_{k=0}^{n-1}( p_{\rm grid}^{k}(\rho_{t})-p_{\rm grid}^{\nu}(\rho_{t}))\Prob_{p_{e}}(|\Gamma|=k)\right|.
\end{align*}
The probability of obtaining a neighbourhood size $k$ close to the expected number of neighbours $\nu=p_{e}(n-1)$ can be obtained from the Chernov bound in Lemma \ref{lem:chernov}, giving $\Prob(|k-\nu|\le t) \ge 1-2\exp(-(n-1) \varepsilon^{2}/p_{e}(1-p_{e}))$, for $\varepsilon\searrow 0$. This leads to the difference being bound by 
\begin{align*}
 &|p_{\rm rg}(\rho_{t})-p_{\rm rand}^{\nu}(\rho_{t})| \le\\
  &\biggl| \sum_{k=0}^{n-1}( p_{\rm grid}^{k}(\rho_{t})-p_{\rm grid}^{\nu}(\rho_{t}))\\
  	&\times 2\exp(-(n-1) \varepsilon^{2}/p_{e}(1-p_{e}))\biggr|.
\end{align*}
Using H\"{o}lder's inequality with the sup and $\ell_{1}$ norms, we find that the above is
\begin{align*}
 \le &\max_{k} |p_{\rm grid}^{k}(\rho_{t})-p_{\rm grid}^{\nu}(\rho_{t})|\\
 &\times  \sum_{k=0}^{n-1}2\exp(-(n-1) \varepsilon^{2}/p_{e}(1-p_{e})).
\end{align*}
The difference $p_{\rm grid}^{k}(\rho_{t})-p_{\rm grid}^{\nu}(\rho_{t})$ is determined by the mismatch between $k$ and $\nu$ and is at most $2p-1$ if $\nu<k$ and $1-2p$ if $\nu \ge k$ for any $r,k$. And so we obtain
\begin{align*}
&|p_{\rm rg}(\rho_{t})-p_{\rm grid}^{\nu}(\rho_{t})|\le  |p-1/2|\\
&\times 2\exp(-(n-1) \varepsilon^{2}/p_{e}(1-p_{e})+\log(n)),
\end{align*}
completing the proof.
\end{proof}
\begin{proof}(Theorem \ref{thm:evo-rg})
In the random graph $G_{\rm rg}(n,p_{e})$ each node has the same number $n-1$ of possible neighbours. Hence, for each node the local rule $\phi$ is determined by $r$ out of $n-1$ possible neighbours being both connected and 1. Lemma \ref{lem:prg} shows that the probability of this binomial process is then $p_{\rm rg}$ in (\ref{eq:pr-rg}). Then by Lemma \ref{lem:prg-ext} for large $n$ the probability $p_{\rm grid}^{\nu}$ converges to $p_{\rm rg}$, and by consequence, the number of active nodes in both processes converges. 
\end{proof}


%

\end{document}